 \documentclass[preprint2,twoside]{myaastex}

\usepackage{amssymb}                       
\usepackage{amsmath}                       
\usepackage{wasysym}
\newcommand{\doverd}[2]{\frac{\partial #1}{\partial #2}}

\newcommand{\superscr}[1]{^\mathrm{#1}}
\newcommand{\gv}[1]{\boldsymbol{#1}}      
\newcommand{\refEqt}[1]{Equation~(\ref{#1})}   
\newcommand{\refeqt}[1]{equation~(\ref{#1})}   

\newcommand{\refeqp}[1]{eq.~[\ref{#1}]}        
\newcommand{\refFgt}[1]{Figure~\ref{#1}}       
\newcommand{\refFgp}[1]{Fig.~\ref{#1}}         
\newcommand{\refTab}[1]{Table~\ref{#1}}        
\newcommand{\refSec}[1]{Section~\ref{#1}}      
\newcommand{\refsec}[1]{\S~\ref{#1}}           
\newcommand{\KDH}{\textsc{Paper~I}}            
\newcommand{\DHK}{\textsc{Paper~II}}           
\newcommand{\MSun}{M_{\odot}}                  
\newcommand{\MEarth}{M_{\oplus}}               
\newcommand{\MJup}{M_{\jupiter}}               
\newcommand{\Mp}{M_\mathrm{p}}                 
\newcommand{\dMp}{\dot{M}_\mathrm{p}}          
\newcommand{\taum}{\tau_\mathrm{M}}            
\newcommand{\Se}{S_\mathrm{p}}                 
\newcommand{\RH}{R_\mathrm{H}}                 
\newcommand{\AU}{\mbox{\textrm{AU}}}
\newcommand{\Msyr}{\mbox{$\MSun\,\mathrm{yr}\superscr{-1}$}}
\newcommand{\dunits}{\mbox{$\mathrm{g}\,\mathrm{cm}\superscr{-3}$}}
\received{year Month day}
\revised{year Month day}
\accepted{year Month day}
\ccc{code}
\cpright{none}{73 AC}

\slugcomment{\small{December 2002}}

\shorttitle{Migration and Accretion of Protoplanets}
\shortauthors{D'Angelo, Kley, \& Henning}

\begin{document}


\title{\textbf{%
       Orbital Migration and Mass Accretion of Protoplanets\\
       in 3D Global Computations with Nested Grids\footnote{%
       To appear in \textsc{The Astrophysical Journal} (v586 n1 March 20,
       2003 issue).
       Also available as ApJ preprint doi: 
       \texttt{10.1086/367555}.}}}


\author{\textsc{Gennaro D'Angelo and Willy Kley}}
\affil{Computational Physics,
       Auf der Morgenstelle 10,
       D-72076 T\"ubingen, Germany}
\email{gennaro@tat.physik.uni-tuebingen.de, 
         wilhelm.kley@uni-tuebingen.de\\[2mm]}
\and
\author{\vspace*{-5mm}\textsc{Thomas Henning}}
\affil{Max-Planck-Institut f\"ur Astronomie,
    K\"onigstuhl 17,
    D-69121 Heidelberg,
    Germany}
\email{henning@mpia-hd.mpg.de\\[-5mm]}




\small

\begin{abstract}
We investigate the evolution of protoplanets with different masses
embedded in an accretion disk, via global 
fully three-dimensional hydrodynamical simulations. 
We consider a range of planetary masses extending from one and a half 
Earth's masses up to one Jupiter's mass, and we take into account
physically realistic gravitational potentials of forming planets.
In order to calculate accurately the gravitational torques exerted 
by disk material and to investigate the accretion process onto the 
planet, the flow dynamics has to be thoroughly resolved on long
as well as short length scales.
We achieve this strict resolution requirement by applying a 
\textit{nested-grid}
refinement technique which allows to greatly enhance the local resolution.
Our results from altogether 51 simulations show that for large
planetary masses, approximately above a tenth of the Jupiter's mass, migration 
rates are relatively constant, as expected in \textit{type~II} migration
regime and in good agreement with previous two-dimensional calculations.
In a range between seven and fifteen Earth's masses, we find a
dependency of the migration speed on the planetary mass that yields
time scales considerably longer than those predicted by linear
analytical theories.
This property may be important in determining the overall orbital evolution of
protoplanets.
The growth time scale is minimum around twenty Earth-masses, but it rapidly
increases for both smaller and larger mass values.
Significant differences between two- and three-dimensional calculations are 
found in particular for objects with masses smaller than ten Earth-masses.
We also derive an analytical approximation for the numerically computed mass 
growth rates.
\end{abstract}




\section{Introduction}
\label{introduction}

Over the past few years the interest of the astronomical community in 
the enigma of planet formation and evolution has been rising as the number 
of observed extrasolar planets has continued to increase.
Today about 100 extrasolar planets are known, mostly orbiting around 
main-sequence stars of solar type.
An always up-to-date status of extrasolar planet detections can be found
at the \textit{Extrasolar Planet Encyclopedia}%
\footnote{\url{http://www.obspm.fr/planets}.}, maintained by J.~Schneider,
or at the \textit{California \& Carnegie Planet Search}%
\footnote{\url{http://exoplanets.org/}.}.

Several of the orbital and physical key properties of planets (e.g., location,
eccentricity, rotation rate, mass) are believed
to originate from the early phases of planet formation, when
the protoplanet is still embedded in the surrounding protostellar 
disk from which it generated.
In particular, the small semi-major axis of several (51 Peg-type)
planets is usually interpreted as a migration process produced by
gravitational torques of the disk material acting on the protoplanets.
In order
to properly take these effects into account, one has to consider
the joint evolution of the circumstellar disk and the embedded planet.

Following the first and mostly linear analytical studies 
\citep{gt1980,papa1984,ward1986}, 
fully non-linear numerical simulations have been performed lately
\citep{kley1999,lubow1999,rnelson2000,papa2001,kley2001,gennaro2002,
tanigawa2002}
in order to achieve a deeper insight into the physical processes governing 
the interactions between a protostellar disk and an embedded protoplanet.
Among the main discoveries we can cite: 
\textit{i}) 
the creation of spiral density perturbations in the disk;
\textit{ii}) 
the formation of a deep annular gap along the orbit of Jupiter-mass planets; 
\textit{iii}) 
an inward migration resulting from the net gravitational torques caused
by inner and outer density wave perturbations;
\textit{iv}) 
the continuation of mass accretion through the gap.

So far most of the computations have investigated mainly
the effects due to large-scale interactions, much larger than the size of
the Roche lobe of the forming planet.
Furthermore, very little is known about the non-linear
effects that such interactions have when low-mass planets are involved.
The reason for this has been primarily the lack of appropriate numerical 
tools.
\pagestyle{myheadings}
\markboth{\hfill Migration and Accretion of Protoplanets \hfill}%
       {\hfill \textsc{G. D'Angelo, W. Kley, \& Th. Henning} \hfill}

In the majority of the previous studies, the disk is modeled as a 
two-dimensional ($r$--$\varphi$) system, by using vertically-averaged 
quantities. Two main arguments lie behind this choice.
First,
on a physical basis, the validity of a two-dimensional (2D) description 
is consistent because the Hill radius of a massive object is larger or
comparable to the disk semi-thickness.
In fact, this basically means that the sphere of gravitational influence 
of the embedded body, i.e., the Hill sphere\footnote{%
The Hill sphere is a measure of the volume of the Roche lobe.
In a purely restricted three-body problem, parcels of matter inside 
the Roche lobe are gravitationally bound to the secondary. Hence, 
they are confined to that volume of space.},
contains the whole vertical extent of the disk.
But this usually implies that the planet must have a mass on the order of 
one Jupiter-mass. 
Second, 
a less massive planet has a weaker impact on the disk, requiring a higher 
resolution to compute properly and highlight its effects. Such requirement 
typically rules out a full three-dimensional treatment. 
Although there is still a lot of information to be gained by performing 
2D simulations (e.g., to study radiative effects or multi-planet systems)
in particular in the case of large and medium mass planets, provided that 
the local resolution around the planet is accurate enough, three-dimensional 
(3D) effects become more and more important as the mass of the simulated
planet is reduced.

Yet, in many instances, the use of the two-dimensional approximation 
is merely dictated by the computational costs of 3D calculations which are 
generally not affordable.
Depending on the resolution one is interested in, 3D runs 
still take more than an order of magnitude of CPU time with respect to
2D runs.
As a proof of the severe limitations posed by fully three-dimensional 
calculations, only very few papers have been published on this issue.

\citet{miyoshi1999} made a comparison study of 2D and 3D disks within  
the framework of the shearing sheet model. Hence, rather than
considering the whole disk, they were restricted to local simulations.
Global three-dimensional simulations were performed for the first time
by \citet[hereafter \KDH]{kley2001} who also measured the gas accretion rate 
onto the planet and the gravitational torques which cause the planet 
to alter its orbit. 
They found that for planetary masses below one half of Jupiter's mass,
the outcomes of 3D calculations start to differ from those of 2D ones.
They also pointed out that to obtain more reliable results, the flow
within the Roche lobe needs to be accurately resolved. 
This was done recently, for an infinitesimally thin disk, by 
\citet[hereafter \DHK]{gennaro2002} who introduced,
for the first time in this context, a \textit{nested-grid} technique in order 
to model in detail a variety of planetary masses, spanning from Earth's to
Jupiter's.
The authors proved such approach to adapt comfortably to these computations 
because global-scale structures as well as small- and very small-scale features
of the flow can be captured simultaneously. 
They demonstrated that disks form around high- and low-mass 
planets and that circumplanetary material can exert very strong torques
on the planet, usually slowing down their inward drifting motion.

In the present paper we intend to combine the fully three-dimensional and 
global treatment of disk-planet interactions with a nested-grid refinement 
technique in order to carry out an extensive study on migration, accretion, 
and flow features around large- and small-size protoplanets.
Thus, the paper comes as an extension to \KDH\ and \DHK.
In addition, here we abandon the standard approach of treating the planet
as a point-mass but rather assume that it has an extended structure.

The outline of the paper is as follows. \refSec{sec:PD} deals with those 
aspects of the physical description that we adopt and which were not already 
specified in \KDH.
We explain how we approximate the protoplanet's structure by using different 
solutions for the gravitational potential.
\refSec{sec:NI} presents a brief overview about the numerical procedures 
employed in this work and describes the technical details of the models.
As for the implementation of the nested grids in three dimensions, 
for brevity we mainly refer to the two-dimensional strategy traced in \DHK.
The various results of our simulations are addressed in \refsec{sec:SR}.
Fluid circulation, gravitational torques, orbital migration, mass accretion 
rates, and how all of them depend on the perturber mass are examined.
A comparison between 2D and 3D models is also carried out, together with
an analysis of some numerical effects.
In \refsec{sec:D} two issues related to 2D and 3D geometry effects are 
discussed in more detail. We finally present our conclusions in \refsec{sec:C}.
\section{Physical Description}
\label{sec:PD}

The nature of most astrophysical objects is such that their behavior can be 
approximated to that of \textit{fluids}. 
This is indeed the case for circumstellar disks, hence we can rely on the
hydrodynamic formalism to describe them.
The equations of motion that govern the evolution of a disk in
a spherical polar coordinate system $\{O; R, \theta, \varphi\}$ are presented
in \KDH\ and therefore, for the sake of brevity, 
we refer the interested reader to it. 
We assume the disk to be a viscous medium and include the 
viscosity terms explicitly by employing a complete stress tensor 
for Newtonian fluids \citep[see e.g.,][Chapter~3]{m&m}.

The set of equations for the hydrodynamic variables
$(\rho, u_R, u_\theta, u_\varphi)$
is written with respect to a reference frame rotating at a
constant rate $\Omega$, around the polar axis $\theta=0$, and whose origin $O$ 
resides in the center of mass of the star--planet system.
The planet is maintained on a fixed circular orbit, lying in the 
midplane of the disk ($\theta=\pi/2$). 
If we let $\Omega$ coincide with the 
angular velocity of the planet $\Omega_\mathrm{p}$, the planet does not move 
within the reference frame.
The assumption that a single protoplanet, not heavier than Jupiter, moves on a 
circular orbit is reasonable because the global effect of the resonances,
arising from disk--planet interactions,
in most of the cases favors an eccentricity damping 
\citep{papa2001,agnor2002}.

Disk material evolves under the combined gravitational action of a star
and a massive body. 
In fact, as long as the inner parts of low-mass protostellar disks 
are concerned, self-gravity can be neglected.
Indicating with $\gv{R}_{\bigstar}$ the radius vector pointing to the position
of the star, the gravitational potential $\Phi$ of the whole system is 
represented by
\begin{equation}
 \Phi = - \frac{G\,M_\bigstar}{| \gv{R} - \gv{R}_{\bigstar} |}
        + \Phi_\mathrm{p},
      \label{phit}
\end{equation}
where $M_\bigstar$ is the stellar mass.
In \refeqt{phit}, the function $\Phi_\mathrm{p}$ identifies the perturbing 
potential of the planet, which we leave unspecified for the moment. 

Since the energy equation is not considered in the present work, we join an
established trend \citep[e.g.,][]{kley1999,lubow1999,miyoshi1999,%
rnelson2000,papa2001,tanaka2002,masset2002,tanigawa2002}
and use a locally isothermal equation of state as closure 
of the hydrodynamic equations
\begin{equation}
 p = c_\mathrm{s}^{2}\,\rho,
      \label{p}
\end{equation}
where the sound speed $c_\mathrm{s}$ equals the Keplerian velocity 
$v_\mathrm{K}$ times the disk aspect ratio $h=H/(R\,\sin{\theta})$.
The length $H$ is the pressure scale-height of the disk, that also
represent its semi-thickness. 
As the ratio $h$ is assumed to be constant,
the disk is azimuthally and vertically isothermal, whereas radially
$T\propto h^2 M_\bigstar/(R\,\sin{\theta})$.
This simplified approach permits to circumvent the difficulties posed by the 
solution of a complete energy equation which nobody has tackled yet. 
In fact this kind of computations would require a length of time which is
presently not affordable. As reference, even without including energetic 
aspects, the CPU-time consumed by our three-dimensional global simulations is 
already between 
ten and twenty times as long as that spent by two-dimensional ones.
An investigation into the effects that may arise in two-dimensional disks 
when an energy equation is also taken into account, 
will be presented in a forthcoming paper.

However, an important issue to improve the physical description of the system 
in the vicinity of the protoplanet is to adopt an appropriate equation of 
state which can account for the protoplanetary envelope. Yet, in our case
this would imply that either $p$ or $\rho$ should be specified in some volume
around the planet.
In order to avoid this, we choose to constrain the local structure by means of
suitable analytic expressions for $\Phi_\mathrm{p}$.
We assume that the protoplanet has a measurable size, i.e., 
it can be resolved by the employed computational mesh.
Within the planetary volume, we approximately take into account the
effects due to self-gravity by imposing a certain gravitational field. 
Since we aim at covering various possible scenarios, we utilize four different 
forms of planet gravitational potential, each representing a protoplanet with 
different characteristics.
It is worthwhile to point out that with this choice none of the hydrodynamic 
variables
$(\rho, u_R, u_\theta, u_\varphi)$ is prescribed in any case. They simply 
evolve in a particular gravitational field. Therefore, planetary 
material is allowed to interact with the surrounding environment so that their 
mutual evolutions are still connected. 

\subsection{Planet Gravitational Potential}
\label{ssec:PGP}
With no exception, both numerical and analytical work that have so far 
investigated the interactions between massive bodies and protostellar 
disks have made the \textit{point-mass} assumption, i.e., the protoplanet has 
a finite mass $\Mp$ but no physical size, as was done in \KDH\ and II.
This property is expressed through the gravitational potential
\begin{equation}
 \Phi^\mathrm{PM}_\mathrm{p} = - \frac{G\,\Mp}%
                                {| \gv{R} - \gv{R}_\mathrm{p} |},
                             \label{phipm}
\end{equation}
where $\gv{R}_\mathrm{p}$ is the radius vector indicating the position of
the planet.

Because of the singularity at $\gv{R}=\gv{R}_\mathrm{p}$, a parameter
$\varepsilon$ is introduced in order to smooth the function over a certain
region. 
If we denote $\gv{S} = \gv{R} - \gv{R}_\mathrm{p}$ the position vector
relative to the planet, the smoothed point-mass potential can be written in 
the following form
\begin{equation}
 \Phi^\mathrm{PM}_\mathrm{p} = - \frac{G\,\Mp}{\varepsilon}%
                     \left[1 + \left(\frac{S}{\varepsilon}\right)^2
                     \right]^{-\frac{1}{2}}.
                 \label{phipms}
\end{equation}
A physical meaning of the smoothing length can be deduced from \refeqt{phipms}.
The potential $\Phi_\mathrm{p}$ enters the Navier-Stokes equations 
through its derivatives, which can be reduced to 
$\partial \Phi_\mathrm{p}/\partial S$
because of the spherical symmetry of the gravitational field. 
Restricting to distances $S < \varepsilon$, a binomial expansion of that
derivative yields
\begin{eqnarray}
   \doverd{\Phi^\mathrm{PM}_\mathrm{p}}{S}%
                     &\simeq& \frac{G\,\Mp}{\varepsilon^2}%
                              \left(\frac{S}{\varepsilon}\right)%
                     \left[1 - \frac{3}{2}\,\left(\frac{S}{\varepsilon}\right)^2
                     \right] \nonumber \\
                     &\approx& \frac{G\,\Mp}{\varepsilon^2}%
                              \left(\frac{S}{\varepsilon}\right).
                 \label{dphids}
\end{eqnarray}

As we will see in \refsec{sssec:HSS}, \refeqt{dphids} can be interpreted as the
sign-reversed gravitational force, per unit mass, exerted by a spherically 
homogeneous medium of radius $\varepsilon$. 
Thus, the smoothing may act as an indicator of the size of the planet.

\begin{figure*}[!t]
\epsscale{2.0}
\plottwo{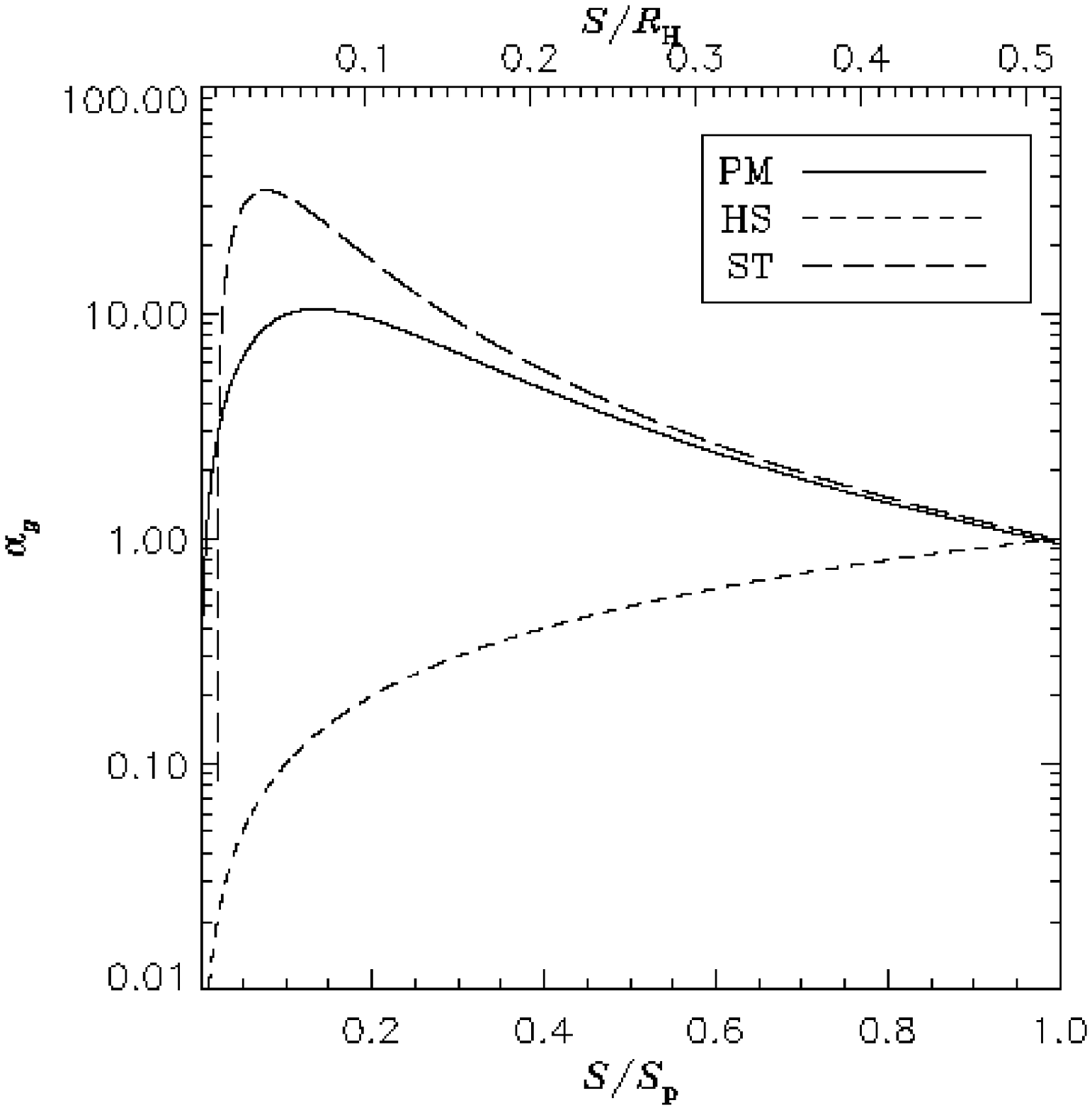}{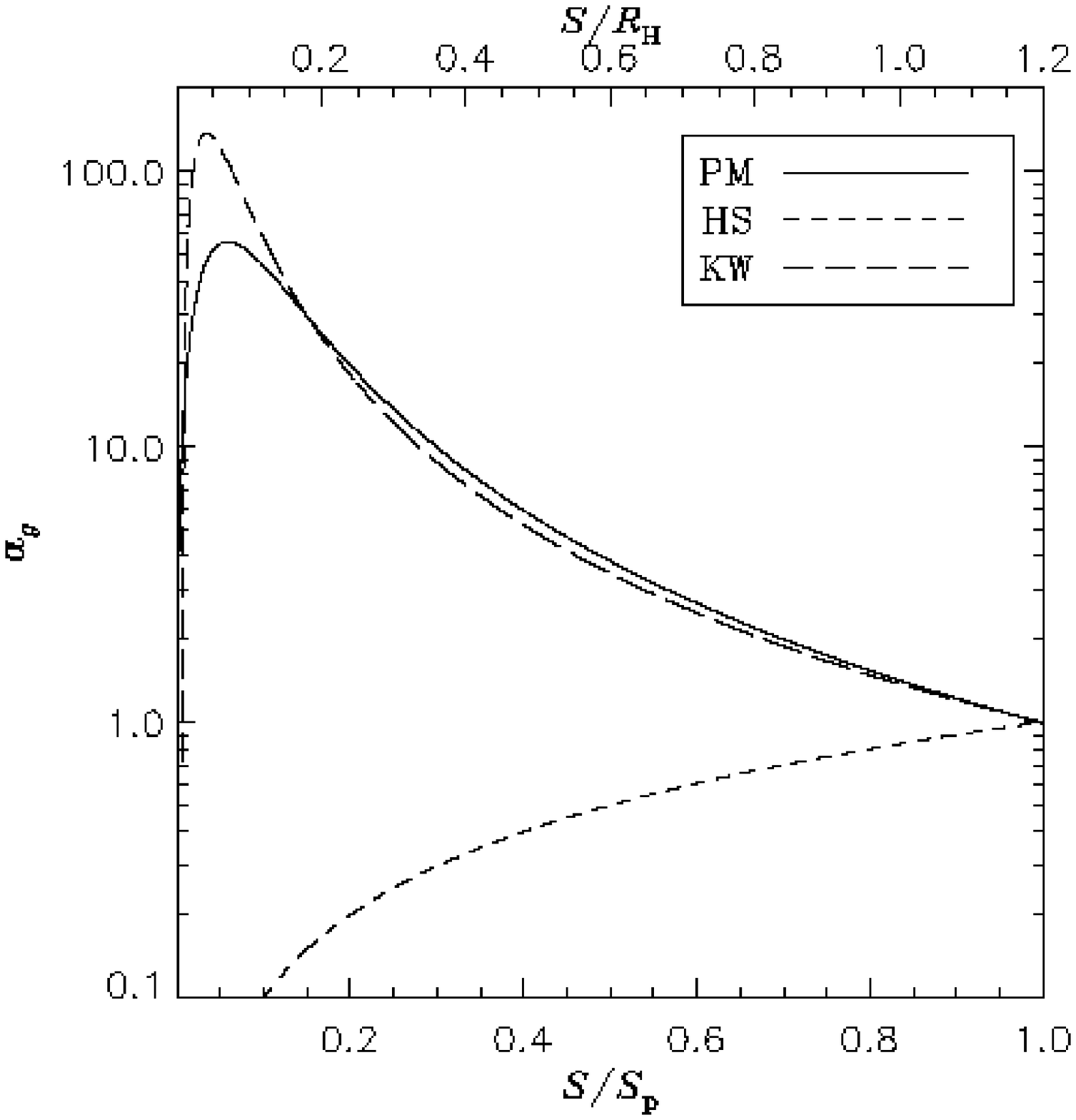}
\caption{%
\textbf{Left panel.} %
Gravitational acceleration ($-\partial \Phi_\mathrm{p}/\partial S$) 
inside a $20\,\MEarth$ planet as generated by the three potential functions 
$\Phi^\mathrm{PM}_\mathrm{p}$, 
$\Phi^\mathrm{HS}_\mathrm{p}$, and
$\Phi^\mathrm{ST}_\mathrm{p}$ 
(eqs.~[\ref{phipms}], [\ref{phihs}], and [\ref{phist}],
respectively). 
Accelerations are normalized to $-G\,\Mp/\Se^2$, where $\Se$ is the envelope
radius.
The core mass (see \refsec{sssec:SS}) is $15\,\MEarth$ while $\Se=0.52\,\RH$. 
The core is supposed to have a density of $5.5\;\dunits$. 
The point-mass potential is smoothed over a length $\varepsilon = 0.1\,\RH$ 
whereas in the Stevenson's potential $\delta = 5\times10^{-2}\,\RH$.
\textbf{Right panel.} %
The same quantity is displayed inside a $90\,\MEarth$ planet but this time 
involving the potential solution $\Phi^\mathrm{KW}_\mathrm{p}$ 
(\refeqp{phikw}). 
The adiabatic
exponent is $\Gamma=1.43$ while the core mass is $60\,\MEarth$.
In all circumstances, the gravitational potential outside the envelope radius
is of the type given in \refeqt{phipms}.
\label{fig:ag}}
\end{figure*}

\refEqt{phipms} may be appropriate to describe the solid core of a protoplanet 
which does not possess any significant envelope.
Indicating with $q$ the planet-to-star mass ratio $\Mp/M_\bigstar$, in these
computations we apply the potential $\Phi^\mathrm{PM}_\mathrm{p}$ in models 
with $q > 3 \times 10^{-5}$, i.e., ten times the Earth's mass ($\MEarth$)
when $M_\bigstar= 1\;\MSun$.
In all cases the smoothing parameter $\varepsilon$ is set to 10\% of 
the Hill radius $\RH$\footnote{The radius of the Hill sphere is 
$\RH=a\,\left(q/3\right)^{1/3}$, where $a$ is the distance of the star from
the planet.
Considering only the linear terms in $q$, this length is 
equal to the distance from the inner Lagrangian point (L1) to the secondary.
Although we named $S$ the distance from the planet, we preferred to keep a
more \textit{familiar} notation to indicate the Hill radius.}.
Along with \refeqt{phipms}, we introduce three alternative expressions 
for $\Phi_\mathrm{p}$, namely the potential of a homogeneous sphere
($\Phi^\mathrm{HS}_\mathrm{p}$); that describing a fully radiative and static
envelope ($\Phi^\mathrm{ST}_\mathrm{p}$); and finally that proper for a
fully convective and static envelope ($\Phi^\mathrm{KW}_\mathrm{p}$). 
A comparative example of their behavior is reported in \refFgt{fig:ag}.

Among the last three solutions, only the gravitational potential generated 
by a homogeneous
sphere has a completely different behavior from that given in \refeqt{phipms}, 
while $\Phi^\mathrm{ST}_\mathrm{p}$ as well as $\Phi^\mathrm{KW}_\mathrm{p}$ 
compare better to a smoothed point-mass potential. This is clearly
seen in \refFgt{fig:ag} and formally shown in \refsec{sssec:SS} and
\refsec{sssec:WS}. 
Thereupon, one should expect that 
computation results should depend only marginally on the adopted 
gravitational potential, except when 
$\Phi_\mathrm{p}=\Phi^\mathrm{HS}_\mathrm{p}$.

\subsubsection{Homogeneous Sphere Solution}
\label{sssec:HSS}
The gravitational potential generated by a homogeneous spherical distribution 
of matter can be calculated in a straightforward way by a direct integration 
of the gravitational force. Thereby, one finds 
\begin{equation}
 \Phi^\mathrm{HS}_\mathrm{p} = \left\{%
            \begin{array}{ll}
             - \frac{G\,\Mp}{2\,\Se}%
             \left[3 - \left(\frac{S}{\Se}\right)^2\right] &
             \, \mathrm{if}\, S \leq \Se               \\
                          &                              \\
             - \frac{G\,\Mp}{S}                        &
             \, \mathrm{if}\, S > \Se
            \end{array},
                           \right.
                 \label{phihs}
\end{equation}
where $\Se$ is the radius of the sphere, i.e., the planet's radius.
No smoothing is needed in this case since the force converges linearly 
to zero as the distance $S$ approaches zero. 
Thus, there is no risk of numerical instabilities.

Strictly speaking, \refeqt{phihs} is valid inside very extended and nearly 
homogeneous envelopes without considerable cores. Then, one may think 
of the functions $\Phi^\mathrm{PM}_\mathrm{p}$ and 
$\Phi^\mathrm{HS}_\mathrm{p}$ as rendering two opposite extreme situations.
Though $\Phi^\mathrm{HS}_\mathrm{p}$ does not represent a very realistic 
scenario of planet formation, for the sake of comparison and completeness,
we will apply this potential to high- as well as low-mass bodies.

\subsubsection{Stevenson's Solution}
\label{sssec:SS}

\citet{stevenson1982} proposed a simplified analytical model of protoplanets
having envelopes with constant opacity and surrounding an accreting solid 
core. He developed a \textit{radiative zero solution} for hydrostatic and 
fully radiative spherical envelopes, which implies that both hydrostatic and 
thermal equilibrium are assumed inside the planet's atmosphere. 
Under these hypotheses, the core can grow up to a \textit{critical mass} whose
value is that beyond which at least one of the two equilibriums ceases to exist
\citep[see discussion in][]{wuchterl1991} and the structure cannot be strictly 
static any longer.
The critical core mass also sets an upper limit to the envelope and total mass
of the planet. It can be proved that, at this critical point,
$M_\mathrm{c}/\Mp = 3/4$, where the mass of the core $M_\mathrm{c}$ is 
influenced by neither the nebula density nor its temperature.

The potential of the gravitational field established by a fully radiative 
envelope can be obtained from the density profile \citep[see][]{stevenson1982}
by applying the Poisson equation.
Since the solid core size is by far below the resolution limit of these 
computations, the form of the Stevenson's potential can be cast in the form
\begin{equation}
 \Phi^\mathrm{ST}_\mathrm{p} = \left\{%
            \begin{array}{ll}
             - \frac{G\,M_{\mathrm{c}}}{\sqrt{S^2 + \delta^2}}%
             - \frac{G\,M_{\mathrm{e}}^{\dag}}{\sqrt{S^2 + \delta^2}} &
             \\
             & \\
             \times\left[1 - \left(\frac{S}{\Se}\right)%
             + \ln{\left(\frac{S}{S_{\textrm{c}}}\right)}\right]
             &\, \mathrm{if}\, S \leq \Se                \\
                      &                              \\
             - \frac{G\,\Mp}{\sqrt{S^2 + \delta^2}}  &
             \, \mathrm{if}\, S > \Se
            \end{array}.
                           \right.
                 \label{phist}
\end{equation}

In \refeqt{phist} we have indicated with $S_\mathrm{c}$ the core radius. 
The quantity $M_\mathrm{e}^{\dag}$ is equal to the planet's envelope 
mass $M_\mathrm{e}=\Mp - M_\mathrm{c}$ divided by $\ln(\Se/S_{\textrm{c}})$. 
The presence of the parameter $\delta$, also in the solution valid outside
the envelope, is necessary for continuity reasons at $S=\Se$.
In these simulations we set $\delta=0.05\,\RH$.

If the core has a density $\rho_\mathrm{c}=5.5\;\dunits$ and accretes at the 
rate of $5\times10^{-7}\;\MEarth\,\mathrm{yr}^\mathrm{-1}$, assuming an
envelope opacity equal to $1\;\mathrm{cm}^2\,\mathrm{g}^{-1}$ and a mean 
molecular weight of $2.2$, the critical total mass is $36\;\MEarth$. 
Hence we will use \refeqt{phist} only for protoplanets whose total 
mass $\Mp$ is less than that value. Furthermore, we will suppose that the
ratio of the total planetary mass to the core mass is the critical one.
Therefore the core mass is always known once the mass $\Mp$ is assigned a 
value. Then, supplying $\rho_\mathrm{c}$, the radius $S_\mathrm{c}$ can be 
fixed.

The effects caused by \refeqt{phist} are actually similar to those
caused by \refeqt{phipms}, as seen in \refFgt{fig:ag}. In a more
formal way, inside of the sphere $S=\Se$, the normalized difference
$\mathcal{R}^\mathrm{ST}$ between the two gravitational fields can be
quantified by the ratio of 
$\left|\partial\Phi^\mathrm{ST}_\mathrm{p}/\partial S -
       \partial\Phi^\mathrm{PM}_\mathrm{p}/\partial S\right|$ to 
$\partial\Phi^\mathrm{PM}_\mathrm{p}/\partial S$, that is 
\begin{eqnarray}
\mathcal{R}^\mathrm{ST}&=&\frac{1}{4\,\ln(\Se/S_{\textrm{c}})}
\nonumber \\
& \times&\left[\!\left(\frac{\delta}{S}\right)^2\! - 
\left(\frac{\delta}{\Se}\right)\!\!\left(\frac{\delta}{S}\right) +
\ln\!\left(\frac{\Se}{S}\right)\!\right].
\label{RST}
\end{eqnarray}
In the above relation, the equality $\varepsilon=\delta$ was imposed. 
$\mathcal{R}^\mathrm{ST}$ is a decreasing function of
$S$. 
Referring to the models addressed in the left panel of
\refFgt{fig:ag}, $\mathcal{R}^\mathrm{ST}\simeq 5\%$ at $S/\Se=1/3$.
This result is only slightly affected by the mass of the planet
because $\Se$ is a slowly varying function of $\Mp$.
\subsubsection{Wuchterl's Solution}
\label{sssec:WS}

Along with the fully radiative envelope, other static solutions were found.
Following the track of Stevenson's arguments, \citet{wuchterl1993} developed
an analytical model for protoplanets with spherically symmetric and fully
convective envelopes. In this case the hydrostatic structure is determined 
by the constant entropy requirement that is appropriate when convection is 
very efficient. Supposing that the adiabatic exponent
$\Gamma = d \ln{p}/d \ln{\rho}$ is constant throughout the envelope, 
integrating the envelope density one finds that a solution for the planet 
gravitational potential is
\begin{equation}
 \Phi^\mathrm{KW}_\mathrm{p} = \left\{%
            \begin{array}{ll}
             - \frac{G\,M_\mathrm{c}}{\sqrt{S^2 + \delta^2}}%
             - \frac{G\,M_\mathrm{e}^{\ddag}}{\sqrt{S^2 +
                 \delta^2}}       & \\
                                & \\
               \times\left[\left(\frac{\zeta}{\zeta -1}\right)\!\!%
               \left(\frac{S}{\Se^\Gamma}\right)\!\!%
             - \left(\frac{1}{\zeta -1}\right)\!\!%
               \left(\frac{S}{\Se^\Gamma}\right)^\zeta - \Pi \right]&
             \\
            & \\
              \quad\quad\quad\quad\quad \mathrm{if}\, S \leq \Se^\Gamma  & \\
                          &                              \\
             - \frac{G\,\Mp}{\sqrt{S^2 + \delta^2}}
             \quad \mathrm{if}\, S > \Se^\Gamma &
            \end{array},
                           \right.
                 \label{phikw}
\end{equation}
where we set $\zeta=(3\,\Gamma - 4)/(\Gamma-1)$ and 
$\Pi=(S_\mathrm{c}/\Se^\Gamma)^\zeta$. 
Moreover, the envelope mass is written as 
$M_\mathrm{e}=(1-\Pi)\,M_\mathrm{e}^{\ddag}$.
The condition for stability of gas spheres ($\Gamma>4/3$) implies 
that $\zeta$ is positive. This particular form of 
$\Phi^\mathrm{KW}_\mathrm{p}$ is obtained by choosing an envelope radius
equal to $\Se^\Gamma=(\Gamma -1)\,R_\mathrm{ac}$.
The length
$R_\mathrm{ac}=G\,\Mp/c^2_\mathrm{s}$ is called 
``planetary accretion radius''. 
Outside the sphere $S=R_\mathrm{ac}$ the thermal energy of the gas is higher
than the gravitational energy binding it to the planet.

As in Stevenson's solution, critical mass values exist for the envelope 
structure to be static.
However, unlike the fully radiative envelope case, now the critical core mass
depends on both the temperature and the density of ambient material.
Furthermore, the critical mass ratio is $M_\mathrm{c}/\Mp = 2/3$
\citep[for details see][]{wuchterl1993}.
Setting $\Gamma$ to $1.43$ and the mean molecular weight to $2.2$, 
if nebula conditions are $T_\mathrm{Neb}=100\;\mathrm{K}$
and $\rho_\mathrm{Neb}=10^{-10}\;\dunits$,
the critical total mass is $\Mp=274\;\MEarth$.

Wuchterl's solution well applies to massive protoplanets since they
are likely to bear convective envelopes.

Concerning the differences between \refeqt{phikw} and \refeqt{phipms},
the situation is alike to that met in \refsec{sssec:SS} (see right panel
of \refFgp{fig:ag}).
In this circumstance, for $S\le\Se$,
the normalized difference can be written as
\begin{eqnarray}
\mathcal{R}^\mathrm{KW}&=&\left(\frac{1}{3}\right)\!\!\frac{1}{1-\Pi}
\!\left\{1\! +\!%
\left(\frac{\zeta}{\zeta -1}\right)\!\!%
\left(\frac{\delta}{\Se}\right)\!\!%
\left(\frac{\delta}{S}\right) \right. \nonumber\\
&\times&\left.\left[1 - \left(\frac{S}{\Se}\right)^{\zeta-1}\right] - 
\left(\frac{S}{\Se}\right)^{\zeta}\right\}.
\label{RKW}
\end{eqnarray}
As before, the equality $\varepsilon=\delta$ is assumed and 
$\mathcal{R}^\mathrm{KW}$ decreases with $S$.
Using the parameters adopted for the models illustrated in the right
panel of \refFgt{fig:ag}, $\mathcal{R}^\mathrm{KW}\simeq 18\%$ at $S/\Se=1/3$.
In the considered range of masses, this number is nearly constant. In
fact, \refeqt{RKW} can be approximated to
\begin{equation}
\mathcal{R}^\mathrm{KW}\approx\left(\frac{1}{3}\right)
\left[1 - 
\left(\frac{S}{\Se}\right)^{\zeta}\right].
\label{RKWapp}
\end{equation}
 
\subsection{Physical Parameters}
\label{ssec:PP}
\begin{deluxetable}{ccccc}
\tablewidth{0pt}
\tablecaption{Planetary Masses and 
             Adopted Gravitational Potential.\label{tbl:MpSp}}
\tablehead{%
\colhead{$\Mp\tablenotemark{a}\;/\MEarth$} & 
\colhead{$q$} & 
\colhead{$\Se/\RH$}  &
\colhead{$\RH/a$}  &
\colhead{Potential}
}
\startdata
  333\phm{.0}   &$1.00\times10^{-3}$ & $0.87$                       & $6.9\times10^{-2}$&
   PM, HS \\
  253\phm{.0}   &$7.56\times10^{-4}$ & $2.30$\tablenotemark{b}         & $6.3\times10^{-2}$&
      KW \\
  166\phm{.0}   &$5.00\times10^{-4}$ & $0.78$, $1.70$\tablenotemark{b} & $5.5\times10^{-2}$&
   PM, KW, HS \\
\phn93\phm{.0}   &$2.83\times10^{-4}$ & $1.20$\tablenotemark{b}         & $4.5\times10^{-2}$&
      KW \\
\phn67\phm{.0}   &$2.00\times10^{-4}$ & $0.70$, $0.96$\tablenotemark{b} & $4.0\times10^{-2}$&
  PM, KW \\
\phn33\phm{.0}   &$1.00\times10^{-4}$ & $0.60$ & $3.2\times10^{-2}$&
  PM, HS \\
\phn29\phm{.0}   &$8.80\times10^{-5}$ & $0.58$ & $3.1\times10^{-2}$&
      ST \\
\phn20\phm{.0}   &$6.00\times10^{-5}$ & $0.52$ & $2.7\times10^{-2}$&
      HS, ST \\
\phn15\phm{.0}   &$4.50\times10^{-5}$ & $0.46$ & $2.5\times10^{-2}$&
      ST \\
\phn12.5        &$3.75\times10^{-5}$ & $0.44$ & $2.3\times10^{-2}$&
      ST \\
\phn10\phm{.0}   &$3.00\times10^{-5}$ & $0.38$ & $2.1\times10^{-2}$&
      PM, HS, ST \\
\phn\phn7\phm{.0} &$2.10\times10^{-5}$ & $0.34$ & $1.9\times10^{-2}$&
      ST \\
\phn\phn5\phm{.0} &$1.50\times10^{-5}$ & $0.29$ & $1.7\times10^{-2}$&
      HS, ST \\
\phn\phn3\phm{.0} &$1.00\times10^{-5}$ & $0.23$ & $1.5\times10^{-2}$&
      ST \\
\phn\phn1.5      &$4.50\times10^{-6}$ & $0.16$ & $1.1\times10^{-2}$&
      ST \\
\enddata
\tablenotetext{a}{Values are rounded to the nearest integer numbers.} 
\tablenotetext{b}{Planetary radius used in the Wuchterl's solution:
               $S_\mathrm{p}=S^\Gamma_\mathrm{p}=(\Gamma-1)\,R_\mathrm{ac}$
               (see \refsec{phikw}).} 
\tablecomments{List of all the simulated planet masses:
            $q=\Mp/M_\bigstar$ is the non-dimensional quantity that enters the
            simulation. Note that $\Mp=333\;\MEarth=1.05\;\MJup$.
            The ratio $\RH/a$ is equal to $\left(q/3\right)^{1/3}$ 
            (see \refsec{ssec:PGP}).
            Unless stated otherwise,
            envelope radii $\Se$ are expressed through a combination of the
            Hill ($\RH$) and the accretion ($R_\mathrm{ac}$) radius of the
            planet, as plotted in \refFgt{fig:env_rad} 
            (courtesy of P.~Bodenheimer). 
            When required, we set
            $M_\mathrm{c}=(2/3)\,\Mp$ for a fully convective envelope and
            $M_\mathrm{c}=(3/4)\,\Mp$ for a fully radiative one.
            The core radius is computed assuming a constant density 
            $\rho_\mathrm{c}=5.5\;\dunits$ in both cases.}
\end{deluxetable}

We consider a protostellar disk orbiting a one solar-mass star. 
The simulated region extends for $2\,\pi$ around the polar axis and, 
radially, from  $2.08$ to $13\;\AU$. 
The aspect ratio is fixed to $h = 0.05$ throughout these computations. As in 
\KDH, the disk is assumed to be symmetric with respect to its midplane.
This allows us to 
reduce the latitude range to the northern hemisphere only, where $\theta$ 
varies between $80\degr$ and $90\degr$. The co-latitude interval includes 
$3.5$ disk scale-heights and therefore it assures a vertical density drop of 
more than six orders of magnitude (see \refsec{ssec:IC}). 
The mass enclosed within this domain is 
$M_\mathrm{D}=3.5\times10^{-3}\;\MSun$, which 
implies, in our case, that a mass equal to $0.01\;\MSun$ is confined inside 
$26\;\AU$.
Disk material is supposed to have a constant kinematic viscosity 
$\nu=10^{15}\;\mathrm{cm}^2\,\mathrm{s}^{-1}$, corresponding to 
$\alpha=4\times10^{-3}$ at the planet's location.

The orbital radius of the planet is
$R_\mathrm{p}=5.2\;\AU$ and
its azimuthal position is fixed to $\varphi=\varphi_\mathrm{p}=\pi$.
We concentrate on a mass range stretching from $1.5\;\MEarth$ to
one Jupiter-mass ($\MJup$), 
implying that the mass ratio $q=\Mp/M_\bigstar\in[4.5\times10^{-6},10^{-3}]$,
if $M_{\bigstar} = 1\;\MSun$. 
A detailed list of the examined planetary masses, along with the adopted 
potential form is given in \refTab{tbl:MpSp}.

The choice of few of the above parameters represents one typical example
during the early phase of planet formation.
Additionally, these simulations offer the good advantage that
the system of equations is cast in a non-dimensional form, thus all of the 
outcomes are ``scale-free'' with respect to $M_\bigstar$, $M_\mathrm{D}$, 
and $R_\mathrm{p}$.

\begin{figure}[!t]
\epsscale{1.0}
\plotone{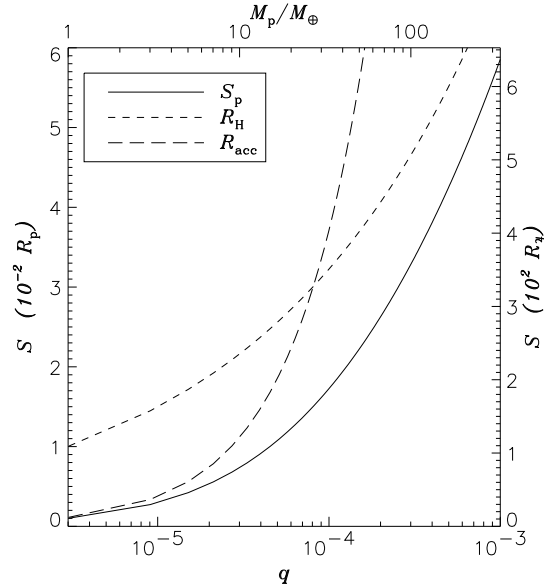}
\caption{%
Envelope radius $\Se$, as provided by P.~Bodenheimer
(2001, private communication), compared to the Hill radius $\RH$ and
the accretion radius $R_\mathrm{acc}$.
In a locally isothermal disk with aspect ratio $h$, 
$\RH$ is larger than $R_\mathrm{ac}$ when $q < h^3/\sqrt{3}$. 
While the scale of the left vertical axis is referred to orbital radius
$R_\mathrm{p}=5.2\;\AU$, the scale on the right one is compared to
the radius of Jupiter $R_{\jupiter}=7.1\times10^{4}\;\mathrm{km}$.
\label{fig:env_rad}}
\end{figure}
According to studies of the early evolution of protoplanets, the Roche lobe 
is usually filled during the growth phase. In such calculations the 
envelope is allowed to extend to either the Hill radius $\RH$ or the accretion 
radius $R_\mathrm{ac}$
\citep{bodenheimer1986,wuchterl1991,tajima1997}.
Except for Wuchterl's solution, where we set
$\Se=\Se^\Gamma=(\Gamma -1)\,R_\mathrm{ac}$,
the estimates of planet radii used in the simulations were provided by
P.~Bodenheimer (2001, private communication). They originate from a 
combination of $\RH$ and $R_\mathrm{ac}$ at an ambient
temperature of $T=100\;\mathrm{K}$
(see \refFgp{fig:env_rad}). 
The values of $\Se$, employed in each model, are also reported in 
\refTab{tbl:MpSp}.

It is worthwhile to stress that the planetary (or more properly, the envelope) 
radius $\Se$
does not represent any real physical boundary but only the distance beyond 
which the planet's potential reduces to the one given in
\refeqt{phipms}, i.e., to a point-mass potential.

\section{Numerical Issues}
\label{sec:NI}

The set of hydrodynamic equations that characterizes the temporal evolution 
of a disk-planet system is solved numerically by means of a finite difference 
scheme provided by an early FORTRAN-coded version of \textsc{Nirvana} 
\citep{ziegler1997,ziegler1998}. The code has been modified and adapted to our
purposes as described in \KDH, \DHK, and references therein. 
In order to investigate thoroughly the flow dynamics in the neighborhood
of the planet, a sufficient numerical resolution is required. We
accomplish that by employing a nested-grid strategy.
This can be pictured as either a set of grids, each hosting an inner one, or a
pyramid of levels: the main grid (level one) includes the whole 
computational domain, while inner grids (higher levels) enclose smaller 
volumes around the planet, with increasing resolution. Levels greater than one
are also called subgrids since they are usually smaller in size.
The linear resolution, in each direction, doubles when passing from a grid to 
the inner one.

The basic principles upon which the nested-grid technique relies and how it is
applied to disk-planet simulations in two dimensions is explained in detail 
in \DHK\ and references therein. 
The extension to the three-dimensional geometry, though requiring some more 
complexity in the exchange of information from one grid level to the 
neighboring ones, is nearly straightforward. 
Coarse-fine grid interaction (which sets the boundary conditions
necessary for the integration of subgrids) 
is accomplished via a direction-splitting
procedure. Thus, with respect to the discussion in \DHK, the addition
of a third dimension just requires the implementation of the third
direction-splitting step in the algorithm.
Fine-coarse grid interaction (which upgrades the solution on finer
resolution regions) involves, in 3D, volume-weighted averages in a
spherical polar topology. These are given in the Appendix. 

\subsection{General Setup}
\label{ssec:GS}

\begin{deluxetable}{ccccc}
\tablewidth{0pt}
\tablecaption{Grid Hierarchies Utilized in the Simulations.\label{tbl:grids}}
\tablehead{%
\colhead{Grid} & 
\colhead{Main Grid Size} & 
\colhead{$ng$}   &
\colhead{Subgrid Size}   & 
\colhead{No. of models}
}
\startdata
G0 & $121\times 13\times 319$ & 4 & $54\times 12\times 48$&   10 \\
G1 & $143\times 13\times 423$ & 4 & $64\times 12\times 64$& \phn5 \\
G2 & $121\times 13\times 319$ & 5 & $54\times 12\times 48$&   19 \\
G3 & $143\times 13\times 423$ & 5 & $64\times 12\times 64$&   12 \\
G4 & $121\times 23\times 319$ & 5 & $54\times 22\times 48$& \phn4 \\
G5 & $133\times 13\times 395$ & 5 & $84\times 16\times 84$& \phn1 \\
\enddata
 
 
 
\tablecomments{Grid sizes are reported as the number of grid points per 
       direction: $N_R\times N_\theta\times N_\varphi$. The third
       column ($ng$) indicates the number of levels within the hierarchy.
       In order to achieve sufficient resolution within the Roche lobe
       of the planet, grids G0 and G1 have been used only for planetary 
       masses in the range $[67\;\MEarth, 1\;\MJup]$.
       Grids are ordered according to their computing time requirements, 
       which grow from top to bottom. The hierarchy G5 has been
       employed to execute the model with $\Mp=12.5\;\MEarth$ (see
       discussion at the end of \refsec{ssec:TM}). }
\end{deluxetable}
For the study of the variety of planetary masses indicated in 
\refTab{tbl:MpSp}, meeting both the requirements of high resolution and 
affordable computing times, we realized a series of six grid hierarchies, 
whose characteristics are given in \refTab{tbl:grids}. 
With mass ratios $q$ larger than $2\times10^{-4}$ ($67\;\MEarth$) only grids 
G0 and G1 are utilized whereas smaller bodies are investigated with the other 
grid hierarchies.
Thereupon, the finest resolution we obtain in the whole set of simulations 
varies form $0.03$ to $0.06\,\RH$.
In all of the models presented here, the planet is centered at the corner
of a main grid cell, which property is retained on any higher hosted subgrid.
As the planet radial distance is $R_\mathrm{p}=a/(1+q)$, we adjust it by 
tuning the value of the star-planet distance $a$.
Adjustments never exceed 0.7\% 
over the nominal value of $R_\mathrm{p}$ 
given in \refsec{ssec:PP}.
Every model is evolved at least till 200 orbits. The evolution of
massive planets ($\Mp\geq 67\;\MEarth$) is followed for 300 to 400
orbital periods because they take longer to settle on a quasi-stationary
state.

Gas accretion is estimated following the procedure sketched in \DHK. 
For better accuracy, mass is removed only from the finest grid level according 
to an accretion sphere radius $\kappa_\mathrm{ac}$ and
an evacuation parameter $\kappa_\mathrm{ev}$.
The former defines the spherical volume which contributes to the accretion
process whereas the latter can be regarded as a
measure of the removal time scale in such volume. 
Two-dimensional simulations showed that the procedure is fairly stable
against these two parameters. We constrain the amount of removed
mass per unit volume not to exceed 1\% 
of that available, as was done in \DHK.
Regarding the extension of the sphere of accretion 
$\kappa_\mathrm{ac}$, we performed simulations using different values, 
as stated in \refTab{tbl:kac}.
Since the planet actually works as a sink, our procedure only 
furnishes upper limits to realistic planetary accretion rates
\citep[see discussion in][]{tanigawa2002}.

However, we also inquire how mass removal can possibly affect gravitational 
torques and, more generally, the dynamics of the flow in the planet 
neighborhood 
by means of models in which accretion is prohibited.
\begin{deluxetable}{cccc}
\tablewidth{0pt}
\tablecaption{Mass Accretion Parameter $\kappa_\mathrm{ac}$.\label{tbl:kac}}
\tablehead{%
\colhead{$\Mp/\MEarth$} &
\colhead{$q$} &
\colhead{$\kappa_\mathrm{ac}/\RH$} & 
\colhead{Accreting Only\tablenotemark{a}}
}
\startdata
333\phm{.0}     &  $1.00\times10^{-3}$ & $0.20$, $0.15$, $0.10$ &   No \\
253\phm{.0}     &  $7.60\times10^{-4}$ & $0.10$                 &  Yes \\
166\phm{.0}     &  $5.00\times10^{-4}$ & $0.10$                 &  Yes \\
\phn93\phm{.0}   &  $2.80\times10^{-4}$ & $0.10$                 &  Yes \\
\phn67\phm{.0}   &  $2.00\times10^{-4}$ & $0.20$, $0.10$         &   No \\
\phn33\phm{.0}   &  $1.00\times10^{-4}$ & $0.20$, $0.10$         &   No \\
\phn29\phm{.0}   &  $8.80\times10^{-5}$ & $0.20$, $0.10$         &   No \\
\phn20\phm{.0}   &  $6.00\times10^{-5}$ & $0.15$, $0.10$         &   No \\
\phn15\phm{.0}   &  $4.50\times10^{-5}$ & $0.10$                 &  Yes \\
\phn12.5        &  $3.75\times10^{-5}$ & $0.10$                 &  Yes \\
\phn10\phm{.0}   &  $3.00\times10^{-5}$ & $0.10$                 &   No \\
\phn\phn7\phm{.0} &  $2.10\times10^{-5}$ & $0.10$                 &  Yes \\
\phn\phn5\phm{.0} &  $1.50\times10^{-5}$ & $0.10$                 &   No \\
\phn\phn3\phm{.0} &  $1.00\times10^{-5}$ & $0.10$                 &  Yes \\
\phn\phn1.5      &  $4.50\times10^{-6}$ & $0.07$                 &  Yes \\
\enddata
 
 
\tablenotetext{a}{``No'' entry stands for the existence of a 
               non-accreting model.}
\tablecomments{The parameter $\kappa_\mathrm{ac}$ represents the radius of the
            accreting region. Within this sphere the mass density is reduced
            by roughly 1\% after every time step.
            The length $\kappa_\mathrm{ac}=0.1\,\RH$ should be small enough
            to make the accretion procedure almost independent of the 
            evacuation parameter $\kappa_\mathrm{ev}$ \citep{tanigawa2002}.
            At $q=3\times10^{-5}$ ($\Mp=10\;\MEarth$) a non-accreting 
            simulation
            was performed with $\Phi_\mathrm{p}=\Phi^\mathrm{HS}_\mathrm{p}$
            (\refeqp{phihs})
            as well as with $\Phi_\mathrm{p}=\Phi^\mathrm{ST}_\mathrm{p}$ 
            (\refeqp{phist}).
            In the case of lowest mass model ($\Mp=1.5\;\MEarth$), we allowed
            the ratio $\kappa_\mathrm{ac}/\Se$ to be less than $0.5$,
            as in all of the other models. For a better evaluation of
            $\dMp$, we used a modified version of the grid system G3
            which contains a sixth level, comprising (approximately) 
            the planet's Hill sphere.}
\end{deluxetable}

\subsection{Boundary Conditions}
\label{ssec:BC}

In order to mimic the accretion of the flow towards the central star,
an outflow boundary condition is applied at the inner radial border of the
computational domain. The outer radial border is closed, i.e., no material 
can flow in or out of it. The same condition exists at the highest latitude
$\theta=80\degr$. Since the disk is symmetric with respect to its midplane
as mentioned in \refsec{ssec:PP}, symmetry conditions are set at 
$\theta=90\degr$. On subgrids, except for the midplane where symmetry 
conditions are applied, boundary values are interpolated
from hosting grids, by means of a monotonised second-order algorithm
(see \DHK\ for details).

The open inner radial boundary causes the disk to slowly deplete during its
evolution. For all the models under study, we observe a depletion rate 
$\dot{M}_\mathrm{D}=-\dot{M}_\bigstar$ measuring $\approx 10^{-8}\;\Msyr$, 
in agreement with the expectations of stationary Keplerian disks:
$\dot{M}_\bigstar=3\,\pi\,\nu\,\Sigma$ \citep{lyndenbell}.

In cases of gap formation, material residing inside the planet's orbit tends 
to drain out of the computational domain.
Since this material transfers angular momentum to the planet, the lack of it 
may contribute to reduce both the migration time scale and the planet's 
accretion rate. To evaluate these effects, a Jupiter-mass model was run with
a closed (i.e., reflective) inner radial border.

\subsection{Initial Conditions}
\label{ssec:IC}

The initial density distribution is a power-law of the distance from the
rotational axis $r=R\,\sin{\theta}$ times a Gaussian in the vertical direction
\begin{equation}
    \rho(t=0)  = \rho_0(r)\,
     \exp\left[-\left(\frac{\cot\theta}{h}\right)^2\right],
    \label{rho0}
\end{equation}
which is appropriate for a thin disk in thermal and hydrostatic vertical 
equilibrium.
The dependency of the midplane value $\rho_0$ with respect to $r$ is such 
that the initial surface density
profile $\Sigma$ decays as $1/\sqrt{r}$, as required by the constant kinematic 
viscosity. However, we also ran a few models where the relation $\rho_0(r)$ is 
such to account for an axisymmetric gap, as often done to speed up the 
computations at early evolutionary times. 

The initial velocity field of the fluid is a purely, counter-clockwise,
Keplerian one corrected by the grid rotation: 
$\gv{u}(t=0)\equiv (0,0,v_{K}-\Omega_\mathrm{p}\,r)$.
Thus, the partial support due to the radial pressure gradient is neglected in 
the beginning.

\section{Simulation Results}
\label{sec:SR}

\subsection{Flow Dynamics near Protoplanets}
\label{ssec:FD}

Two-dimensional computations have shown that a circumplanetary disk
forms around Jupiter-type planets, extending over the size of its 
Roche lobe \citep{kley1999,lubow1999,tanigawa2002}. 
The numerical experiments conducted in \DHK\ proved this characteristic 
to belong not only to massive bodies but also to protoplanets 
as small as $3\;\MEarth$. The authors demonstrated that the flow
of such disks is approximately Keplerian down to distances $\sim 0.1\,\RH$
from the planet.
One of the main features of these disks is a two-arm spiral shock wave 
whose opening angle (that between the wave front and the direction
toward the planet) is an increasing function of the mass ratio $q$ and, below 
$\Mp=67\;\MEarth$, it is roughly given by $\arctan(\mathcal{M})$, 
in which $\mathcal{M}=|\gv{u}|/c_\mathrm{s}$ 
is the Mach number of the circumplanetary flow
\citep{wash2002}.
The spiral patterns shorten and straighten as the perturber mass decreases. 
Eventually, for even smaller masses
they disappear and are not observable anymore when one Earth-mass 
is reached.

\begin{figure*}
\epsscale{0.8}
\begin{center}
\mbox{%
\plotone{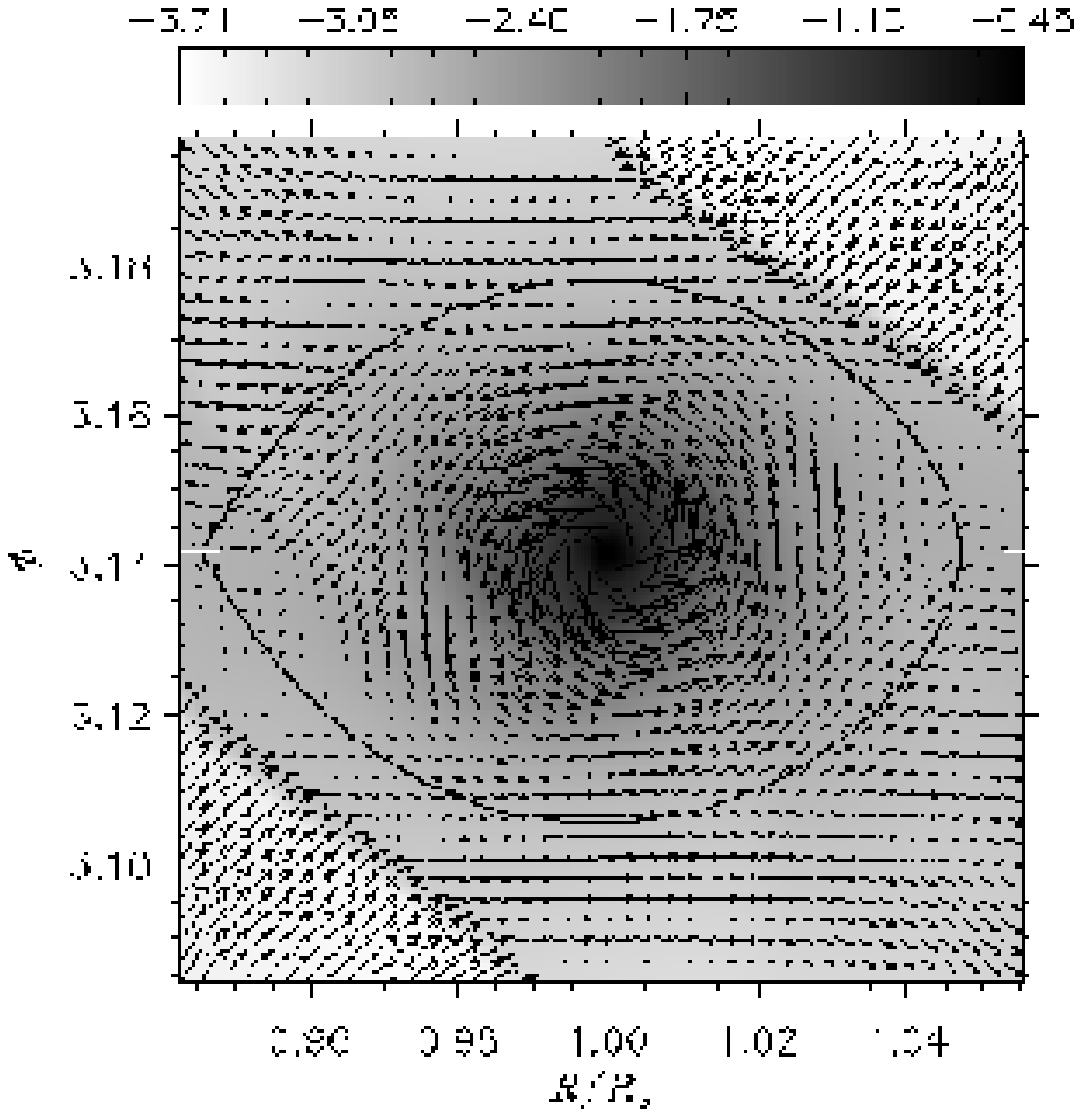} 
\plotone{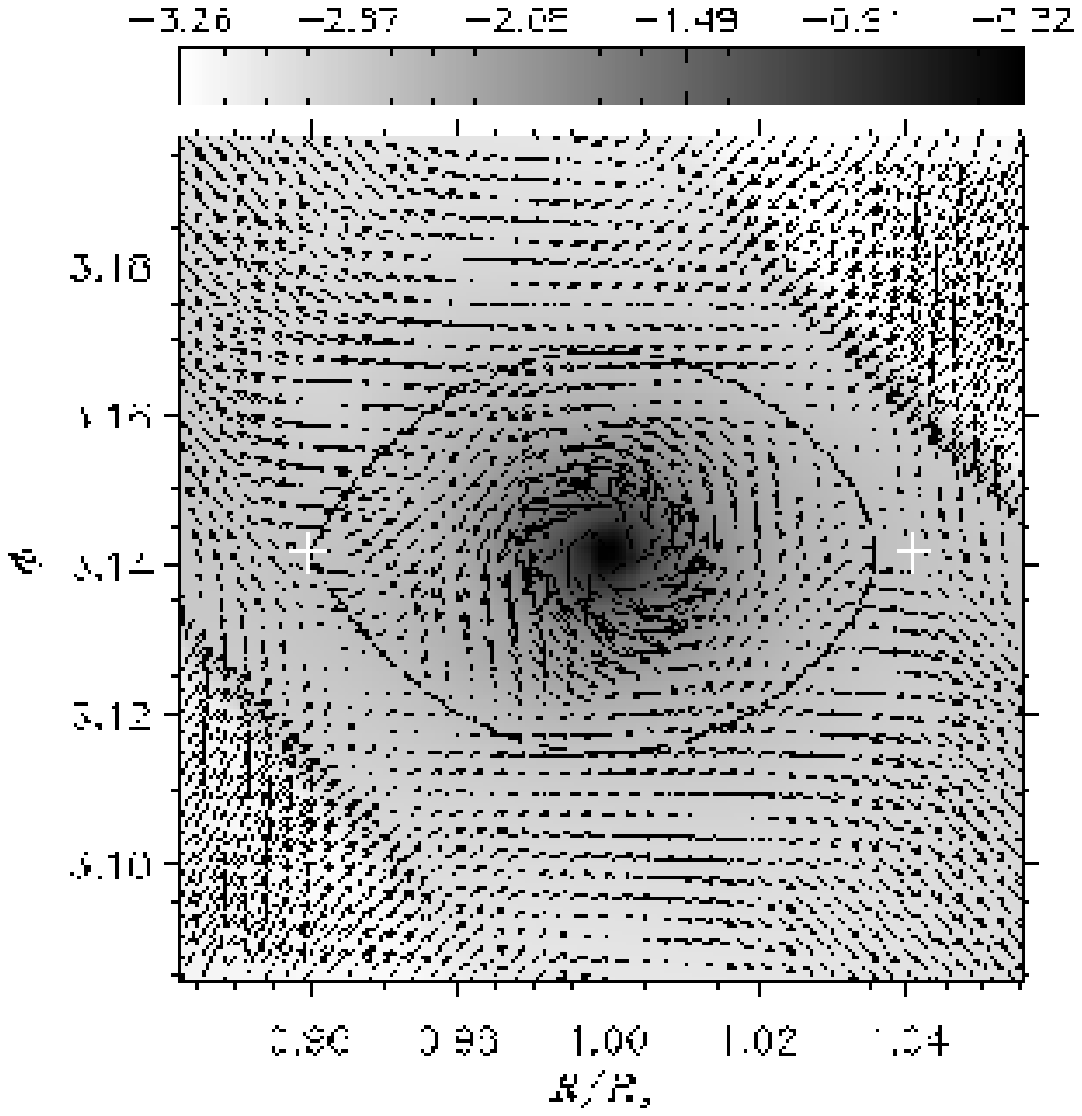}}
\mbox{%
\plotone{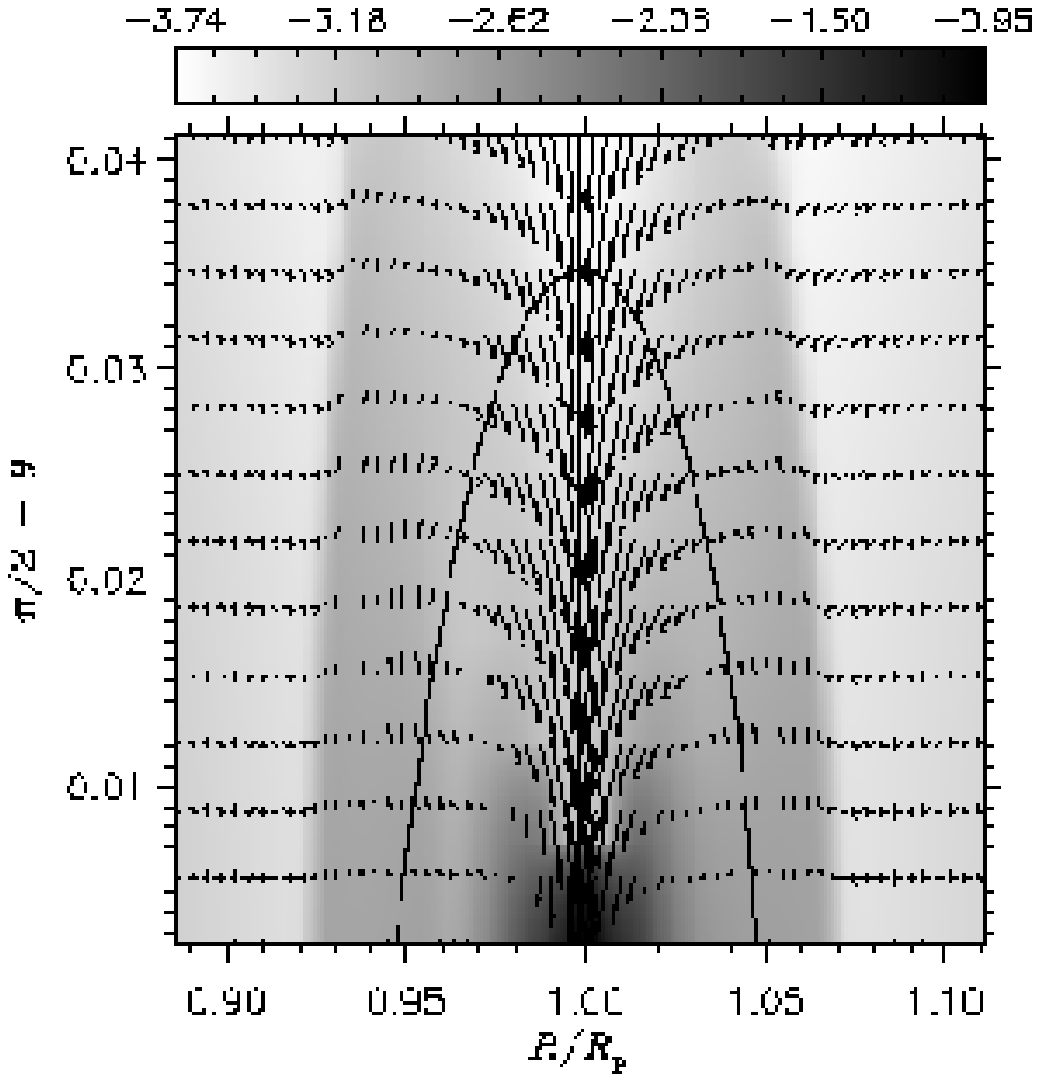} 
\plotone{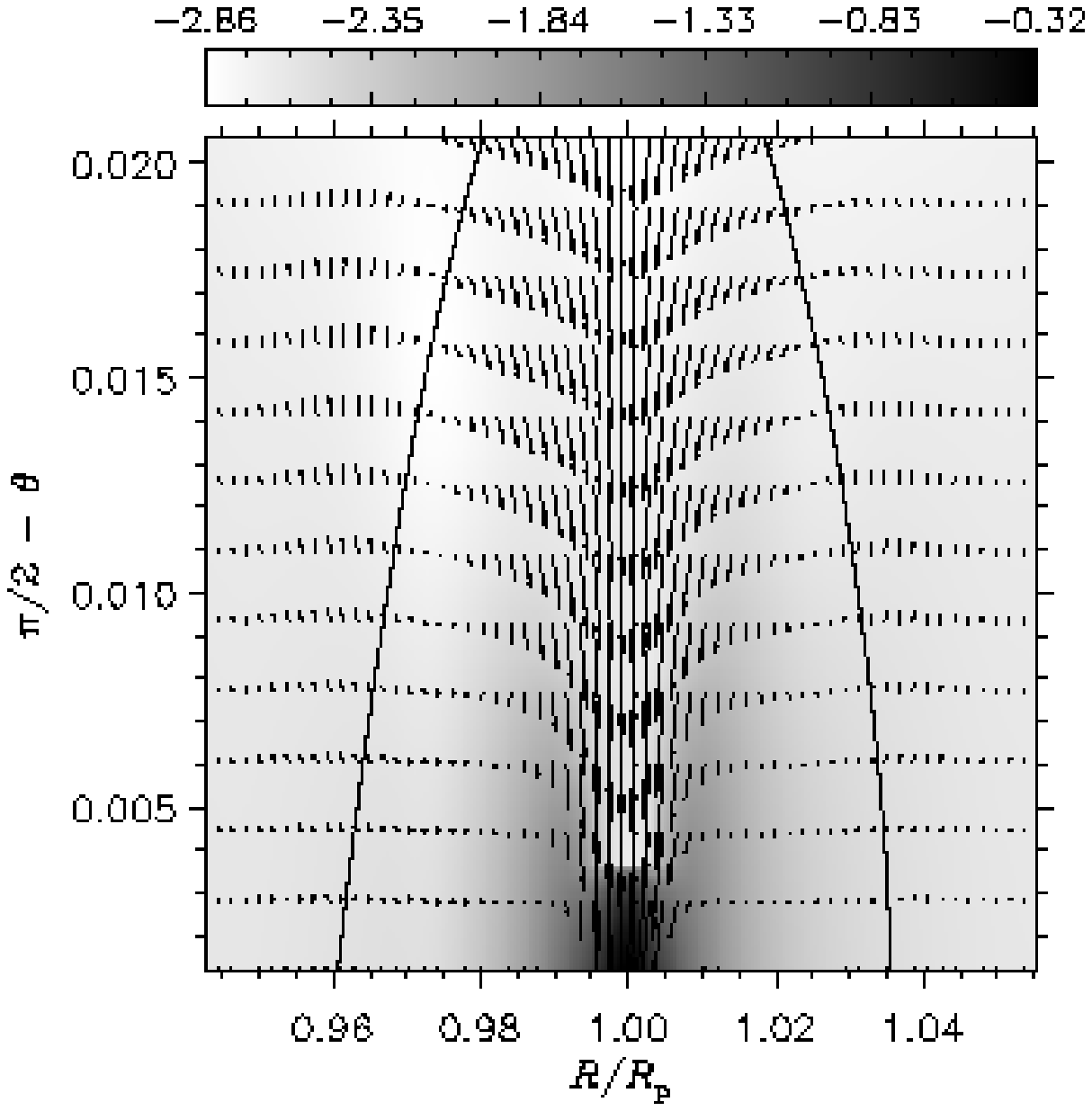}}
\mbox{%
\plotone{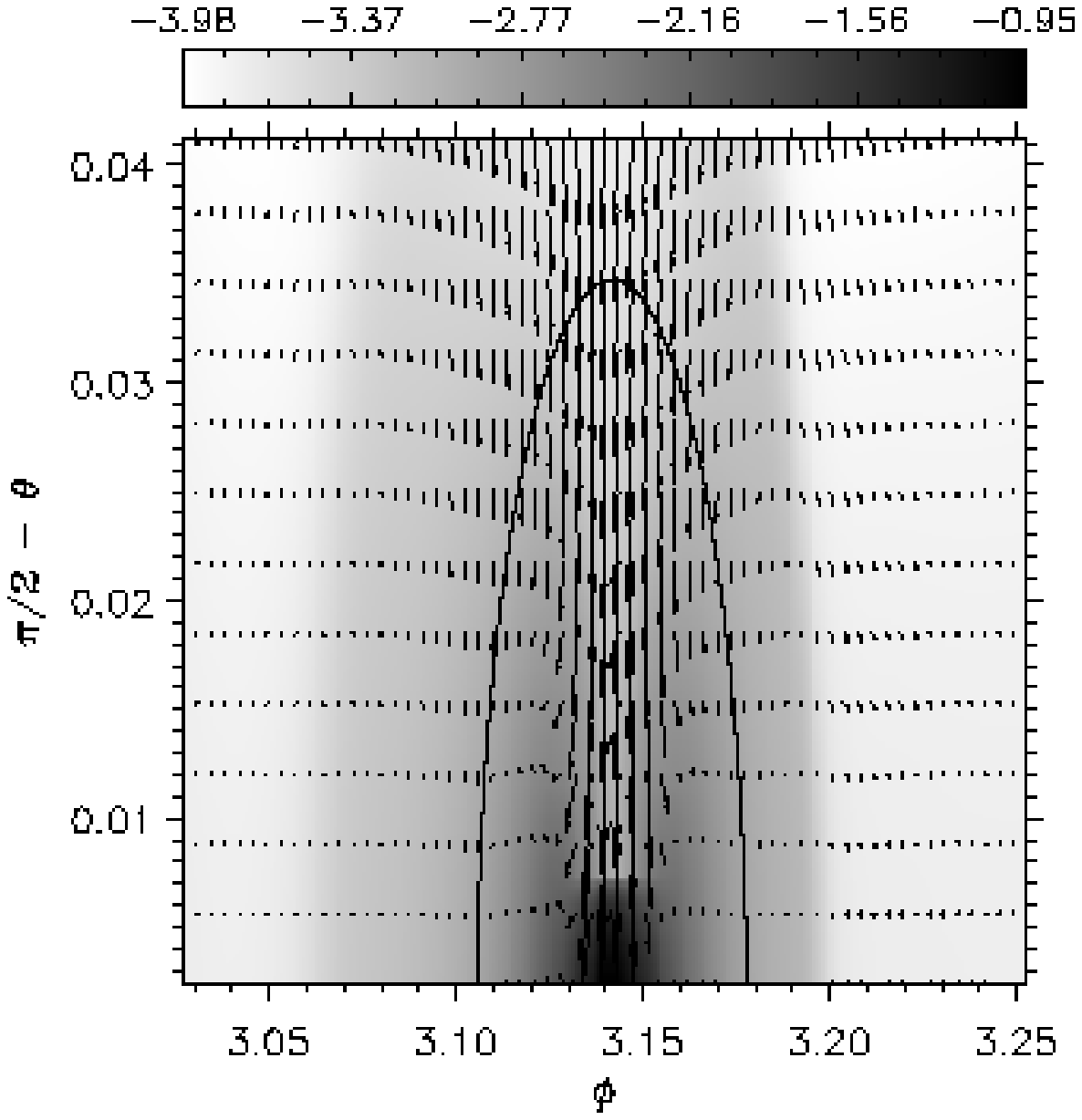} 
\plotone{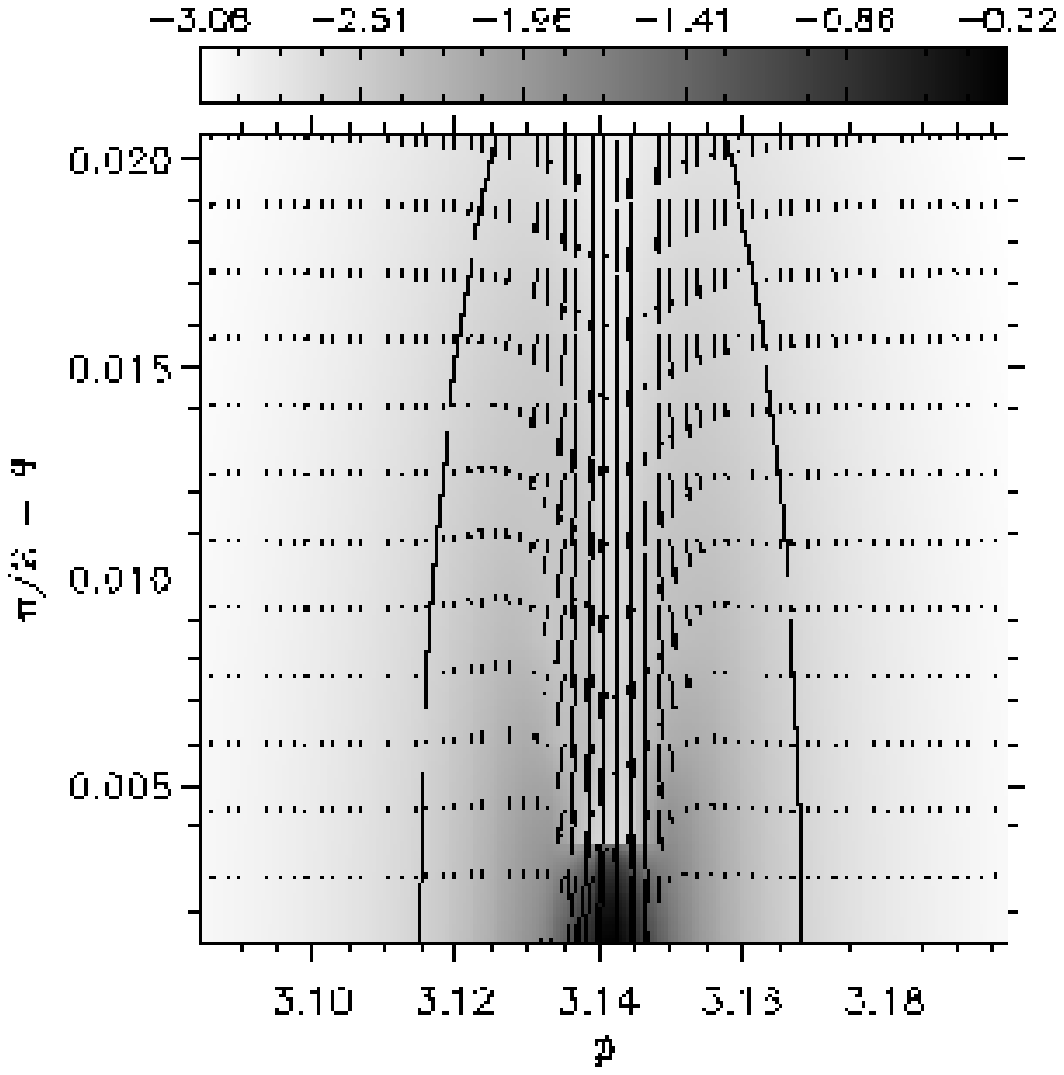}}
\end{center}
\caption{%
       Density slices with an overplotted two-component velocity field.
       In order to comprise the whole density range, the color scale
       represents the logarithm of the density ($\log{\rho}$).
       The evolutionary time is 200 orbital periods. 
       Slices are cut at the planet location: $\theta=\pi/2$ (\textbf{top}),
       $\varphi=\varphi_\mathrm{p}$ (\textbf{middle}), 
       and $R=R_\mathrm{p}$ (\textbf{bottom}).
       Both models are accreting and executed with Wuchterl's potential. 
       In physical units, $\rho=10^{-2}$ is equivalent to
       $4.22\times10^{-11}\;\dunits$.
       \textbf{Left panels}. Planet with a total mass $\Mp=166\;\MEarth$ and
       a core mass equal to $2/3\,\Mp$ (critical core mass).
       \textbf{Right panels}. Planet with $\Mp=67\;\MEarth$ and a critical 
       core mass. 
       The curve drawn in each panel indicates the trace of the Roche lobe.
       In the left and right top panels, the flow reaches velocities equal to
       $\sim 3$ and $\sim 2\;\mathrm{km}\,\mathrm{s}^{-1}$, respectively.
       In the other panels, maximum velocities are on the order of
       $\sim 4\;\mathrm{km}\,\mathrm{s}^{-1}$.
\label{fig:pic1}}
\end{figure*}
\begin{figure*}
\epsscale{0.8}
\begin{center}
\mbox{%
\plotone{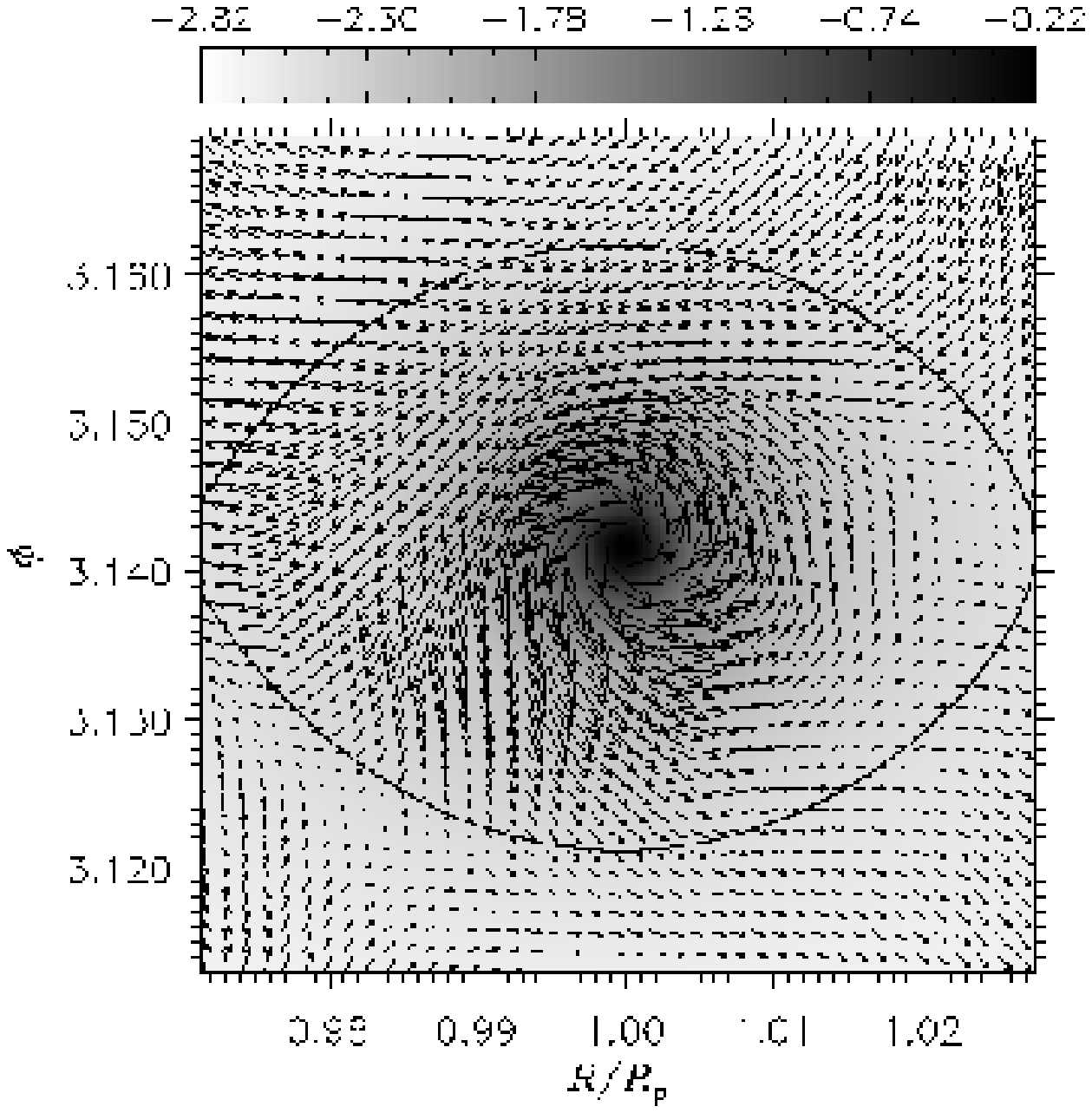}\hfill%
\plotone{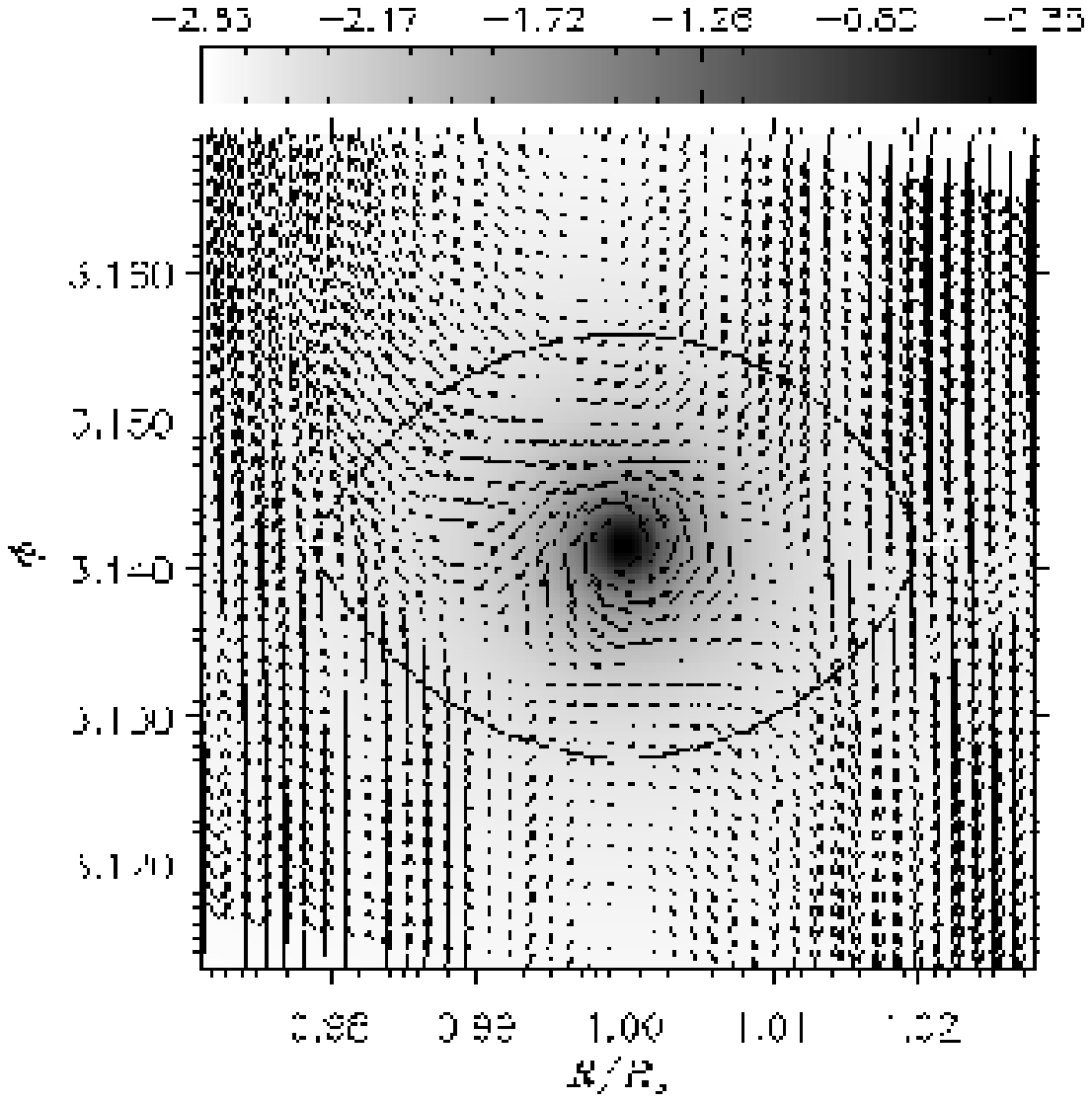}}
\mbox{%
\plotone{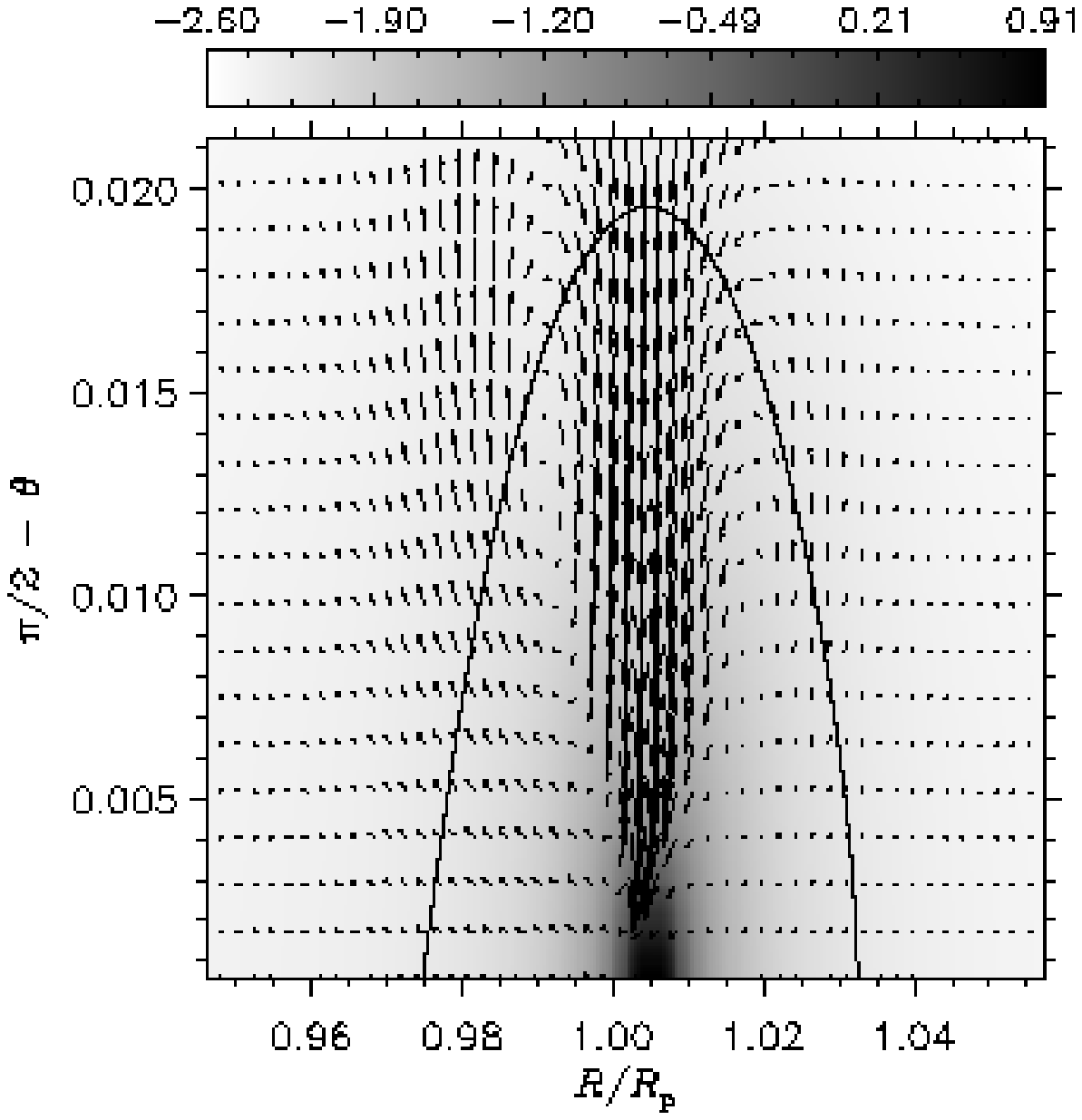}\hfill%
\plotone{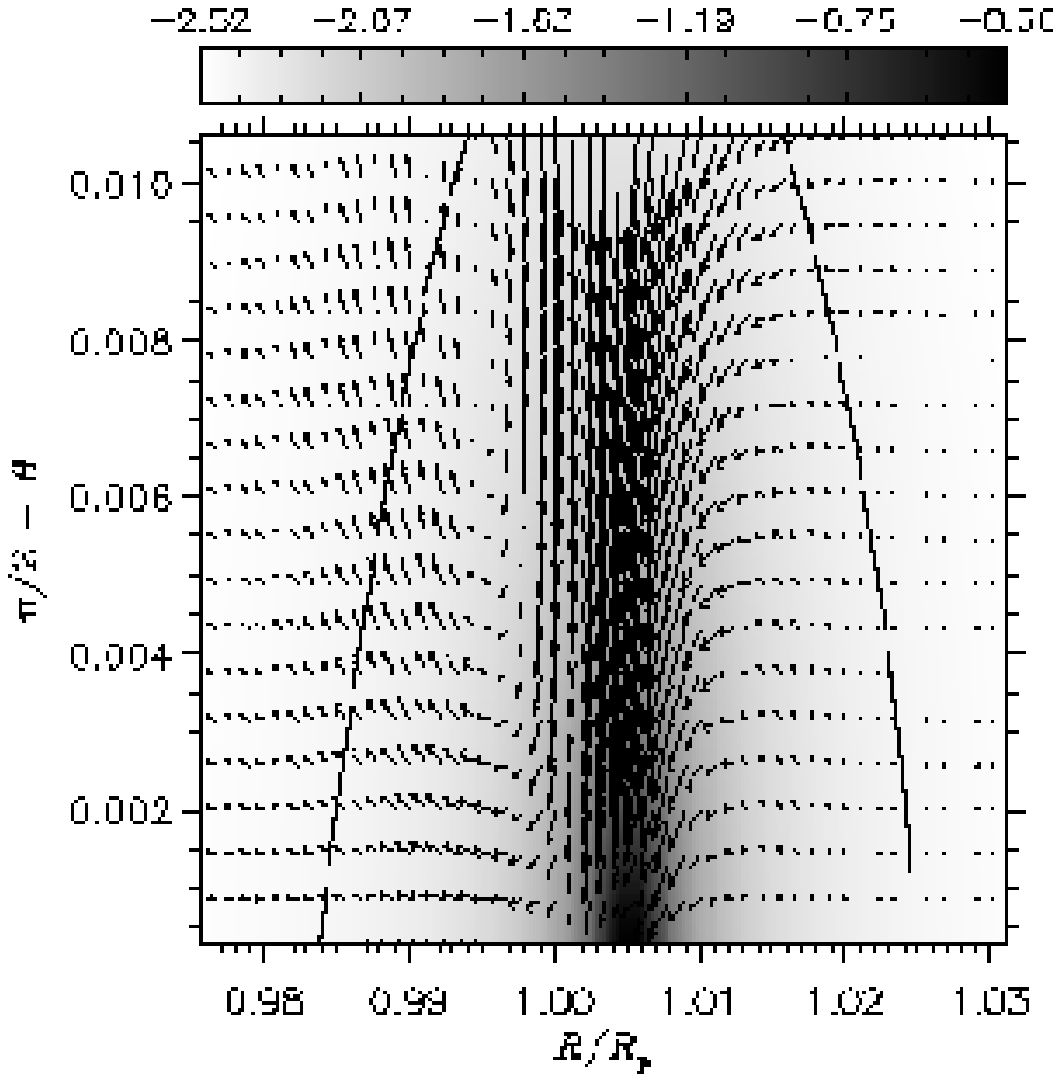}}
\mbox{%
\plotone{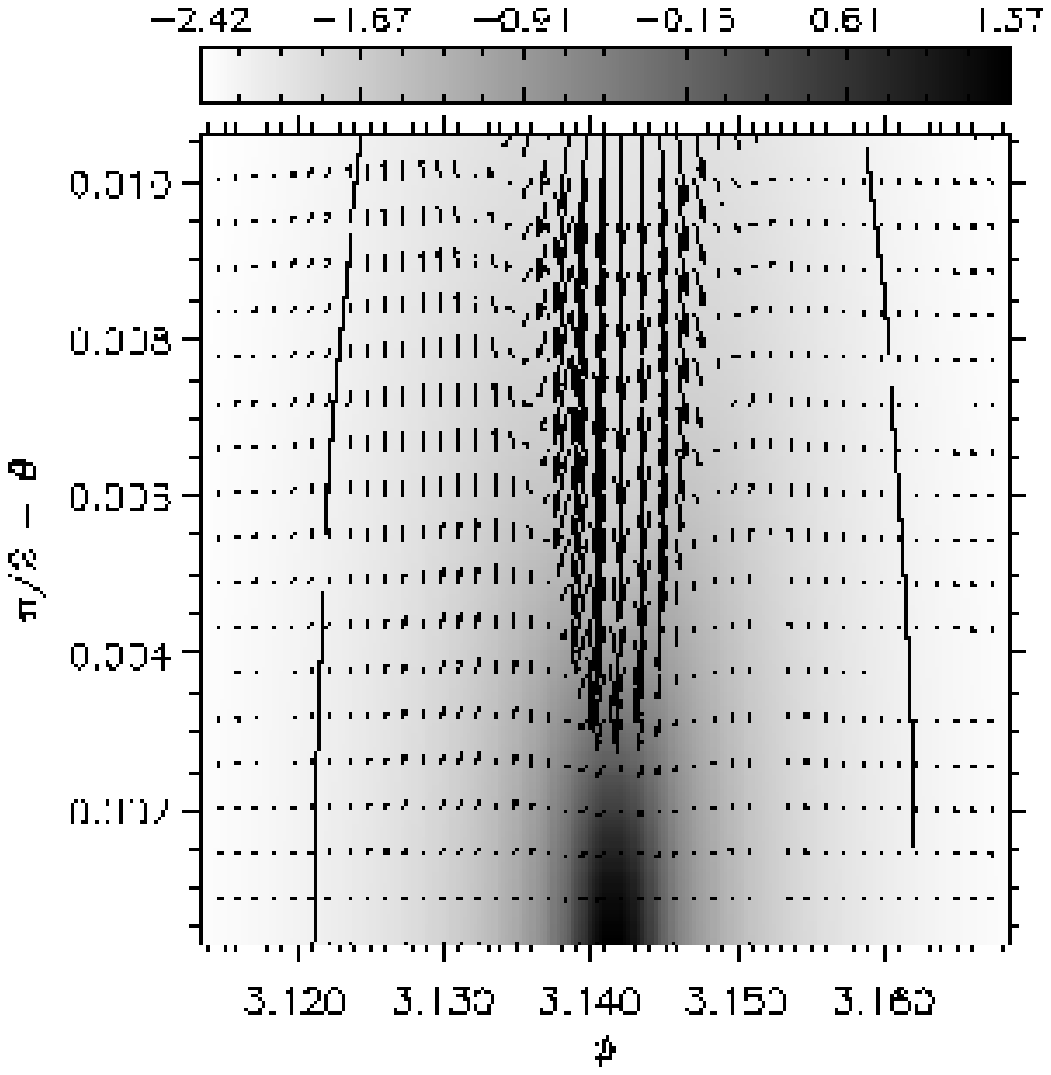}\hfill%
\plotone{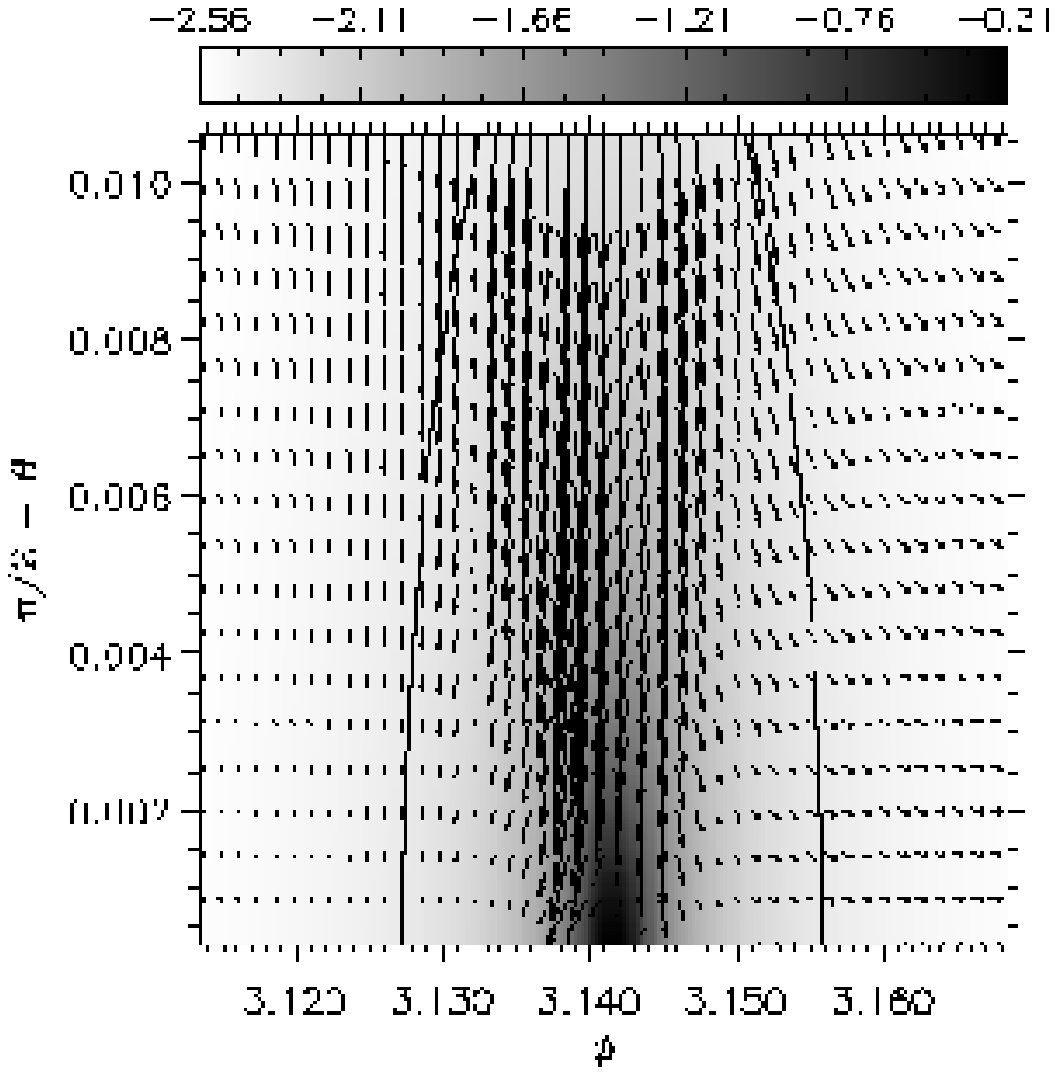}}
\end{center}
\caption{%
       Density slices of the same type as in \refFgt{fig:pic1}, illustrating
       low-mass accreting planets after 200 orbits.
       In these simulations, Stevenson's potential is employed.
       In physical units, the density scale is as in \refFgt{fig:pic1}.
       The density value $\rho=10^{-2}$ corresponds to
       $4.22\times10^{-11}\;\dunits$.
       \textbf{Left panels}. Planet with $\Mp=29\;\MEarth$ and a critical core
       mass $M_\mathrm{c}=21.7\;\MEarth$. Maximum flow speeds,
       within $S=\RH$, are $\sim1\;\mathrm{km}\,\mathrm{s}^{-1}$ in each of
       the three panels. 
       \textbf{Right panels}. Planet with $\Mp=10\;\MEarth$ and a core mass
       and $M_\mathrm{c}=7.5\;\MEarth$. 
       Inside of the Hill sphere, maximum velocities are on the order
       of $0.2\;\mathrm{km}\,\mathrm{s}^{-1}$.
\label{fig:pic2}}
\end{figure*}

Three-dimensional simulations shed new light on these circumplanetary
disks, demonstrating that they can behave somehow differently from 
what depicted by two-dimensional descriptions. 
Differences become more marked as the mass of the embedded protoplanet is
lowered.
A major point is that 
spiral waves are not so predominant as they are in 
the 2D geometry. This is clearly 
seen in the top rows of \refFgt{fig:pic1} and \ref{fig:pic2}, where the 
midplane ($\theta=\pi/2$) density is displayed for four different planetary 
masses.
The double pattern of the spiral is still visible around a $67\;\MEarth$
protoplanet (\refFgp{fig:pic1}, top right panel). 
However, when considering a planet half
of that size, spiral traces are too feeble to be seen on the image
(\refFgp{fig:pic2}, top left panel).
Such an occurrence was to be expected since the energy of the flow is not only
converted into the equatorial motion but can be also transferred to the 
vertical motion of the fluid.
It was already known from wave theories for circumstellar disks 
\citep{lubow1981,lubow1993,ogilvie1999} and related numerical calculations
\citep{makita2000} that the three-dimensional propagation of spiral 
perturbations may be significantly different from that obtained in two 
dimensions
because of the existence of vertical resonances.
Furthermore, \citet{miyoshi1999} already noticed in their shearing sheet
models the weakening effect of the finite disk thickness upon the
formation of spiral waves around embedded protoplanets. 
Hence, we may argue that the averaging of the pressure and the gravitational
potential, which is accomplished in an infinitesimally thin disk,
enhances spiral features in disks.

The remainder of this section is devoted to a general description of the 
vertical circulation of the material in the vicinity of the planet.
We start inspecting what happens in the slice $\varphi=\varphi_\mathrm{p}$,
i.e., in the $R$--$\theta$ surface containing the planet
(see middle panels in \refFgp{fig:pic1} and \ref{fig:pic2}). 
The first thing to note is that the 
material above the equatorial plane moves toward the planet with 
a negligible meridional component, in fact $|u_\theta|\ll u$ 
(where $u=|\gv{u}|$). 
Therefore matter is nearly confined to $\theta$-constant surfaces. 
This circumstance favors the use of 2D outcomes as predictions of 3D 
expectations \citep{masset2002}.
However this turns out to be true only far away from the planet. In fact,
as the fluid enters a certain region around the Hill sphere, 
its dynamics changes drastically. 
The beginning of this zone is marked by two shock fronts, which actually 
develop well outside the Roche lobe of the restricted three body problem. 
The distance of the shock fronts from the perturber, if compared to $\RH$,
shrinks from $2.61$ to $1.16$ as $\Mp$ is increased from $10\;\MEarth$ to
$1\;\MJup$. Generally, shocks are not placed symmetrically with respect
to the planet.
Past these shocks, material is deflected upward and then it
recirculates downward, while approaching the radial position of the
planet (see middle panels in \refFgp{fig:pic1} and \ref{fig:pic2}). 

At $R\approx R_\mathrm{p}$, matter suffers an unbalanced gravitational 
attraction by the planet and accelerates downward
towards it. Velocities are supersonic,
reaching a Mach number $\mathcal{M}\simeq 8$, when 
$67\;\MEarth\lesssim\Mp\leq 1\;\MJup$,
and  $\mathcal{M}\simeq 2$ if $\Mp=20\;\MEarth$.
They become subsonic for planetary masses between $10$ and $1.5\;\MEarth$, 
ranging from 30\% to 5\% of the local sound speed.
Because of the sinking material, at $\theta=\pi/2$ the flow field is 
slightly horizontally divergent from the planet location.

As one can judge from \refFgt{fig:pic2} (middle panels), the flow gets less 
and less symmetric, with respect to $R_\mathrm{p}$, as the planet mass 
is reduced. Recirculation persists before the planet ($R < R_\mathrm{p}$) but 
vanishes behind it ($R > R_\mathrm{p}$). Any symmetry disappears
starting from $\Mp=5\;\MEarth$ downwards.

Along the azimuthal direction (slice $R=R_\mathrm{p}$, i.e., in the surface
$\varphi$--$\theta$ containing the planet), the bottom panels of 
\refFgt{fig:pic1} and \ref{fig:pic2} show an even more complex situation.
Below $166\;\MEarth$,
the region of influence of the planet appears to be more comparable in size 
with the Roche lobe. However, apart from that, the general behavior of
the flow differs from case to case, having in common a rapid descending
motion when the material lies above the planet.
Around Jupiter-sized planets some 
kind of weak, non-closed, recirculation may be seen. 
This flow feature is still present at both sides
of a $29\;\MEarth$ planet (bottom left panel, \refFgp{fig:pic2}), whereas
it tends to vanish in models with lower $q$-ratios. 

\subsubsection{Non-accreting Protoplanets}
\label{sssec:NaP}

\begin{deluxetable}{cccccccc}
\tablecolumns{8}
\tablewidth{0pt}
\tablecaption{Mass enclosed within the Envelope Radius.\label{tbl:noac}}
\tablehead{%
\colhead{}  &  \colhead{} & \multicolumn{2}{c}{No Accretion} &   \colhead{}   &
\multicolumn{3}{c}{Accretion} \\
\cline{3-4} \cline{6-8} \\
\colhead{$\Mp$} & \colhead{} &
\colhead{$\widehat{M}_\mathrm{e}$} & \colhead{$\widehat{\rho}_\mathrm{e}$} & 
\colhead{}   & 
\colhead{$\widehat{M}_\mathrm{e}$\tablenotemark{a}} &
\colhead{$\widehat{M}_\mathrm{e}$\tablenotemark{b}} &
\colhead{$\widehat{M}_\mathrm{e}$\tablenotemark{c}}
}
\startdata
  67 & & $7.07\times10^{-1}$ & $9.37\times10^{-11}$ & 
        & $1.07\times10^{-1}$ & \nodata               & $7.67\times10^{-2}$ \\
  29 & & $2.74\times10^{-1}$ & $1.45\times10^{-10}$ & 
        & $6.49\times10^{-2}$ & \nodata               & \nodata \\
  20 & & $2.26\times10^{-1}$ & $2.43\times10^{-10}$ & 
        & $3.68\times10^{-2}$ & $3.18\times10^{-2}$ & \nodata \\
  10 & & $1.29\times10^{-2}$ & $7.12\times10^{-11}$ & 
        & $9.76\times10^{-3}$ & \nodata               & \nodata \\
\phn5 & & $2.24\times10^{-3}$ & $5.56\times10^{-11}$ & 
        & $2.17\times10^{-3}$ & \nodata               & \nodata \\
\enddata
 
 
\tablenotetext{a}{From models with $\kappa_\mathrm{ac}=0.10\,\RH$.} 
\tablenotetext{b}{From models with $\kappa_\mathrm{ac}=0.15\,\RH$.} 
\tablenotetext{c}{From models with $\kappa_\mathrm{ac}=0.20\,\RH$.} 
\tablecomments{All masses are relative to the Earth's mass. The mean density
            $\widehat{\rho}_\mathrm{e}$ within the planet's radius is
            expressed in cgs units.
            We consider only  
            simulations which were run with the Stevenson's potential.
            The mass $\widehat{M}_\mathrm{e}$ is evaluated assuming
            a disk mass $M_\mathrm{D}=3.5\times10^{-3}\;\MSun$.
            Hence, the values in the Table do not account for the disk 
            depletion rate $\dot{M}_\mathrm{D}$, which is on the order of 
            $3\times10^{-3}\;\MEarth\,\mathrm{yr}\superscr{-1}$.}
\end{deluxetable}
Here we should dedicate some attention to the differences existing between
accreting and non-accreting protoplanets. Since the gas is locally isothermal,
pressure is proportional to the density according to \refeqt{p}. Because of the
mass removal, density nearby the planet is lower in accreting models
than it is in non-accreting ones. In \refTab{tbl:noac} the mass 
$\widehat{M}_\mathrm{e}$ enclosed within the envelope radius $\Se$ is quoted 
for the two sets of models, along with the mean density 
$\widehat{\rho}_\mathrm{e}$. These values demonstrate that the amount of
material contained in the volumes of non-accreting planets can be considerably
larger than in the other case (even 6 times as much).
Since the pressure must converge in the two cases, when the distance from 
the planet $S\gg\Se$, a higher 
mean pressure in the envelope intuitively implies a larger magnitude of the 
pressure gradient inside this region.

\begin{figure*}
\epsscale{0.8}
\begin{center}
\mbox{%
\plotone{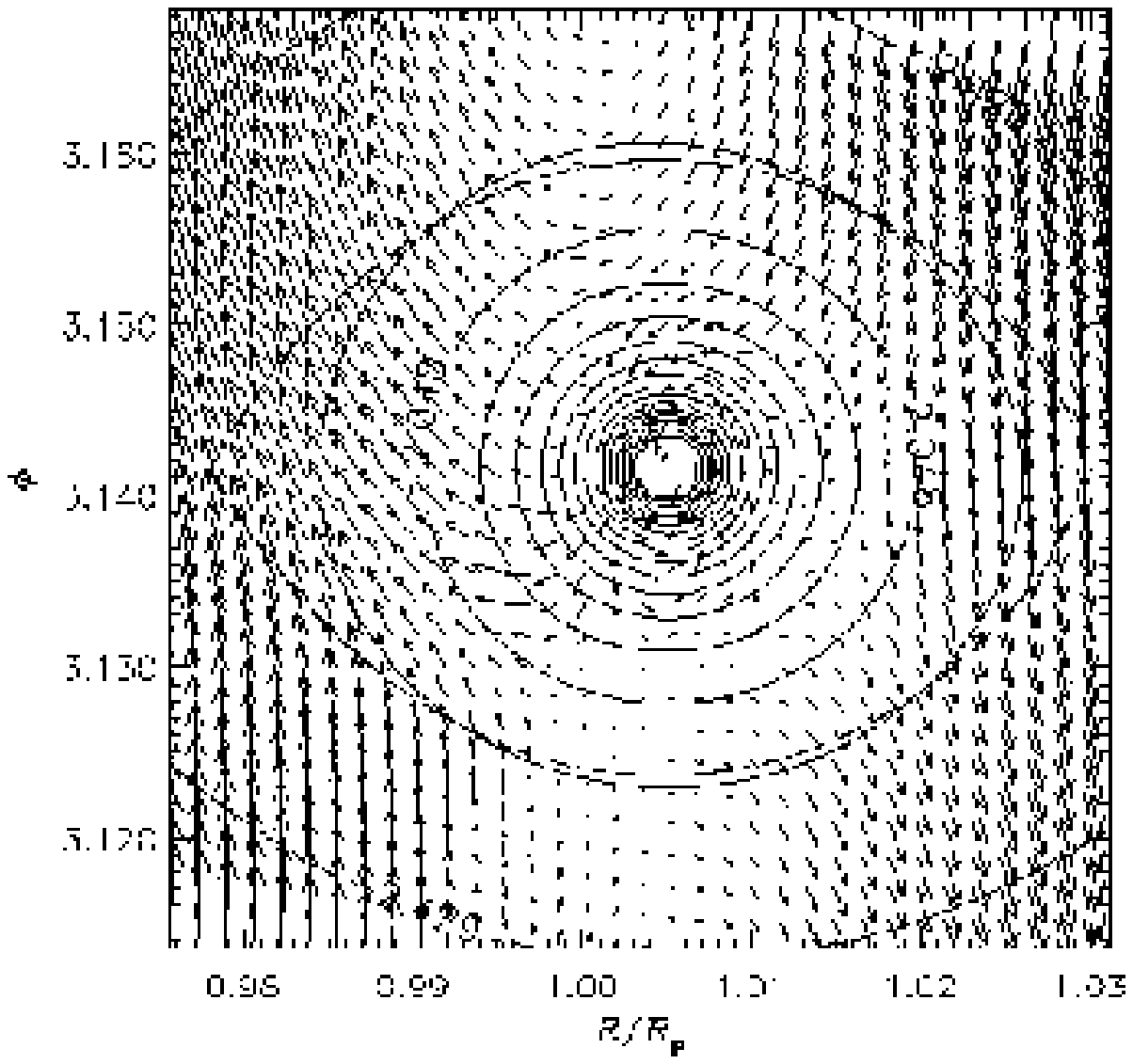}\hfill%
\plotone{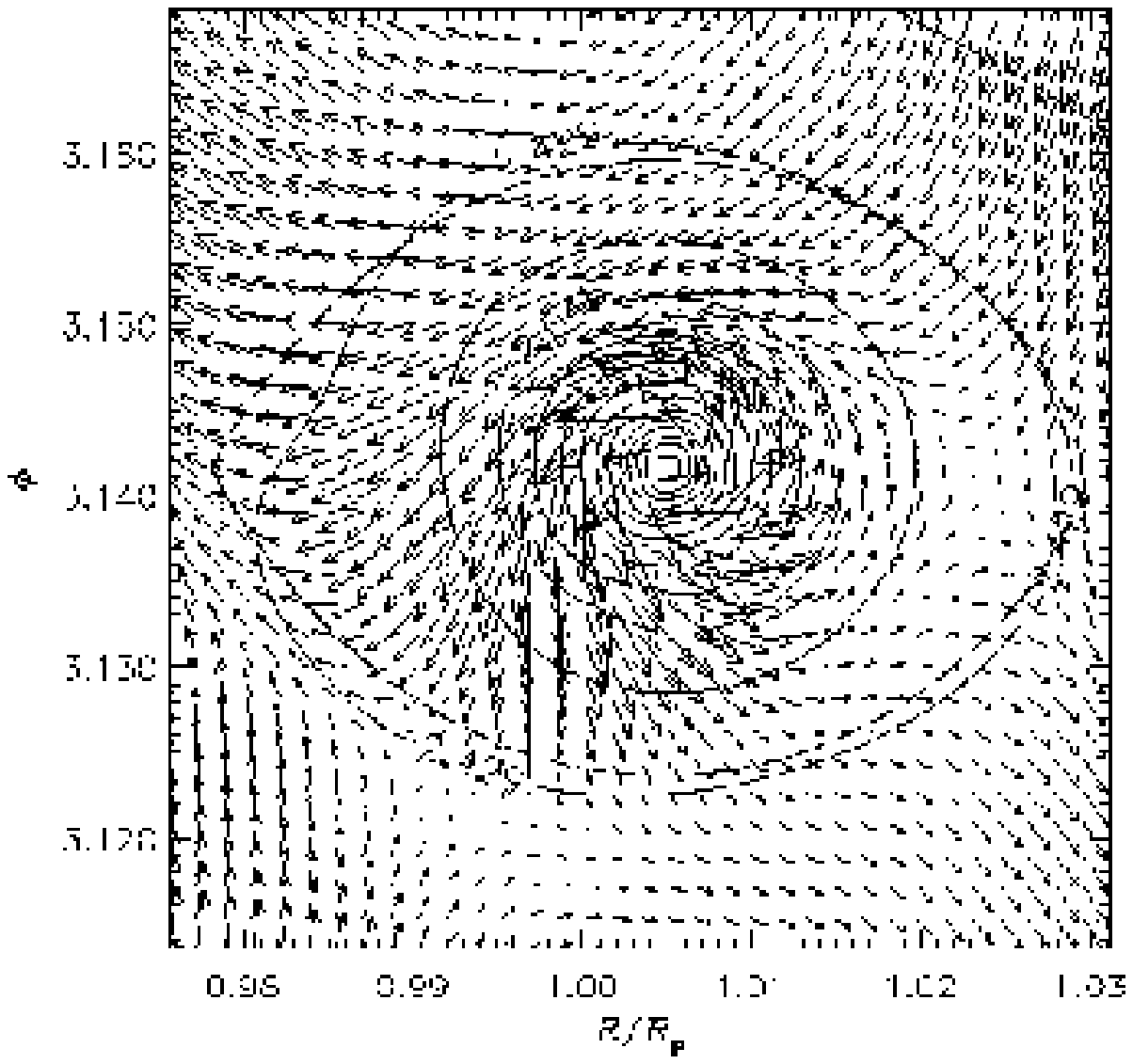}}
\mbox{%
\plotone{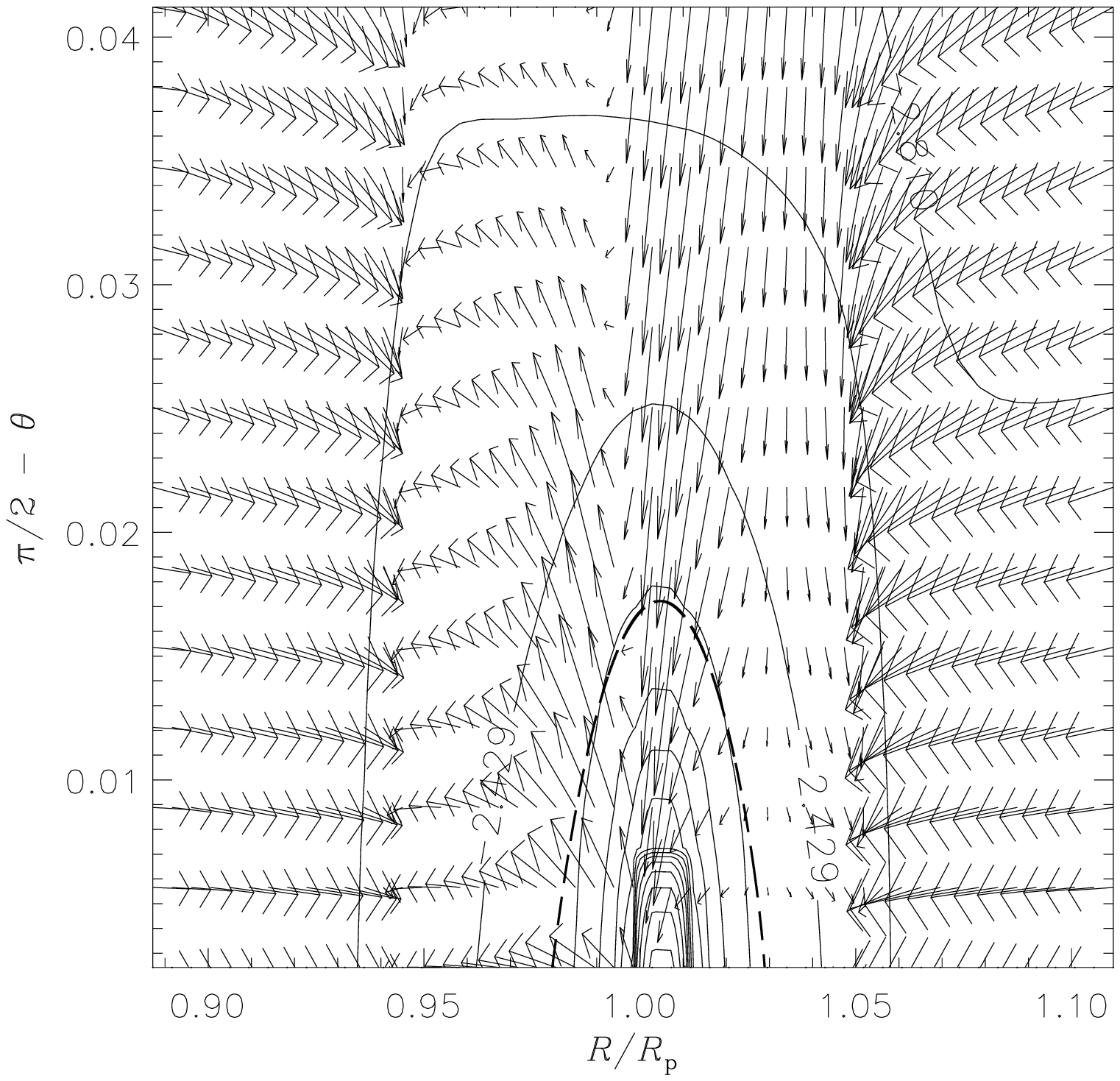}\hfill%
\plotone{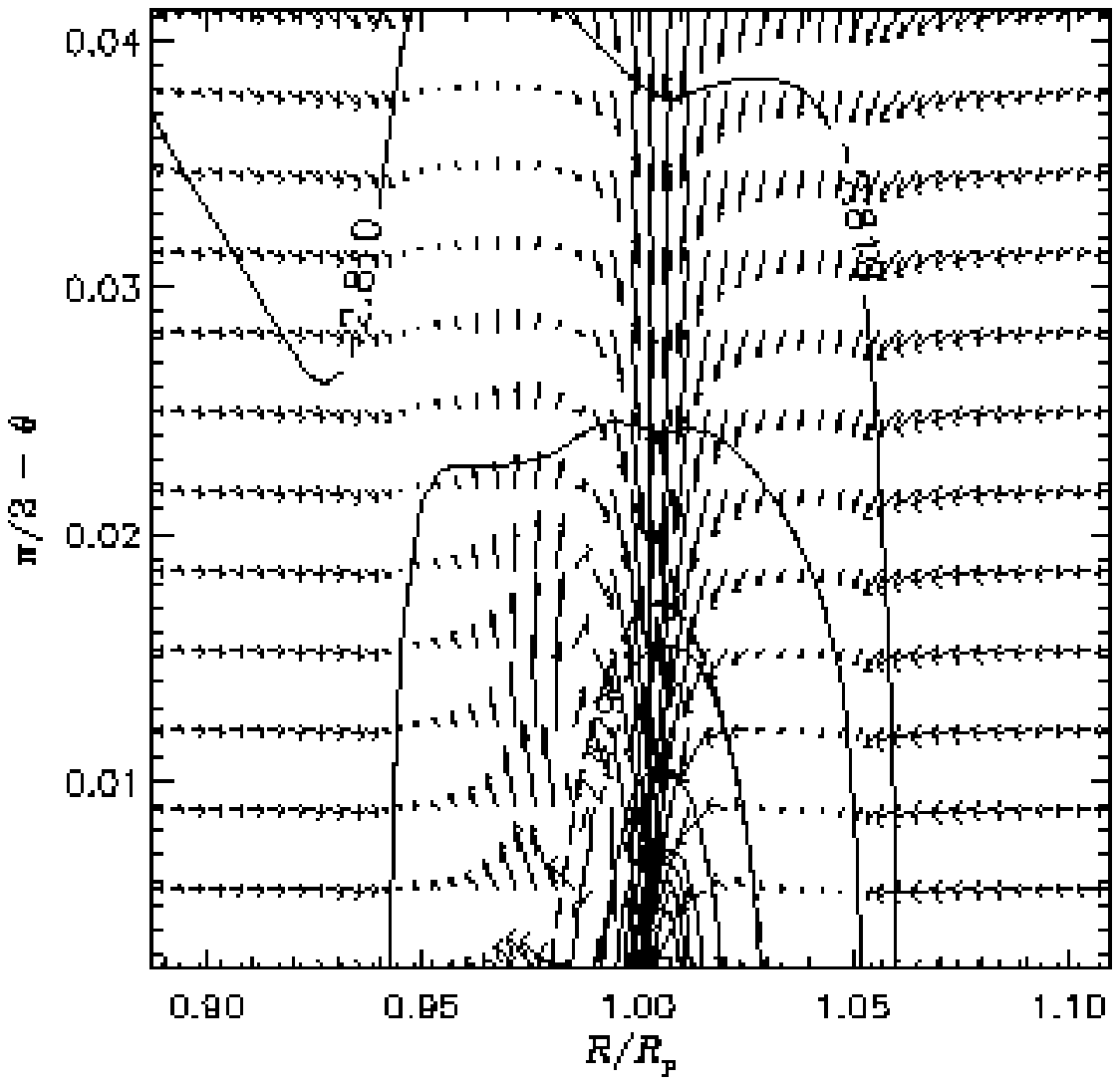}}
\mbox{%
\plotone{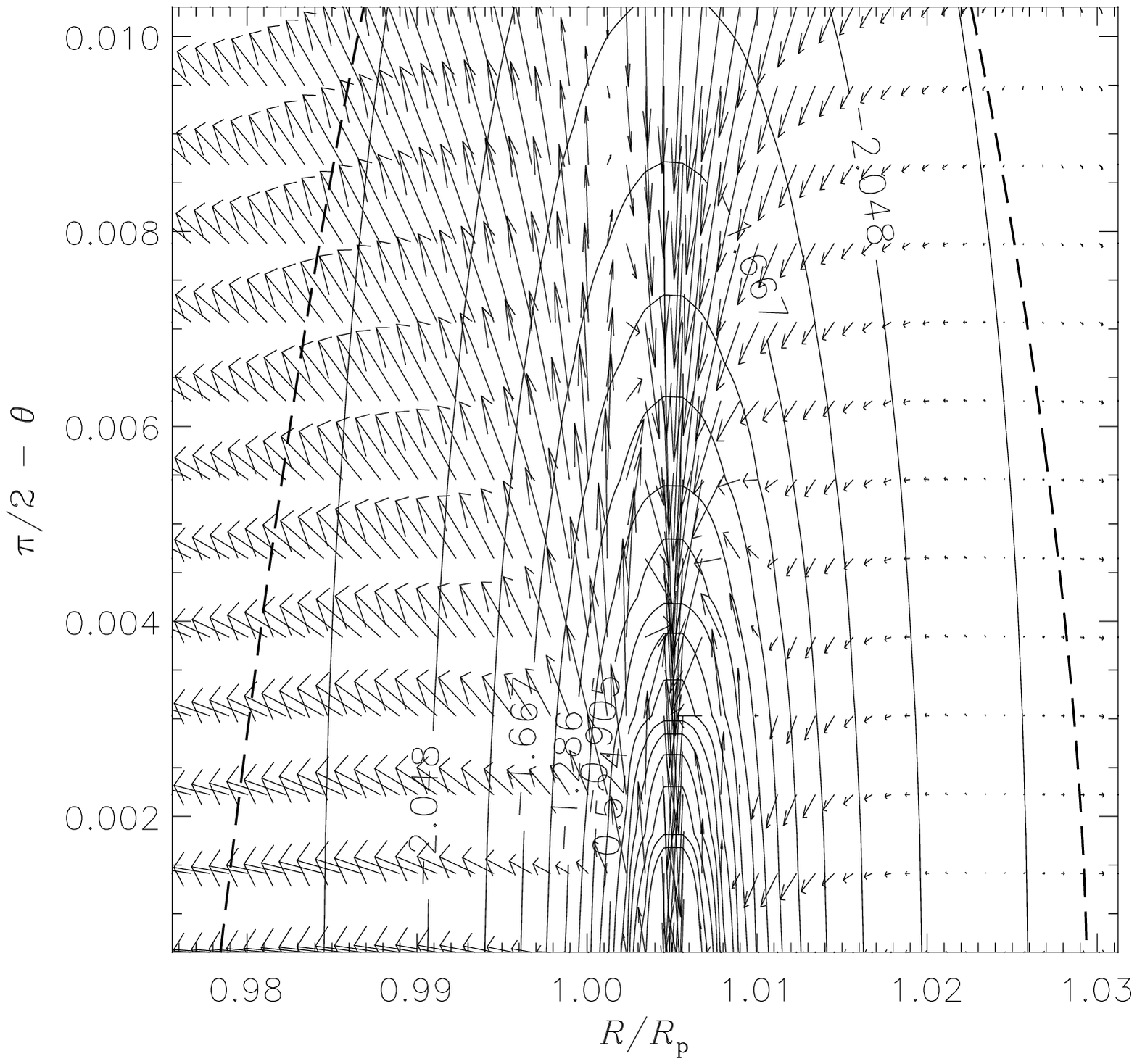}\hfill%
\plotone{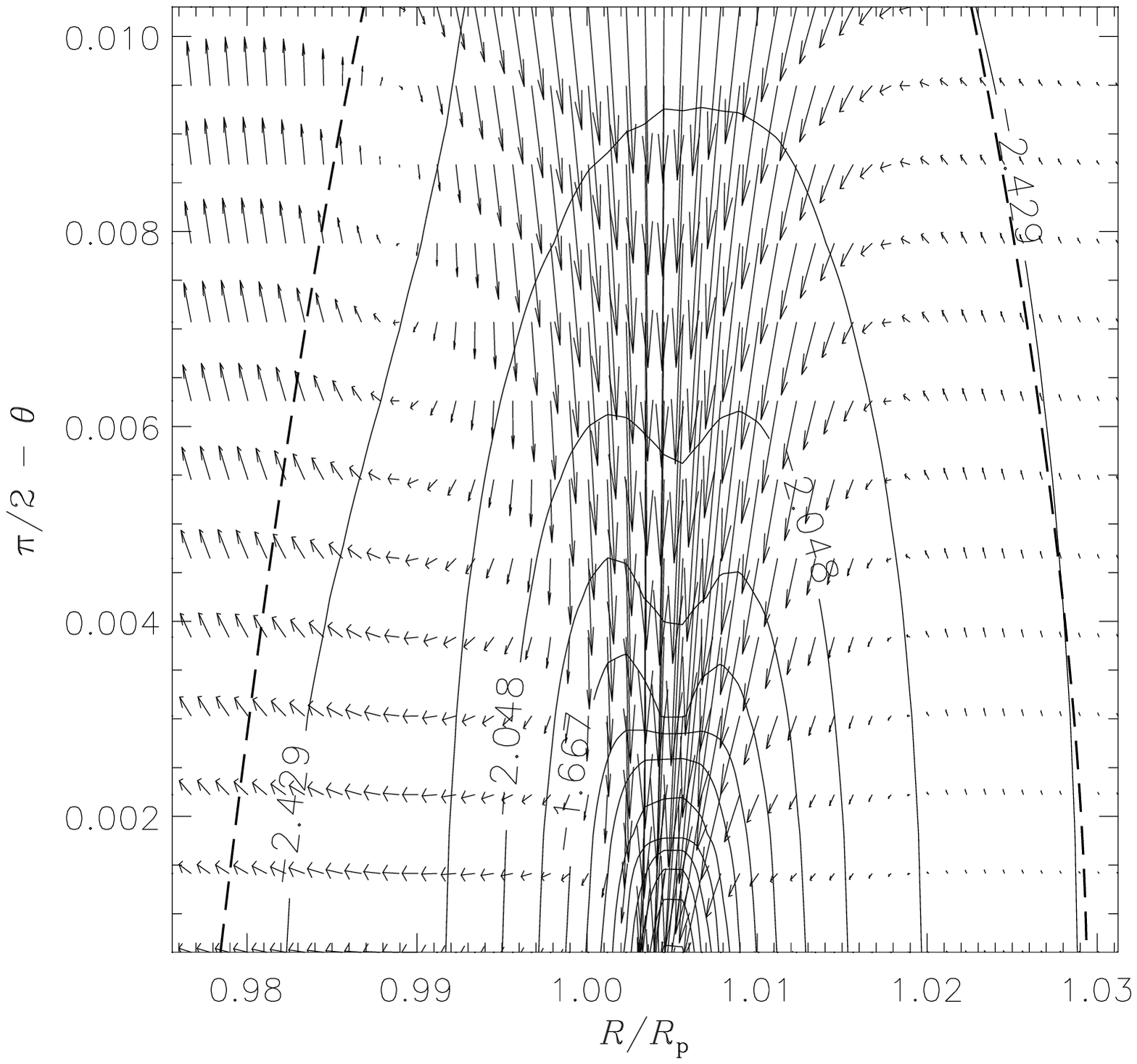}}
\end{center}
\caption{%
       Two-component velocity field on the equatorial plane (top row )
       and on the slice $\varphi=\varphi_\mathrm{p}$ (middle and bottom
       rows) for a non-accreting, $20\;\MEarth$ planet 
       (\textbf{left panels}),
       and an accreting planet of the same mass (\textbf{right panels}) with
       an accreting sphere radius $\kappa_\mathrm{ac}=0.15\,\RH$.
       Lines of equal $\log{\rho}$ are also drawn. 
       The dashed line represents the Roche lobe of the restricted three-body 
       problem.
       In the accreting model, the equatorial flow ($\theta=\pi/2$) in
       the planet's envelope has a Mach number $\mathcal{M}\lesssim
       1.2$ whereas, for the 
       meridional flow, $\mathcal{M} \lesssim 2$. 
       In the non-accreting computation
       the fluid travels subsonically in both the midplane and the
       meridional slice ($\mathcal{M} \lesssim 0.2$).
\label{fig:noac}}
\end{figure*}

As an example, we illustrate in \refFgt{fig:noac} the velocity field in two 
perpendicular slices $\theta=\pi/2$ (i.e., the equatorial plane) and 
$\varphi=\varphi_\mathrm{p}$ (i.e., the surface $R$--$\theta$ passing through
the planet),
in order to show how the enhanced density values affect the local circulation. 
The targeted body has $\Mp=20\;\MEarth$ because, among the eight 
available models 
for which accretion is not considered (see \refTab{tbl:kac}), this is the 
one that suffers an orbital migration substantially different from accreting
counterpart models. From the isodensity lines displayed in \refFgt{fig:noac},
one can infer that matter is spherically distributed around the non-accreting
protoplanet (left panels). For this reason the net torque arising within a 
region of radius $\approx \RH$ is nearly zero. 
This does not happen in the other case because the symmetry is not so strict.

The upper right panel clearly indicates the existence of
a rough balance between gravitational and centrifugal force, with the pressure 
gradient playing a marginal role in opposing the planet potential gradient.
On the other hand, from the circulation in the upper left panel of 
\refFgt{fig:noac} one can deduce that the pressure gradient is no longer 
negligible compared to the potential gradient and can therefore 
counterbalance its effects.

Moreover, the flow above the disk midplane (center and lower left panels) 
suggests that gas is ejected at $R<R_\mathrm{p}$.
Such phenomenon must be ascribed to the pressure gradient as well,
since the fluid opposes any further compression. 
These qualitative arguments will be quantitatively 
corroborated in \refsec{sec:D}, where we will show that the increased 
amount of matter causes the envelope to be pressure supported.

We mention in the caption of \refFgt{fig:noac} that the flow may
travel supersonically within the atmospheric region, though
both Stevenson's and Wuchterl's gravitational potential rely on the
hypothesis of quasi-hydrostatic equilibrium.
Such discrepancy can be attributed to the rate of mass removal from the
innermost parts of the planet's envelope, which is not considered in
the derivation of those analytic solutions.
In fact, while supersonic speeds have been measured in some
accreting models, the flow is always subsonic in the envelope of
non-accreting planets.
This can be also understood with simple arguments. 
We said before in this section that mass accretion is responsible for the
Keplerian-like circulation around protoplanets, in the disk midplane. 
Hence, the local (equatorial) Mach number should be on the order of 
$\sqrt{q\,R_\mathrm{p}/S}/h$. If evaluated at
$S\approx\Se$, this quantity yields $\mathcal{M}\approx1.3$ for a
$20\;\MEarth$ body, which is
comparable to the value reported in the caption of \refFgt{fig:noac}.
Along the vertical direction, if the velocity is approximated to that
of a spherically accreting flow then
$u_{\theta}\approx\dMp/(4\,\pi\,S^2\,\rho)$. 
At $S=\kappa_{\mathrm{ac}}=0.1\,\RH$, that relation gives a
(meridional) Mach number $\mathcal{M}\approx 1.8$ when applied to the
accreting model shown in \refFgt{fig:noac}.
Also this number is similar to the measured value.
\subsection{Gravitational Torques and Orbital Migration}
\label{ssec:TM}

Gravitational torques are believed to be responsible for 
the migration of protoplanets from their initial formation sites. 
In this work torques are directly estimated from the gravitational
force
exerted by each fluid element of the circumstellar disk on the
planet. When computing the gravitational force, we consider
the density solution on the finest available grid level. 
This procedure permits to obtain higher accuracy because of the
increasing resolution of hierarchy levels.

Since the planet is an extended object, torques acting on each of its
portions should be calculated and then added up to give the 
total torque vector, whose most general expression\footnote{We would
like to stress that in this form all of the gravitational
contributions (due to star, the planet, disk self-gravity, etc.) are
implicitly enclosed in the differentials $dM_\mathrm{D}(\gv{R})$ and 
$dM_\mathrm{p}(\gv{S})$.} is
\begin{eqnarray}
\gv{\mathcal{T}_\mathrm{D}}& =& 
           \int_{M_\mathrm{D}}\int_{M_\mathrm{p}}
          (\gv{R}_\mathrm{p} \gv{+} \gv{S})  \nonumber \\
           &\gv{\times}& \frac{G\,dM_\mathrm{D}(\gv{R})\,dM_\mathrm{p}(\gv{S})}%
           {|\gv{R} - \gv{R}_\mathrm{p} - \gv{S}|^{3}}\,\gv{R}\,.
\label{Ttot} 
\end{eqnarray}
\refEqt{Ttot} explicitly states that circumstellar material can alter
both the orbital and rotational spins of a protoplanet.
However, here we shall confine our study to variations of the planet's
orbital angular momentum because the evaluation of the rotational spin
requires a rigorous treatment of the envelope self-gravity.
This is not done here, as stated in \refsec{sec:PD} (see \refeqp{phit}).
Thereby, we can proceed as if the whole 
planetary mass were concentrated in its geometrical center 
$\gv{R}=\gv{R}_\mathrm{p}$ and integrate the force 
over the whole disk domain, excluding the volume
$\mathcal{V}_\mathrm{p}$ occupied by the planetary envelope.
Yet, outside of such volume the gravitational potential is always that
of a point-mass object (see the behavior of eqs.~[\ref{phipms}], [\ref{phihs}],
[\ref{phist}], and [\ref{phikw}], for $S>\Se$), therefore \refeqt{Ttot}
simplifies and becomes
\begin{equation}
\gv{\mathcal{T}_\mathrm{D}} = 
           \gv{R}_\mathrm{p} \gv{\times}
           \int_{\gv{R}\notin\mathcal{V}_\mathrm{p}}
           \nabla\Phi^\mathrm{PM}_\mathrm{p}\,\rho(\gv{R})\,dV(\gv{R}).
\label{simpleTtot} 
\end{equation}

The orbital angular momentum of a protoplanet can be affected
only by the $z$-component (that parallel to the polar axis) of the
torque vector $\gv{\mathcal{T}_\mathrm{D}}$.
To avoid useless distinctions, we indicate this component as
$\mathcal{T}_\mathrm{D}$.
The sign of $\mathcal{T}_\mathrm{D}$ determines the gain (positive
torques) or loss (negative torques) of orbital spin. Larger spins 
correspond to more distant orbits.
Since torques generally change on time scales on the order of $\sim50$
revolutions, we work with their final magnitudes. 
We usually observe a slow decay of $|\mathcal{T}_\mathrm{D}|$ with time
(see also the end of \refsec{ssec:MA}).

Two-dimensional computations reveal that torques exerted by circumplanetary
material may amount to a fair fraction of the total torque, unless a
suitable smoothing length (usually of the size of the Hill radius) is used 
in the planet gravitational potential $\Phi_\mathrm{p}$. 
This is caused by the
high surface densities reached around the planet and the lack of a 
vertical torque
decay which naturally occurs in three dimensions (see \refsec{sec:D}).
In fact, in 3D, we observe that torques arising from locations close
to the planet do not play such an important role as they do in 2D.

Analyzing the relative strength of torques exercised by different disk
portions, it turns out that in the mass range 
$q\in[2\times 10^{-4},1\times 10^{-3}]$, 
dominating negative torques arise from distances 
$S\gtrsim 1.2\,h\,R_\mathrm{p}$, where $h$ is the disk aspect ratio. 
Below $33\;\MEarth$, the largest net contributions are generated by 
material lying between $S\simeq 0.6\,h\,R_\mathrm{p}$ and 
$S\simeq 2.2\,h\,R_\mathrm{p}$.
Therefore we can conclude that predominant torques are exerted
at distances from the planet comparable with the Hill radius. 
Not more than 10\% of $\mathcal{T}_\mathrm{D}$ is built up by matter 
located within $S\approx 0.6\,h\,R_\mathrm{p}$.

Apart from the $20\;\MEarth$ protoplanet, the total torque evaluated in 
non-accreting models does not deviate considerably from that estimated 
in accreting ones, independently of the used potential. 
In fact,
simulations based on the potentials $\Phi^\mathrm{PM}_\mathrm{p}$,
$\Phi^\mathrm{HS}_\mathrm{p}$, and $\Phi^\mathrm{ST}_\mathrm{p}$ 
(eqs.~[\ref{phipms}], [\ref{phihs}], and [\ref{phist}], respectively) 
supply values of
$\mathcal{T}_\mathrm{D}$ which differ by less than 40\%.
This circumstance may signify that, whether or not a protoplanet is still
accreting matter from its surroundings, this is not generally  crucial to the 
gravitational torques by the circumstellar disk. 
Thereby, being an exception, the case $\Mp=20\;\MEarth$ deserves some 
comments. 
For such mass,
the torque integrated over the first two grid levels 
($S\gtrsim 2.2\,h\,R_\mathrm{p}$) yields a positive
value for both accreting and non-accreting planets. When adding 
the contributions from the third and forth level 
($0.6\,h\,R_\mathrm{p} \lesssim S \lesssim 2.2\,h\,R_\mathrm{p}$), 
the torque experienced by the planet lowers but, while it becomes negative in 
the accreting case, it still remains positive in the non-accreting 
counterpart. It is especially matter residing between $1$ and $2\,\RH$
from the planet that builds up the difference.
As material in the uppermost grid level ($S\lesssim 0.6\,h\,R_\mathrm{p}$)
of the non-accreting simulation does not exert any significant torques 
(some little negative contribution is indeed measured in the accreting model), 
$\mathcal{T}_\mathrm{D}$ keeps the positive sign 
although, in magnitude, it is eleven times as small as 
that evaluated in the accreting case.
This phenomenon of torque reversal for non-accreting planets with masses
of about $20\;\MEarth$ may be related to the very long
migration time scales obtained for fully accreting models having masses 
$\Mp\approx 10\,\MEarth$ (see below, and \refFgp{fig:tau_m}).

Conservation of orbital angular momentum implies that a protoplanet has
to adjust its orbital distance from the central star because of external 
torques exerted by the disk. If the orbit remains circular, the time scale
over which this radial drift motion happens is inversely proportional to 
$\mathcal{T}_\mathrm{D}$, according to the formula:
\begin{equation}
\tau_\mathrm{M}=\frac{a}{|\dot{a}|}
               =\frac{\Mp\,a^2\,\Omega_\mathrm{p}}{2\,%
                |\mathcal{T}_\mathrm{D}|}.
\label{taum}
\end{equation}
In \refeqt{taum} we indicated with $\Omega_\mathrm{p}$ and $a$ the planet's
angular velocity and its distance from the star, respectively.
Linear, analytical theories  \citep[e.g.,][]{ward1997} provide two separate 
regimes governing the migration of low- (\textit{type~I}) and 
high-mass (\textit{type~II}) protoplanets.
Both migration types predict that the planet moves toward the star.
More recent studies by \citet{masset2001} and \citet{tanaka2002} have 
reconsidered the role of co-orbital corotation torques and proved that they
can be very effective in opposing Lindblad torques. 
Hence, they can significantly slow down inward migration. 
Two-dimensional results presented in \DHK\ well fit to these predictions.
A further reduction of the migration speed is expected from a full 3D
treatment of torques, as also derived numerically by \citet{miyoshi1999} and 
theoretically predicted by \citet{tanaka2002}.

\begin{figure*}[!t]
\epsscale{1.5}
\plotone{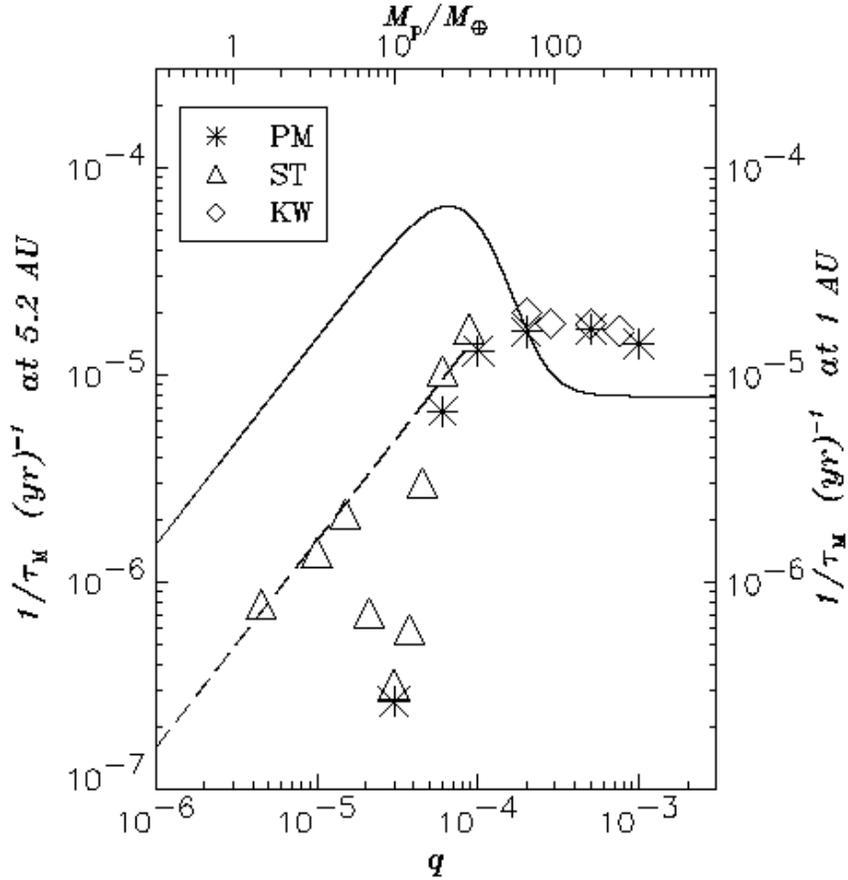}
\caption{%
       Migration time scale versus the mass ratio $q=\Mp/M_\bigstar$.
       Outcomes from simulations carried out with the various forms of 
       gravitational potential are marked with different symbols.
       At $q=10^{-4}$ and $6\times10^{-5}$, calculations carried out
       with the point-mass potential were initiated with a density gap
       (see \refsec{ssec:IC}).  
       To avoid confusion, migration rates obtained from models with 
       $\Phi_\mathrm{p}=\Phi_\mathrm{p}^\mathrm{HS}$ (\refeqp{phihs})
       are quoted in \refTab{tbl:taumhs}.
       The solid line represents the theoretical prediction by 
       \citet{ward1997}, which
       does not include corotation torques and 3D effects. Both are indeed
       accounted for in the analytical model by \citet{tanaka2002} 
       (dashed line). 
       The parameter 
       $\pi\,a^2\,\Sigma_{\mathrm{p}}/M_\mathrm{D}$, needed to draw the
       analytic curves,
       was retrieved from the initial surface density profile (see
       \refsec{ssec:IC}).
       The scale on the right vertical axis gives the migration rates of a
       protoplanet orbiting at $1\;\AU$ from the primary.
       \label{fig:tau_m}}
\end{figure*}
\refFgt{fig:tau_m} shows our estimates for the migration time scale 
$\tau_\mathrm{M}$ as computed for models having different masses and
in which  
the potential solutions $\Phi^\mathrm{PM}_\mathrm{p}$,
$\Phi^\mathrm{KW}_\mathrm{p}$, and $\Phi^\mathrm{ST}_\mathrm{p}$ were
adopted.
We compare these values with the two analytical theories developed by 
\citet{ward1997} (solid line) and \citet{tanaka2002} (dashed line). 
The first of them comprises both migration regimes, though accounting only 
for Lindblad torques.
The second theory is limited to type~I migration, albeit it treats both
Lindblad and corotation torques. Moreover, the first is explicitly 
two-dimensional whereas the second is applicable in two as well as three 
dimensions.

As one can see from \refFgt{fig:tau_m}, numerical results are very similar
for $\Mp\geq 67\;\MEarth$, yielding
$\tau_\mathrm{M}\approx 5\times 10^{4}\;\mathrm{years}$,
whatever of the four gravitational potential is used 
(see also \refTab{tbl:taumhs}).
While this time scale is consistent with Ward's 1997 description if 
$\Mp= 67\;\MEarth$, for more massive planets
it is nearly two times as short as the theoretical prediction.
The depletion of the disk inside the planet's orbit is probably responsible
for part of the discrepancy (see \refsec{ssec:NE}), because
Ward's theory assumes a disk with a constant unperturbed surface density.
In the type~I regime, our numerical experiments with 
$\Phi_\mathrm{p}=\Phi_\mathrm{p}^\mathrm{ST}$ (\refeqp{phist}) 
well reproduce the behavior of the analytical curve 
when $\Mp=30$, $20$, $5$, $3.3$, and $1.5\;\MEarth$. 

Computations executed with $\Phi_\mathrm{p}=\Phi_\mathrm{p}^\mathrm{HS}$ 
(\refeqp{phihs}) probably
underestimate the magnitude of differential torques because of the much 
weaker gravitational field.
Nevertheless, for Jupiter-mass and Earth-mass protoplanets, 
migration times yielded by these 
models well compare to those displayed in \refFgt{fig:tau_m}, as proved by
the values reported in \refTab{tbl:taumhs}. 
\begin{deluxetable}{cccccc}
\tablecolumns{5}
\tablewidth{0pt}
\tablecaption{Migration Times obtained with the Homogeneous
              Sphere Potential.\label{tbl:taumhs}}
\tablehead{%
\colhead{} & \colhead{}  &  \colhead{} & 
               \multicolumn{3}{c}{$\tau_\mathrm{M}\;[\mathrm{years}]$} \\
\cline{4-6} \\
\colhead{$\Mp/\MEarth$} & \colhead{$q$} & \colhead{} &
\colhead{Accretion} & \colhead{}   & \colhead{No Accretion}
}
\startdata
333 &$1.0 \times 10^{-3}$ &  & $5.04 \times 10^{4}$ &  
      & $4.80 \times 10^{4}$ \\
166 &$5.0 \times 10^{-4}$ &  & $4.76 \times 10^{4}$ &  & \nodata \\
\phn33 &$1.0 \times 10^{-4}$ &  & $2.10 \times 10^{5}$ &  
      & $2.54 \times 10^{5}$\\
\phn20 &$6.0 \times 10^{-5}$ &  & $5.74 \times 10^{5}$ &  & \nodata \\
\phn10 &$3.0 \times 10^{-5}$ &  & $5.26 \times 10^{5}$ &  
      & $4.78 \times 10^{5}$ \\
\phn\phn5 &$1.5 \times 10^{-5}$ &  & $6.19 \times 10^{5}$ &  & \nodata \\
\enddata
 
\tablecomments{Migration rates from accreting and non-accreting models 
               in which $\Phi_\mathrm{p}=\Phi_\mathrm{p}^\mathrm{HS}$ 
               (\refeqp{phihs}). Time scales are sensibly different
               from those
               indicated in \refFgt{fig:tau_m} only at 
               $\Mp=33$, $20$, and $10\;\MEarth$.}
\end{deluxetable}

Significant deviations from the linear estimate of \citet{tanaka2002} 
are observed in the mass interval $[7,15]\;\MEarth$, where 
the migration time is longest at $10\;\MEarth$. For this planet
$\tau_\mathrm{M}$, estimated with Stevenson's as well as the
point-mass potential,
is thirty times as long as the theoretical description
by \citet{tanaka2002}
predicts. This depends on the strong positive torques arising at
$S > 2\,h\,R_\mathrm{p}$ which are not efficiently contrasted by negative ones 
generated inside $S \simeq h\,R_\mathrm{p}$.
However, for this particular planetary mass, we obtain discrepant estimates 
from computations performed with different resolutions.
In fact, the simulation carried out with the grid G2 yields a positive
total torque acting on the planet, i.e., it predicts an outward migration,
whereas models based on the higher resolution hierarchies G3 and G4 provide a
negative total torque. Yet,
the absolute value of $\mathcal{T}_\mathrm{D}$ evaluated with grid G4 
is a seventh of that achieved with grid G3.

Since we believe that gravitational torques are accounted for in a more 
accurate fashion by hierarchy G3 than by grid G4 because of the arguments 
in \refsec{ssec:NE}, we may rely more on the outcome of the
hierarchy G3 (shown \refFgp{fig:tau_m}) rather than on the other two. 
We note that such migration time ($\tau_\mathrm{M}=3.3\times 10^6$ years) 
is roughly the double of that supplied 
by the $\Mp=10\;\MEarth$, non-accreting model.

We have further inquired into the matter by running a simulation with
$\Mp=12.5\;\MEarth$ and $\Phi_\mathrm{p}=\Phi^\mathrm{ST}_\mathrm{p}$
(see \refTab{tbl:MpSp}). Based on the experience acquired with models
executed with grids G3 and G4, we set up the high-resolution hierarchy
G5 (see \refTab{tbl:grids}).
This grid is designed to better resolve those regions responsible for
the strongest torque contributions in the mass interval
$[7,15]\;\MEarth$. As seen in \refFgt{fig:tau_m}, the resulting
migration rate follows the trend established by the other assessments
in this range of masses.

The interesting property that computed migration time scales are very
long for $10\;\MEarth$ planets may be caused by non-linearity
effects. We note, in fact, that in numerical simulations the first traces
of a trough in the density structure is observed around the same
value of $\Mp$ and 
gap formation starts when disk-planet interactions become non-linear.
However,
this issue needs to be addressed more thoroughly with future computations.

\subsection{Mass Accretion}
\label{ssec:MA}

\begin{figure*}[!t]
\epsscale{1.5}
\plotone{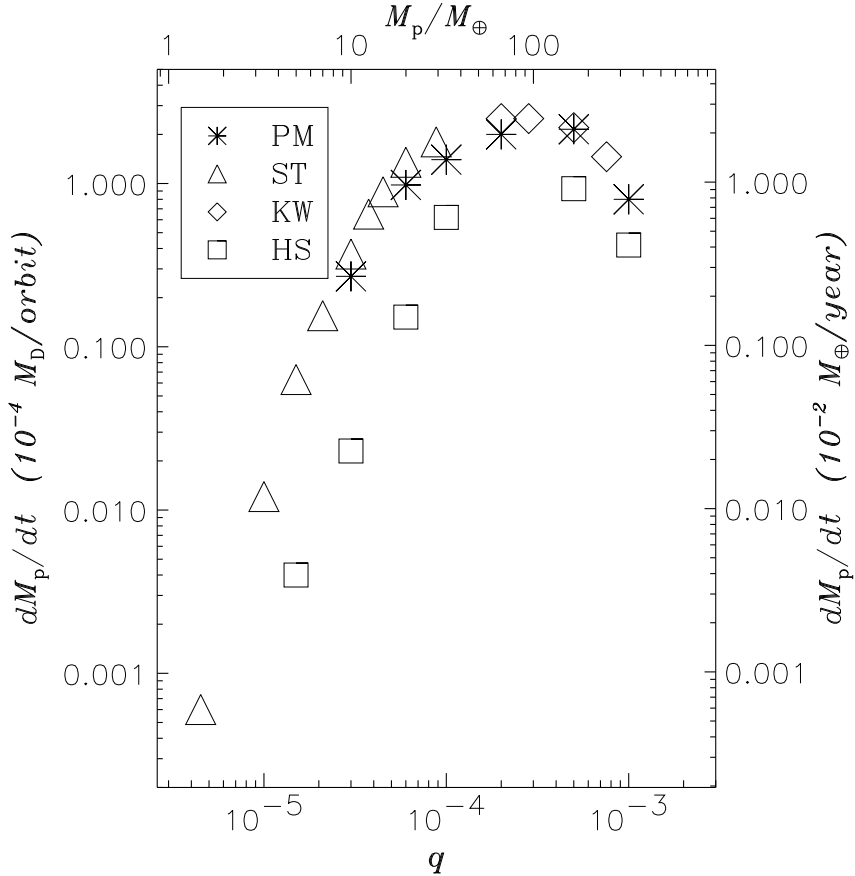}
\caption{%
       Planet's accretion rate as function of the normalized planet mass $q$.
       Different symbols stand for the different forms of gravitational 
       potential $\Phi_\mathrm{p}$ adopted in the computations.
       Apart from those models run with the homogeneous sphere potential
       $\Phi_\mathrm{p}^\mathrm{HS}$ (see \refeqp{phihs}), 
       mass is removed from a volume, centered on the planet, with radius 
       $\kappa_\mathrm{ac}=0.1\,\RH$ (see \refTab{tbl:kac} for some details
       concerning the simulation with $\Mp=1.5\;\MEarth$).
       Models for which
       $\Phi_\mathrm{p}=\Phi_\mathrm{p}^\mathrm{HS}$ have 
       $\kappa_\mathrm{ac}$ equal to $0.2\,\RH$ if $\Mp>20\;\MEarth$
       and to $0.15\,\RH$ if $\Mp=20\;\MEarth$; 
       otherwise $\kappa_\mathrm{ac}$ is set to $0.1\,\RH$.
       \label{fig:mpdot}}
\end{figure*}
Three-dimensional computations of one Jupiter-mass bodies provide 
estimates of the mass accretion rate $\dMp$ on the same order
of magnitude as those obtained by two-dimensional ones (see \KDH).
Two-dimensional calculations performed by the authors reveal a maximum 
of the accretion rate, as function of the mass, around $0.5\;\MJup$ 
(see \DHK). 
Yet, those estimates appear surprisingly high in the very low mass
limit. Part of the reasons may lie in the assumed flat geometry
which cannot account for the vertical density stratification.
The present simulations overcome this restriction, hence they permit 
to evaluate also the effects due to the disk thickness.

The values of $\dMp$ is plotted against the planetary mass in 
\refFgt{fig:mpdot}. As comparison, estimates relative to models with
different gravitational potential solutions are shown. 
The overall behavior of the data points resembles that reported in \DHK, 
with a peak around $0.3\;\MJup$. 
For $\Mp=1\;\MJup$ the agreement between two and three-dimensional models 
is very good and not much discrepancy is seen down to $\Mp=20\;\MEarth$,
since values are comparable within a factor 3 (see \refsec{ssec:C2D}).
Below this mass, however, the accretion rate rapidly declines,
which drop is not observed in 2D outcomes.
In fact, one can infer from \refFgt{fig:mpdot} that the dynamical range of 
$\dMp$ stretches for more than two orders of magnitude. By using
model results obtained applying the point-mass, Stevenson's, and 
Wuchterl's potential (eqs.~[\ref{phipms}], [\ref{phist}], and [\ref{phikw}],
respectively) 
the following approximate relation can be found:
\begin{equation}
\log\left[\frac{\dMp}{\MEarth/yr}\right] 
\simeq b_0 + b_1\,\log{q}\, + b_2\,\left(\log{q}\right)^{2},
\label{fitmpdot}
\end{equation}
whose coefficients are $b_0=-18.47\pm 0.76$, $b_1=-9.25\pm 0.38$, 
and $b_2=-1.266\pm 0.046$.
\refEqt{fitmpdot} holds as long as the mass ratio 
$q\in[4.5\times10^{-6},10^{-3}]$ or,
for a one solar-mass star, when $1.5\;\MEarth\le \Mp \le 1\;\MJup$.
Such an equation can be applied to scenarios studying the global long-term 
evolution of young planets.

Calculations in which the homogeneous sphere potential 
$\Phi^\mathrm{HS}_\mathrm{p}$ (\refeqp{phihs}) is adopted yield
accretion rates substantially lower (from $3$ to $15$ times) than those 
achieved when the other potential forms are employed.
This is due to the weak gravitational attraction this potential exerts
within the planet's envelope. As proved by \refFgt{fig:ag}, the gravitational 
field can be $100$ times as small as that established by
the other three potential functions for $S\leq \Se$.
Also in this circumstance, a relation similar to \refeqt{fitmpdot} exists
for which the coefficients
are $b_0=-19.42\pm 2.68$, $b_1=-9.96\pm 1.41$, and $b_2=-1.42\pm 0.18$.

While the accretion rate is fairly stable with time for masses below 
$30\;\MEarth$, it keeps reducing for higher masses. 
Between $67\;\MEarth$ and $0.8\;\MJup$, $\dMp$ 
drops by 10 to 20\% during the last $50$ orbits of the simulations.
This is an indication of a deepening gap and a depleting disk.
As for the dependency upon the accretion volume, 
from our numerical experiments it is found that doubling the radius 
$\kappa_\mathrm{ac}$, the accretion rate grows at most by 30\%. The smaller
the planet mass, the less sensitive $\dMp$ is to the parameter
$\kappa_\mathrm{ac}$.

\subsection{Comparison with 2D Models}
\label{ssec:C2D}
\begin{figure*}[!t]
\epsscale{2.0}
\plottwo{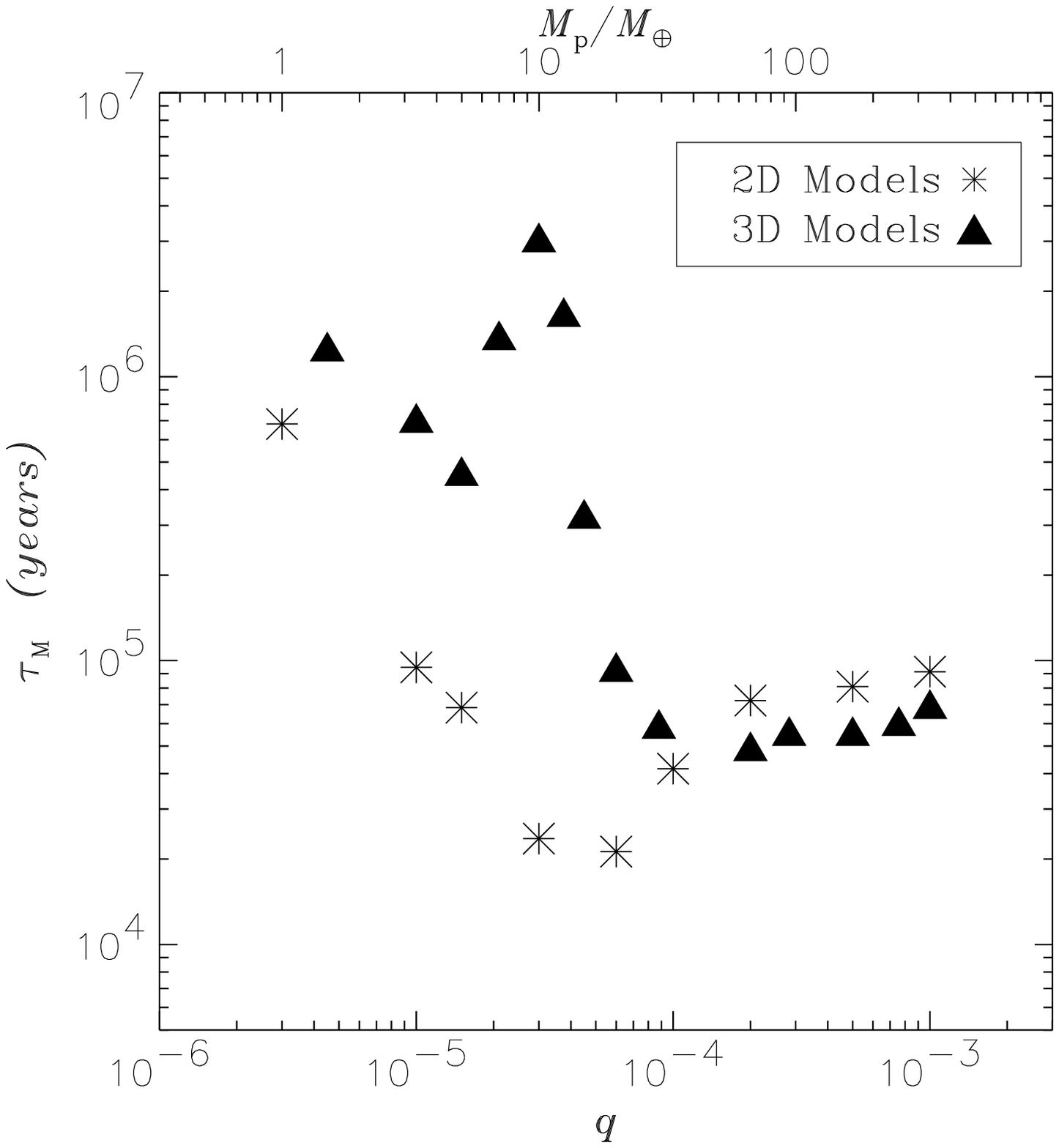}{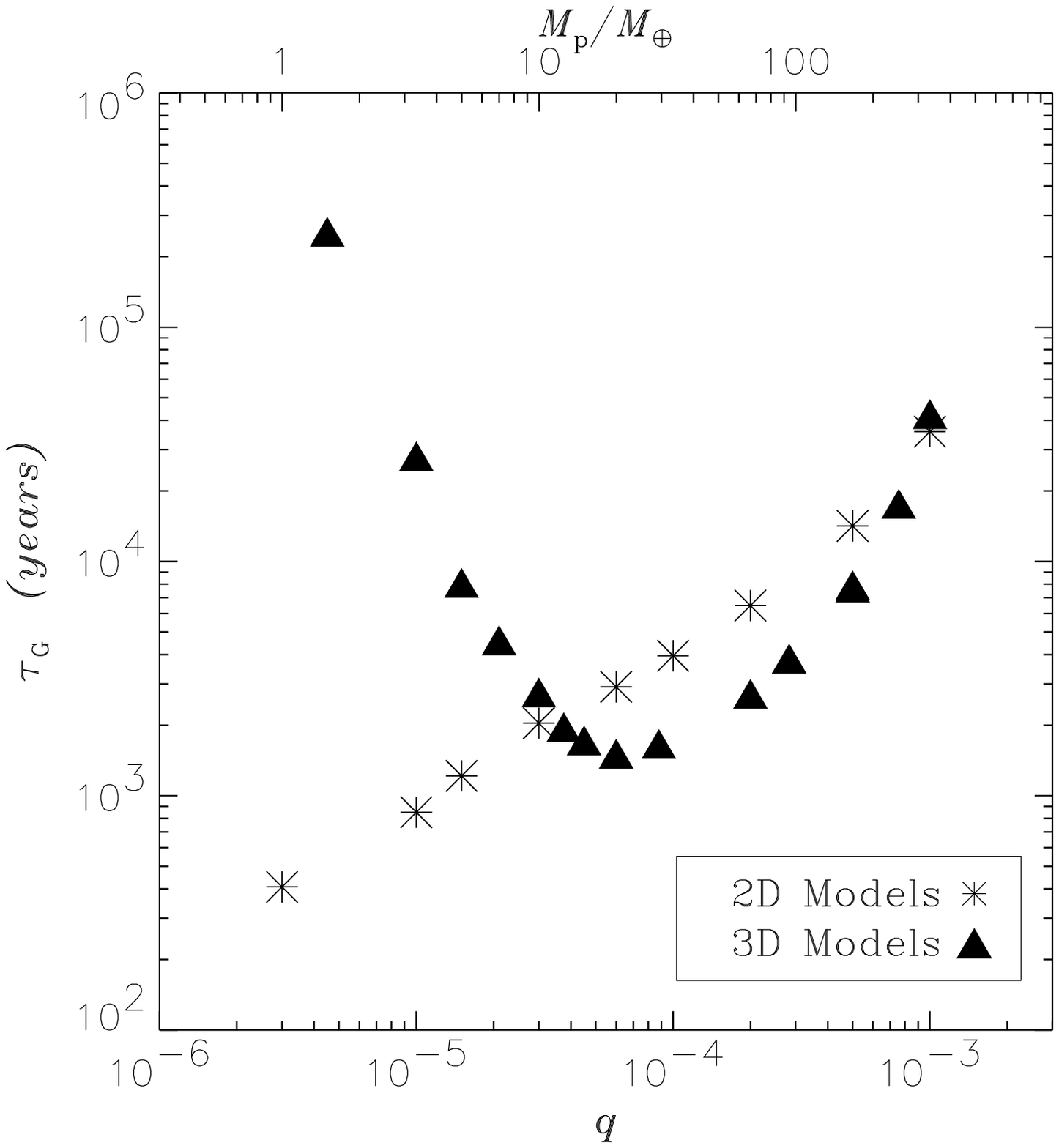}
\caption{%
       \textbf{Left panel}. Migration rates as evaluated in two-dimensional
       (\DHK) and three-dimensional models (this paper).
       \textbf{Right panel}. Using the same sources as in the left panel,
        a comparison of the growth time scales 
        $\tau_\mathrm{G}\equiv \Mp/\dMp$, between 2D and 3D outcomes, 
        is shown. 
        Filled triangles, in both panels, indicate results obtained from 
        three-dimensional models in which the planetary gravitational
        potential is $\Phi_\mathrm{p}^\mathrm{PM}$, 
        $\Phi_\mathrm{p}^\mathrm{ST}$, or $\Phi_\mathrm{p}^\mathrm{KW}$.
 \label{fig:2D3D}}
\end{figure*}

In this Section we aim at comparing the migration time scale as well
as the planet's accretion rate obtained in this work with those presented
in \DHK.
However,
while two-dimensional estimates of $\dMp$ (Fig.~25 in \DHK) are directly 
comparable to those plotted here in \refFgt{fig:mpdot}, the time scales 
$\tau_\mathrm{M}$ shown in Figure~20 of \DHK\ are not completely consistent 
with those in \refFgt{fig:tau_m}. Therefore, they need to be corrected.

This is because in the present study torques are integrated all over the
disk domain excluding the planet envelope, i.e., 
the sphere of radius $S=\Se$. 
Instead, in \DHK\ the excluded region has a radius $\approx 0.1\,\RH$ 
(for details see \DHK, \S~5.4).
From \refTab{tbl:MpSp} one can see that, above $33\;\MEarth$, the envelope 
radius $\Se$ can be much larger than a tenth of the Hill radius.

\refFgt{fig:2D3D} illustrates the migration rate $\tau_\mathrm{M}$ 
(left panel) and the growth time scale $\tau_\mathrm{G}\equiv \Mp/\dMp$ 
(right panel) as computed in the two geometries.
Orbital migration estimated by means of 3D simulations is slower than that
evaluated in 2D calculations only below $33\;\MEarth$ ($q=10^{-4}$).
As for planet's accretion, the most important difference is the rapid
drop, for $\Mp<10\;\MEarth$, observed in disks with thickness.
The larger values of three-dimensional estimates, measured in the range
$10\;\MEarth\lesssim\Mp\lesssim1\;\MJup$, are due to the gap which is 
not so deep as it is in two-dimensional models, hence the average density 
around the planet is higher. 
This occurrence can be partly attributed to gravitational potential
effects that, 
as we mentioned in \refsec{ssec:FD}, are intensified by the flat geometry
approximation.

\subsection{Numerical Effects}
\label{ssec:NE}

In \DHK\ it was found that, upon increasing the smoothing parameter, there is 
a reduction of the torques' mismatch, over a region around the planet whose
linear size is comparable with the double of the smoothing length. 
A $33\;\MEarth$ model was run with a point-mass potential without any
kind of softening. This is possible because none of the hydrodynamical 
variables is placed at a cell corner, where the planet dwells.
A similar simulation was performed applying a grid dependent smoothing of the
type described in \DHK.
Resulting migration time scale and mass accretion are not
significantly affected by the smoothing choice.

As for the consequences of the circumstellar disk depletion, inside of
the 
planet's orbit, we ran a Jupiter-mass model in which both inner and
outer radial 
borders were closed. Since more material is available in the disk portion
$R<R_\mathrm{p}$ (roughly twelve times as much), one should expect larger 
values for both $\dMp$ and $\taum$. 
Indeed, accretion is two times as much as that calculated in the model with 
open inner border.
Positive torques arising from the inner disk are also stronger and
$\mathcal{T}_\mathrm{D}$ is reduced by 50\%, i.e., the migration time scale is 
two times as long.

When simulating a Jupiter-size body embedded in a disk with no initial
gap, a density indentation is
gradually carved in. In order to skip the gap formation phase, an approximate 
analytical gap is sometimes imposed in the initial density distribution
\citep[see][]{kley1999}.
We performed three computations adopting this choice.
In these cases, a partial shrinking and refilling of the analytical gap is 
observed. Besides, material drains out of the inner radial border faster 
than it does in our standard models (no initial gap). 
Hence, the inner disk depletion is intensified. 
With respect to standard models, we measure smaller accretion rates and longer
migration time scales. 
Discrepancies in both quantities stay below 20\%, after 200 
orbits.
However, since the model outcomes indicate a tendency to converge as the
evolution proceeds, a more appropriate comparison should be made after
a long-term evolution.

\subsubsection{Grid Resolution}
\label{sssec:GR}

Hardly any hydrodynamic calculation is strictly resolution independent.
Thus, for completeness we analyze in this section how our estimates
on migration and accretion vary because of different hierarchy resolutions.
Two tests are presented for each of the quantities $\dMp$ and 
$\mathcal{T}_\mathrm{D}$.
They are computed with the aid of grid systems G3 and G4 and then results
are tested against those calculated with the less resolved hierarchy G2
(see \refTab{tbl:grids}).
In this way we aim at checking finite resolution effects in the radial
and azimuthal directions and, separately, those in the meridional direction.
In fact, G3 and G2 have the same number of grid points in the vertical
direction but $\Delta R[\mathrm{G3}]=0.82\,\Delta R[\mathrm{G2}]$ and
$\Delta\varphi[\mathrm{G3}]=0.75\,\Delta\varphi[\mathrm{G2}]$.
In the other test case (G4 against G2), $R$ and $\varphi$ gridding is 
unchanged while the number of latitude grid points is nearly doubled.

\begin{figure*}[!t]
\epsscale{2.0}
\plottwo{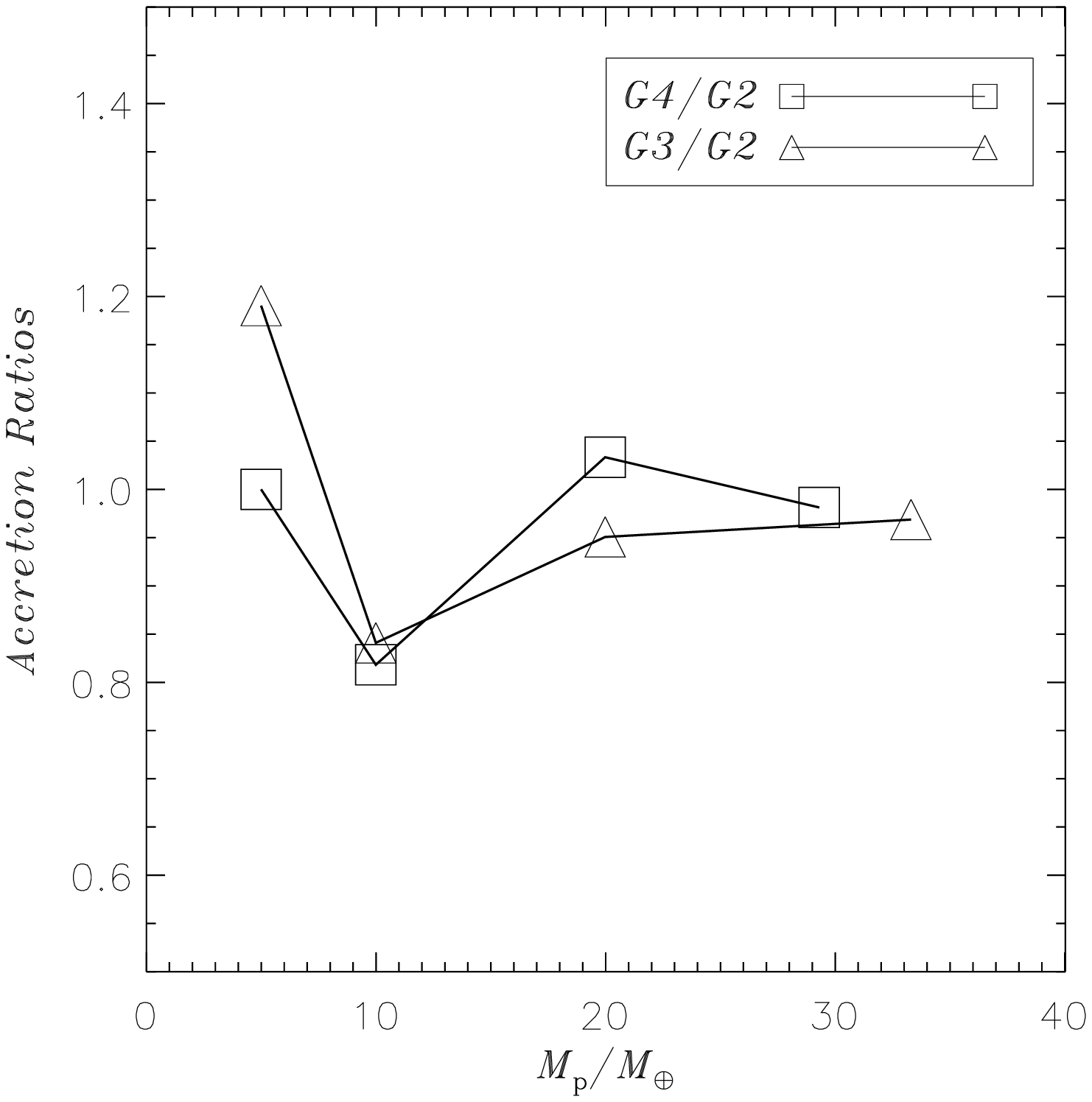}{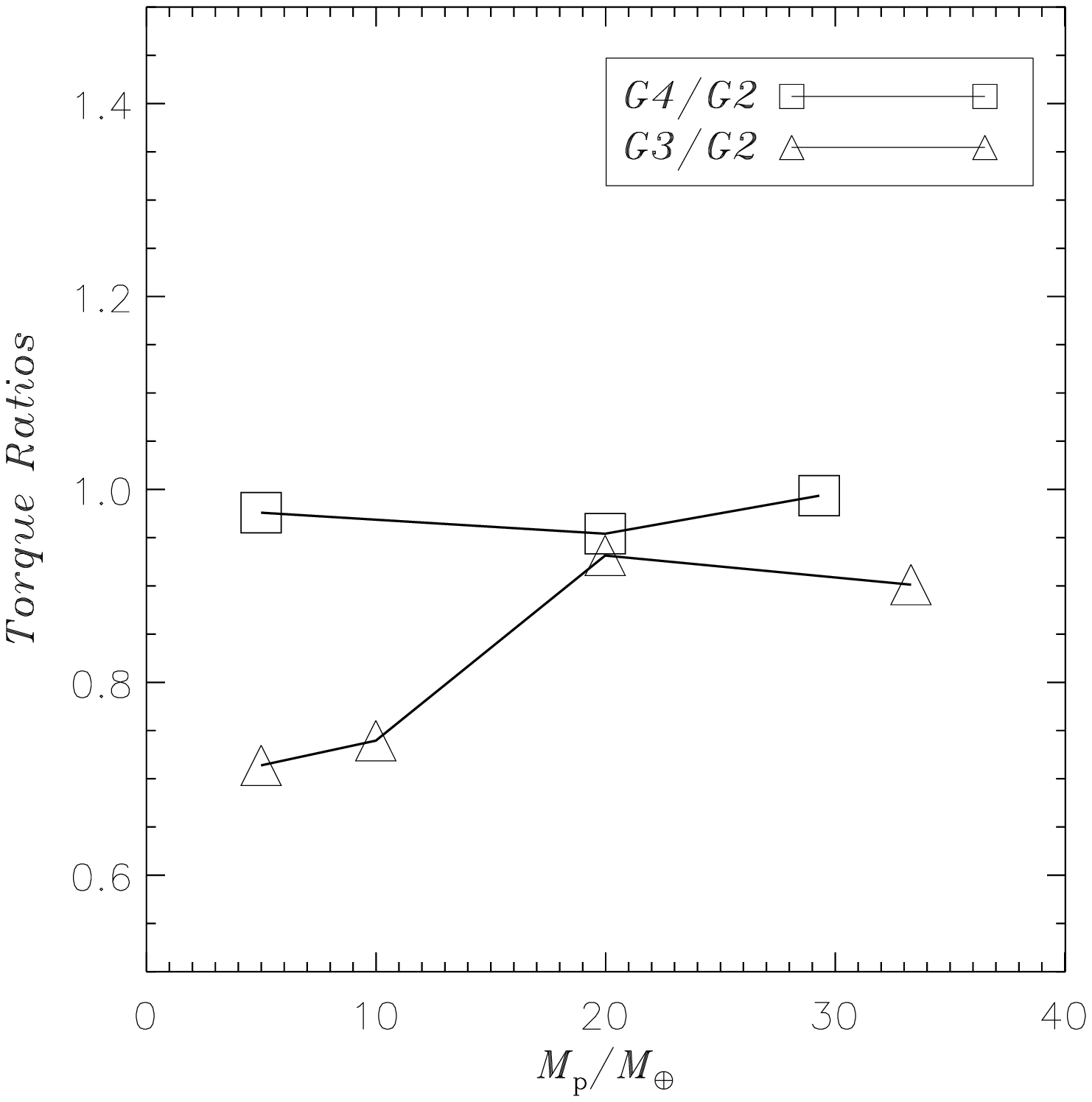}
\caption{%
       \textbf{Left panel}. Comparison of the accretion rate $\dMp$ 
       as computed on grid systems with different resolutions. 
       Squares indicate that the ratios $\dMp[\mathrm{G4}]/\dMp[\mathrm{G2}]$ 
       are drawn, whereas triangles refer
       to the ratios $\dMp[\mathrm{G3}]/\dMp[\mathrm{G2}]$ (see text).
       \textbf{Right panel}. The same type of comparison for the total torque 
       $\mathcal{T}_\mathrm{D}$ experienced by the protoplanet. 
       Squares and triangles have the same meaning as before.
       At $\Mp=10\;\MEarth$, we compare results from models with the 
       homogeneous sphere potential given in \refeqt{phihs}, when available
       (for this particular mass value, see the discussion in \refsec{ssec:TM}).
 \label{fig:gcomp}}
\end{figure*}
With regard to the mass accretion rate (\refFgp{fig:gcomp}, left panel) 
differences are below 20\%. 
Though the number of grid cells in the accretion sphere
is enlarged by a factor either $1.7$ or $2$, no systematic tendency seems 
to arise from this test.
Something
different happens to the total torque (right panel). In fact, while the 
increased resolution in the latitude direction $\theta$ does not play any 
considerable role, the
larger number of grid points in $R$ and $\varphi$ causes a reduction of
the total torque magnitude between 20 and 30\%.
This is not at all unexpected.
On one hand circumstellar disk spirals are better captured by
a finer gridding in the radial and azimuthal dimensions. Thus, Lindblad
torques are accounted for in a more accurate fashion. The Figure proves
this to be especially true when the ratio $q$ is small, because of the
diminishing wave amplitudes.
On the other hand, due to the vertical exponential drop of the density
and the lack of temperature stratification, disk layers above the midplane 
do not contribute very much to $\mathcal{T}_\mathrm{D}$. 
A finer resolution along the vertical dimension cannot sensitively modify 
the total torque outcome.

\section{Discussion}
\label{sec:D}

Here we devote some further comments to the differences between
accreting and non-accreting protoplanets and then to the effects
of the vertical density structure
on the gravitational torques acting on embedded objects.

\subsection{Pressure Effects in Protoplanetary Envelopes}
\label{ssec:PEPE}

\begin{figure*}[!t]
\epsscale{1.0}
\mbox{%
\plotone{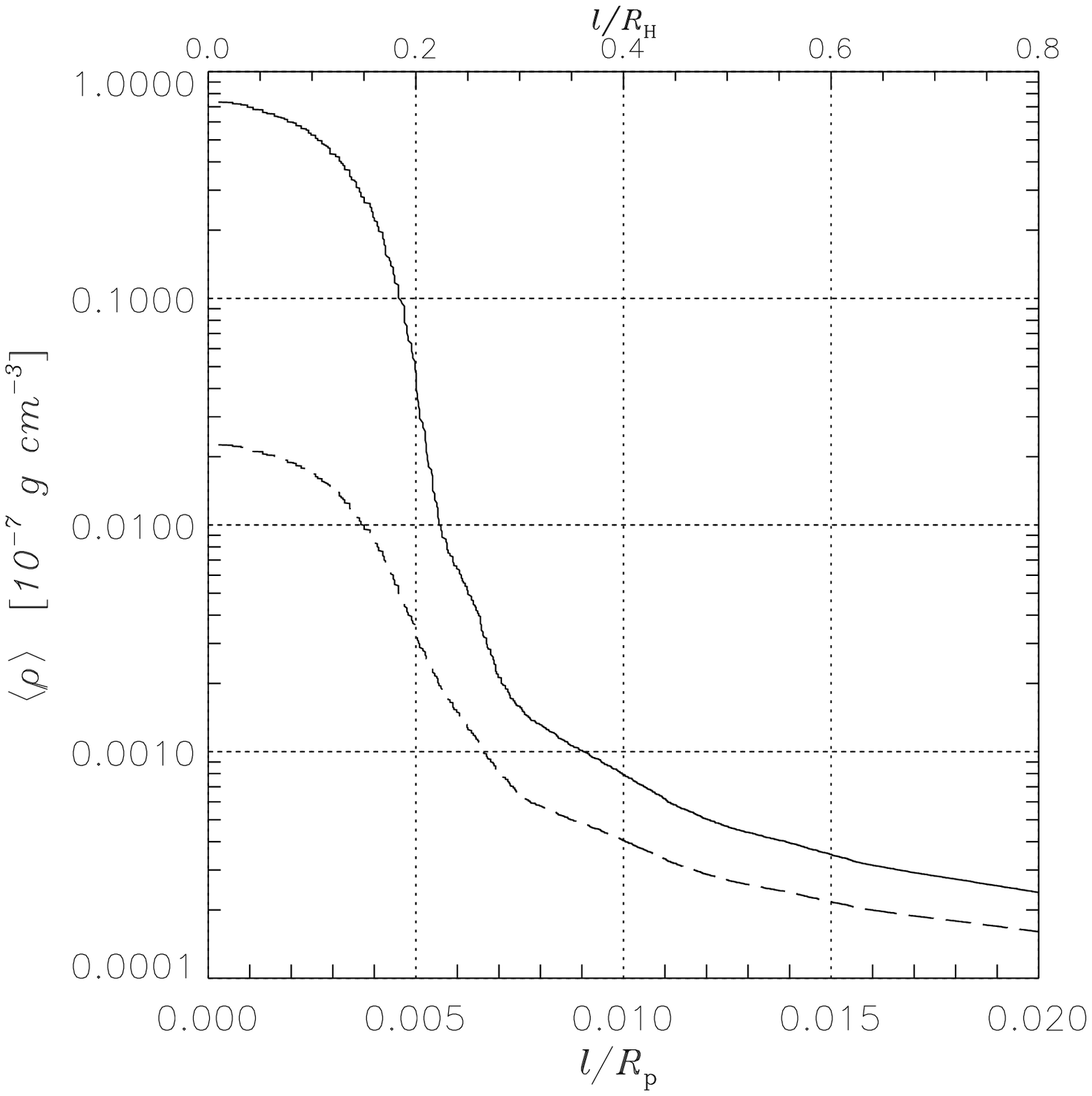}\hfill%
\plotone{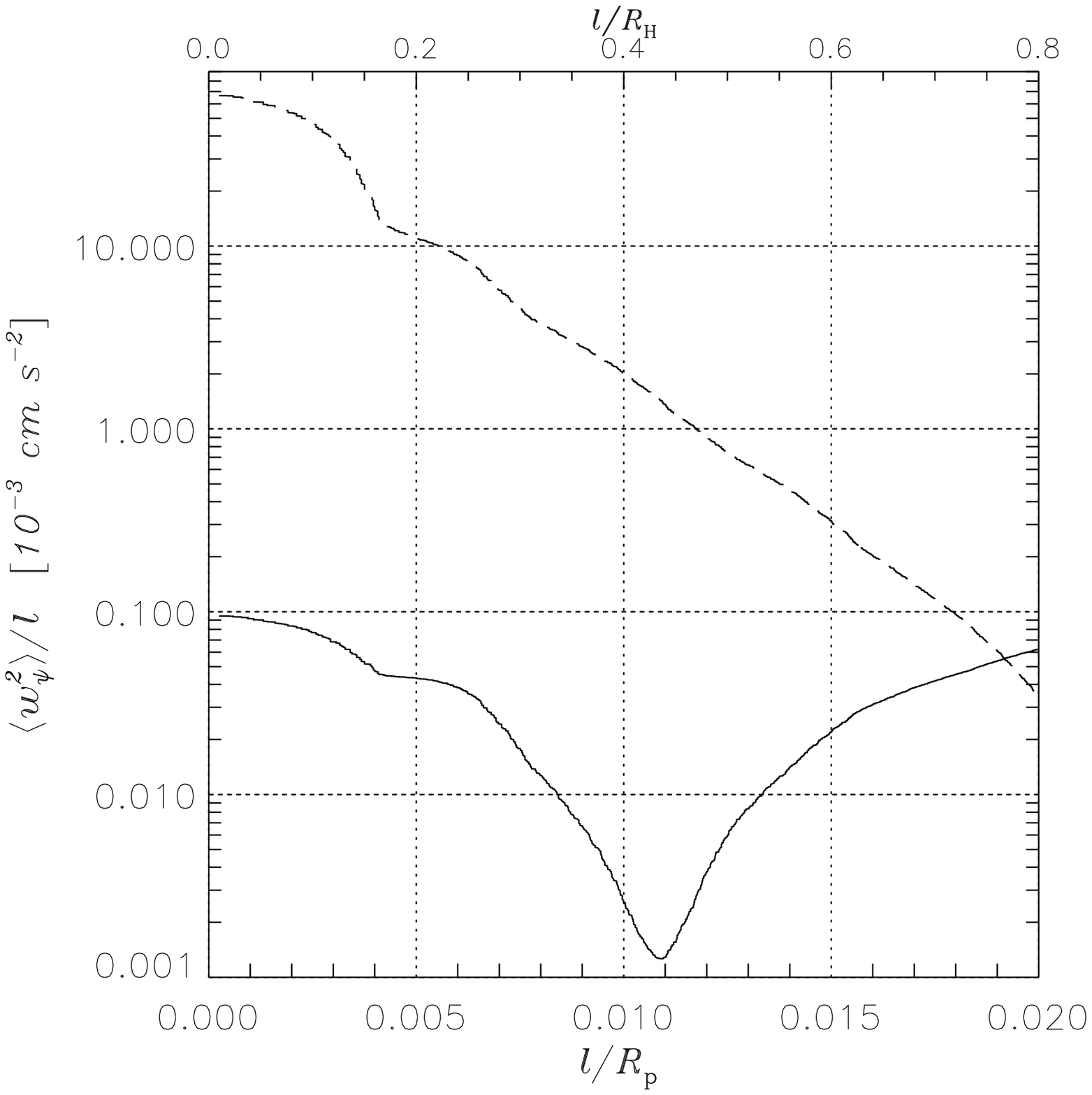}}
\mbox{%
\plotone{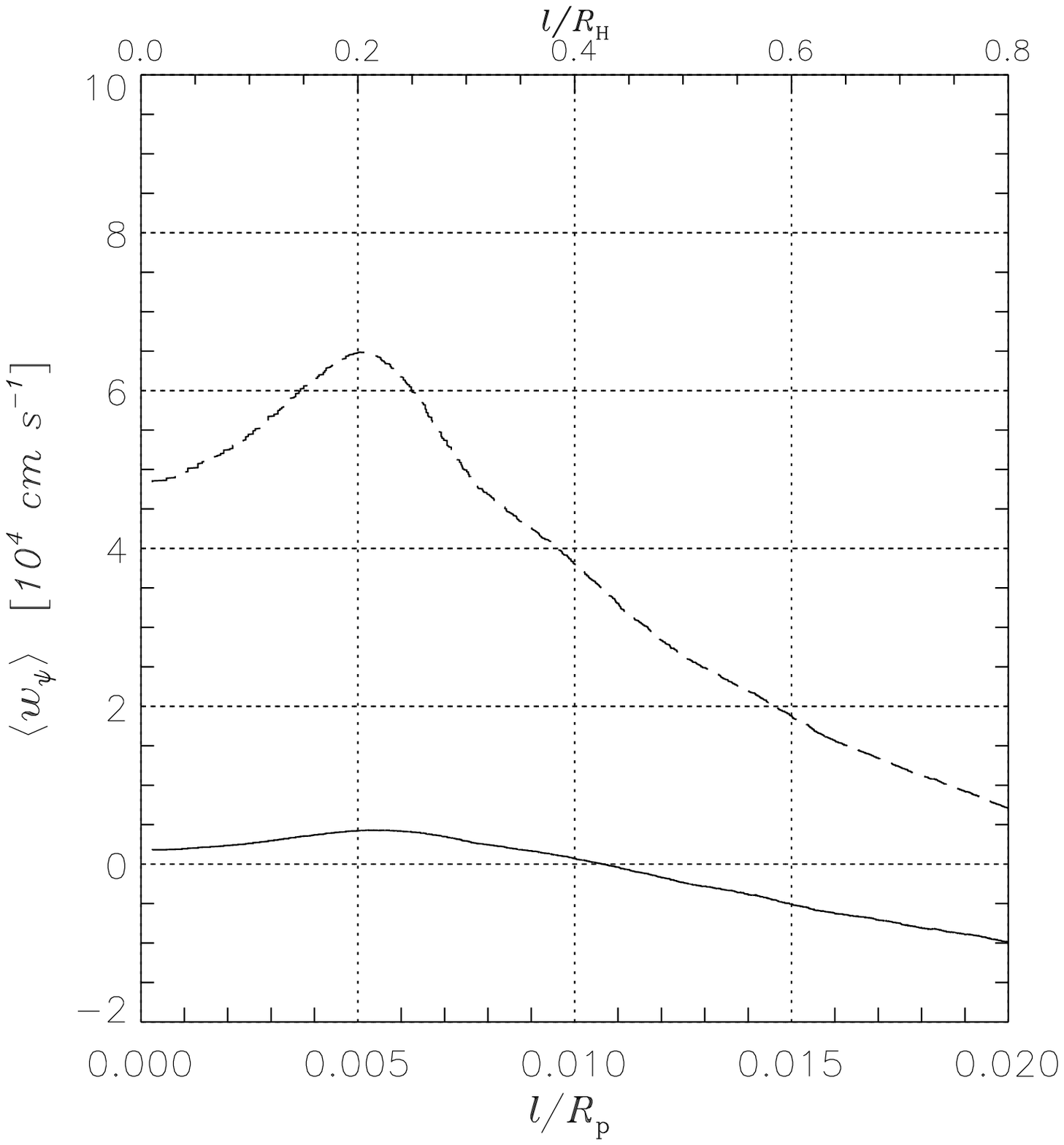}\hfill%
\plotone{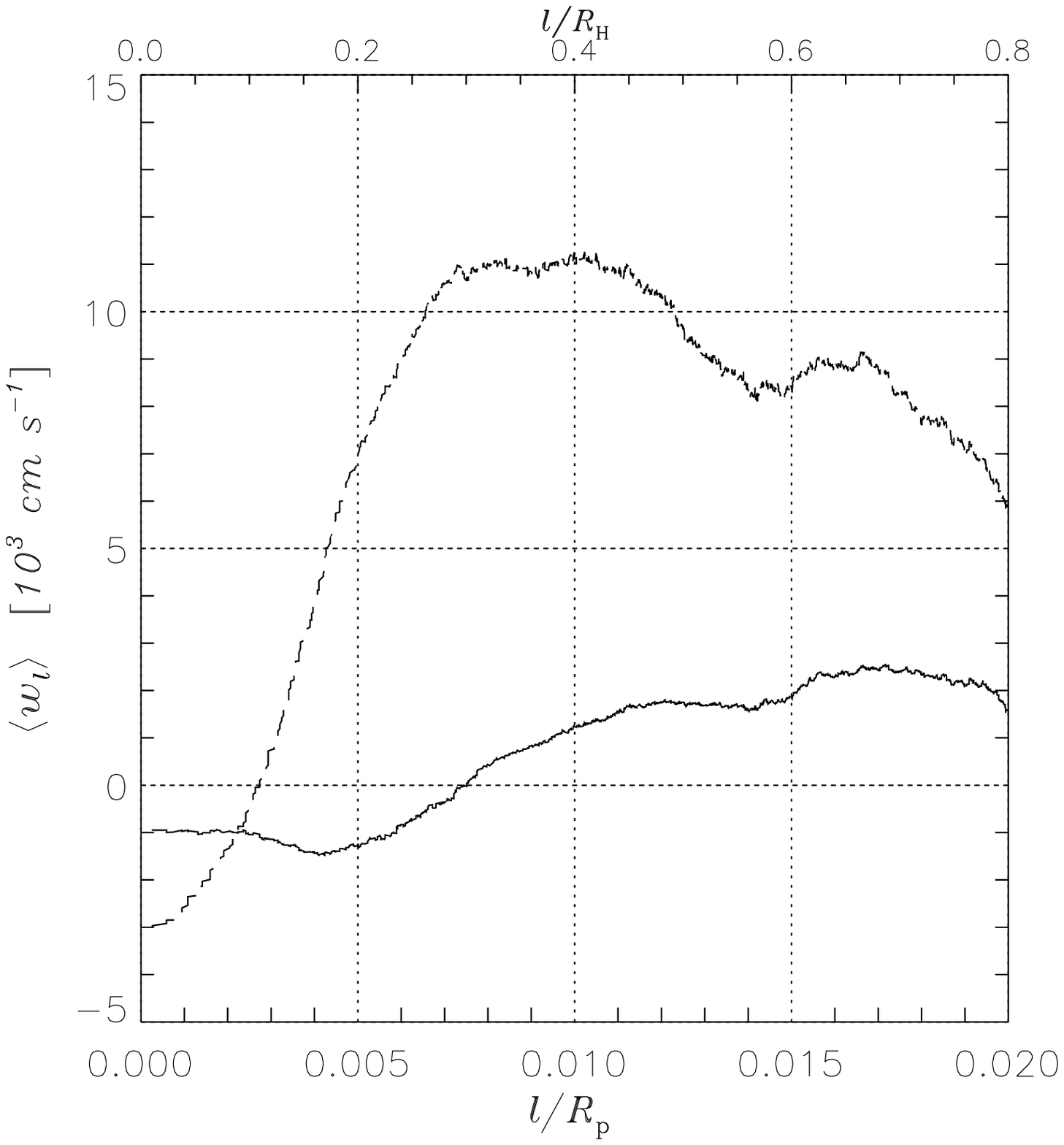}}
\caption{%
         Midplane quantities ($z=0$ or $\theta=\pi/2$) azimuthally
         averaged around the protoplanet for two simulations in which
         $\Phi_\mathrm{p}=\Phi_\mathrm{p}^\mathrm{ST}$ and $\Mp=20\;\MEarth$. 
         The solid line belongs to a 
         non-accreting model, the dashed line to an accreting one.
         \textbf{Top-left}: mass density.
         \textbf{Top-right}: centrifugal acceleration.
         \textbf{Bottom-left}: rotational velocity.
         \textbf{Bottom-right}: velocity component along the radial
         distance $l$ (see \refsec{sec:D}).
         All of the four quantities indicate that the envelope is
         mostly pressure supported in the non-accreting case, whereas
         it is mainly centrifugally supported in the other.
 \label{fig:avq}}
\end{figure*}

For the purpose of carrying out a local analysis in the vicinity
of accreting and non-accreting protoplanets, we introduce a cylindrical 
coordinate reference system
$\{O^\prime; l, \psi, z\}$ with its origin $O^\prime$ coinciding with the 
planet position and the $z$-axis perpendicular to the disk midplane. 
Hence, we will have that $S^2 = l^2 + z^2$. The longitude angle $\psi$ is
counterclockwise increasing, and $\psi=\pi$
points toward the star. Supposing that the flow nearby the planet is 
stationary, neglecting fluid advection and viscosity, 
the Navier-Stokes equation for the radial momentum reads:
\begin{equation}
 \frac{w^2_\psi}{l} = \frac{\partial \Phi_\mathrm{p}}{\partial l} 
                  + \frac{1}{\rho}\,\frac{\partial p}{\partial l},
 \label{wpsi}
\end{equation}
where $w_\psi$ is the azimuthal velocity component around the planet.
Excluding the particular situation represented by a homogeneous sphere 
(\refeqp{phihs}),
the first term on the right hand side of \refeqt{wpsi} is positive
(see \refFgp{fig:ag}). Recalling \refeqt{p}, we see that the second term is 
proportional
to the density gradient, which is negative, and therefore it reduces the
centrifugal acceleration $w^2_\psi/l$.
In \refFgt{fig:avq} we show some quantities, at $z=0$, 
averaged over the angle $\psi$,
regarding the same simulations addressed in \refsec{sssec:NaP}
($\Mp=20\;\MEarth$ with $\Phi_\mathrm{p}=\Phi_\mathrm{p}^\mathrm{ST}$).
From the top left panel one can realize that the mean density is indeed higher
in the non-accreting case (solid line) than it is in the accreting case 
(dashed line), as it was argued in \refsec{sssec:NaP} 
from the values listed in \refTab{tbl:noac}.
In order to evaluate how much the pressure gradient affects the left hand side
of \refeqt{wpsi} in both cases, we plot the average of such quantity 
($\langle w_\psi^2 \rangle/l$) in the top
right panel of \refFgt{fig:avq}. The centrifugal acceleration is much 
smaller in the envelope of the non-accreting model (solid line) than it is 
in that of the accreting one. 
Such circumstance is a clear indication that the envelope is 
pressure supported in the first case.
The behavior of the averaged velocities 
$\langle w_\psi \rangle$ and $\langle w_l \rangle$ is shown 
in the two bottom panels. 
As expected in a pressure dominated flow, the magnitude of both 
velocity components is smaller in the non-accreting model (dashed lines).

\subsection{Torque Overestimation in 2D Geometry}
\label{ssec:TO2D}

Gravitational torques exerted by a three-di\-men\-sional disk onto a 
medium- or low-mass protoplanet are weaker than those generated 
by a two-dimensional disk. \citet{miyoshi1999} state that the total
torque $\mathcal{T}_\mathrm{D}$ in 3D is $0.43$ times as small as that
in 2D.
Something similar was found by \citet{tanaka2002}.
Our fully non-linear calculations predict that low-mass protoplanets have a
migration rate, at least,
an order of magnitude less in disks with thickness than they have
in infinitesimally thin disks.
One of the main reasons for that relies upon the vertical decay of the density,
as one can demonstrate easily with a simplified approach. 

Let $t_z$ be the $z$-component of the gravitational torque exerted 
by a column of mass $\Sigma\,l\,dl\,d\psi$, located at distance $l$ from the 
planet (see \refsec{ssec:PEPE}). The surface density is defined as $\Sigma=\int \rho\,dz$.
If $f_\mathrm{g}$ is the force exerted by such mass distribution, projected on 
the equatorial plane, then we can write
\begin{equation}
t_z = R_\mathrm{p}\,f_\mathrm{g}\,\sin{\psi}.
\label{t_z}
\end{equation}
The ratio of $t_z[\mathrm{3D}]$ to $t_z[\mathrm{2D}]$ is therefore equal
to 
\begin{equation}
\chi=\frac{f_g[\mathrm{3D}]}{f_g[\mathrm{2D}]}
   = \frac{l^3}{\Sigma}\int_{-\infty}^{+\infty} 
     \frac{\rho}{(l^2 + z^2)^{3/2}}\,dz.
\label{chigen}
\end{equation}
Since $f_g[\mathrm{3D}]$ and $f_g[\mathrm{2D}]$ are coherent in sign,
$\chi$ is also equal to the ratio of $|t_z[\mathrm{3D}]|$ to 
$|t_z[\mathrm{2D}]|$.
In order to quantify this quantity, we can 
assume a Gaussian mass density profile with a scale-height $H$, 
which is appropriate as long as no deep gap has formed.
Thus
\begin{equation}
\chi=\frac{l^3}{\sqrt{2\pi}\,H}\int_{-\infty}^{+\infty} 
    \frac{\exp{\left(-\frac{z^2}{2\,H^2}\right)}}{(l^2 + z^2)^{3/2}}\,dz.
\label{chi}
\end{equation}

\begin{figure}[!t]
\epsscale{1.0}
\plotone{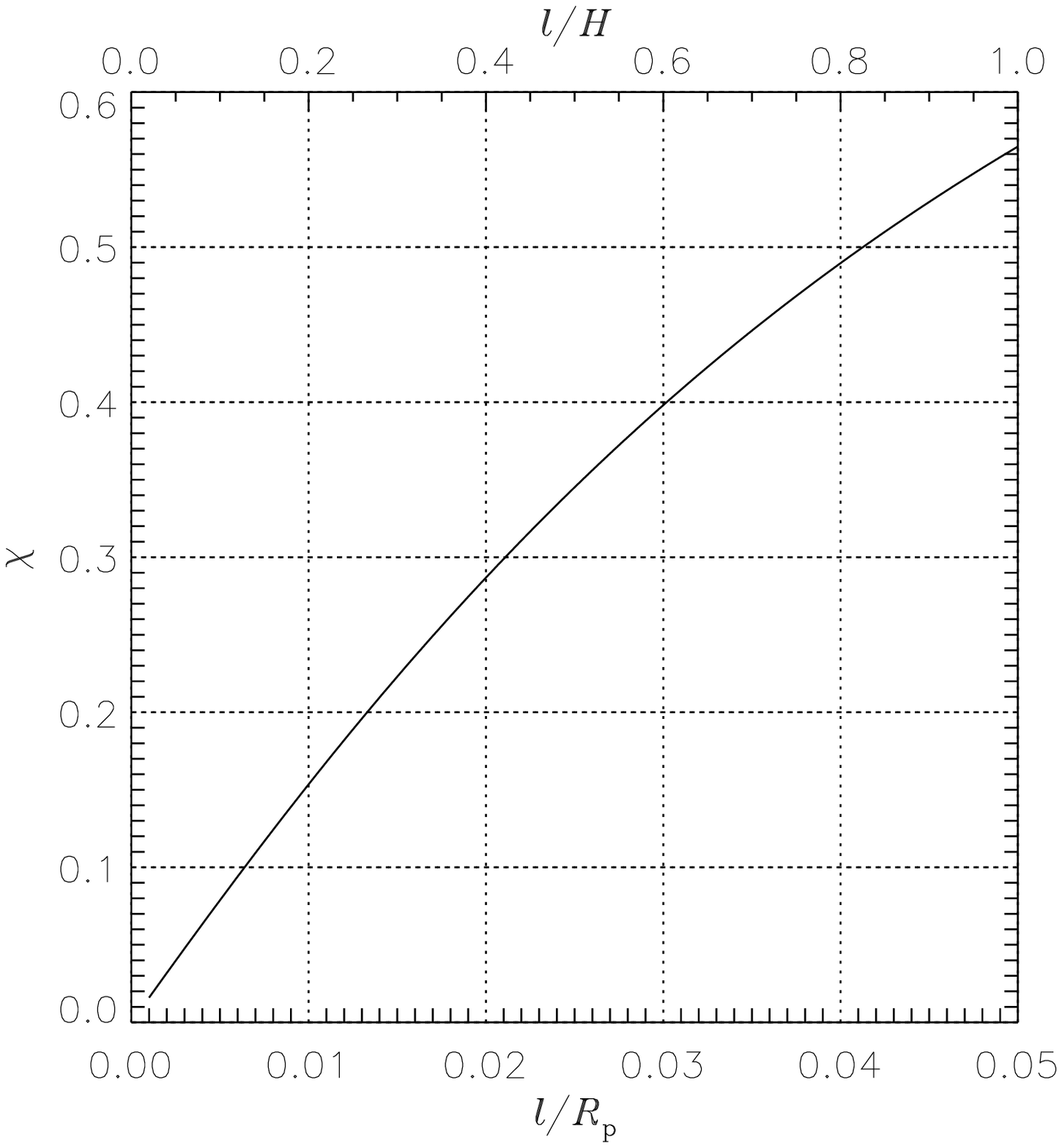}
\caption{%
       Ratio of the tree-dimensional to the two-dimensional torque
       exerted by a column of material lying at a midplane distance 
       $l$ from the protoplanet (see \refsec{sec:D}).
       The upper x-axis is in units of the disk semi-thickness
       at the planet location: $H= h\,R_\mathrm{p}$. 
\label{fig:chi}}
\end{figure}

The ratio $\chi$ as function of $l$ is plotted in \refFgt{fig:chi} and it
evidences how a two-dimensional geometry overestimates the magnitude of
gravitational torques acting on the protoplanet. 
In the limit $l^2 \gg z^2$,
$\chi$ converges to 1, which proves that only torques arising from
locations near to the planet ($l\lesssim H$) are magnified.

Though larger torque magnitudes do not necessarily imply faster migration 
speeds, they can favor a larger mismatch between negative and positive 
torques and therefore shorter $\tau_\mathrm{M}$.

\section{Conclusions}
\label{sec:C}

On the background of the numerical computations of disk-planet interaction 
presented in \KDH\ and II, in this paper we combine the full 3D geometry
of a circumstellar disk with a nested-grid technique in order to investigate
in detail flow dynamics, orbital decay, and mass accretion of protoplanets 
in the mass range $[1.5\;\MEarth, 1\;\MJup]$. 
Besides, we overcome the point-mass assumption by employing analytic 
expressions of the gravitational potential derived from simple theoretical 
models of protoplanetary envelopes.
Each of them applies to distinct physical situations: when the envelope
mass is negligible with respect to the core mass; when the envelope is
homogeneous and much more massive than the core; 
when the envelope is fully radiative, and finally when it is fully convective.

Through a series of 51 simulations, we inspect the evolution and differences
of protoplanets represented by the aforementioned gravitational potentials.
We analyze the behavior of both accreting and non-accreting objects. 
Furthermore, we evaluate physical and numerical effects due to our standard 
set-up of the models.
The computations clearly show that to accurately determine the early physical 
evolution of planets three-dimensional effects have to be taken into account.

\noindent
The main results of our studies can be summarized as:
\begin{enumerate}
\item Above the disk midplane the flow is nearly laminar only far away 
      from the planet.
      The region of influence of the planet extends well outside the Hill
      sphere and its boundaries are marked by vertical shock fronts.
      Past the shock, matter is deflected upward and then downward.
      In some cases, a closed recirculation is also observed. 
      In the disk midplane, spiral waves  around the planet are not as strong 
      and tight as they appear in two dimensions because of wave deflection in
      the vertical direction.
\item In the mass range of their applicability, Stevenson's and Wuchterl's
      gravitational potentials produce flow structures, close to the planet,
      similar to those determined by a smoothed point-mass potential.       
      Migration times and accretion rates are alike.
      In contrast models with the (unrealistic) potential of a homogeneous 
      sphere yield different dynamics though, as for $\tau_\mathrm{M}$ and 
      $\dMp$, not much difference is observed for Jupiter and Earth-size 
      bodies.
\item Since the numerical accretion procedure might be considered somewhat 
      arbitrary, we ran several models in which the protoplanet does not 
      accrete at all.
      Non-accreting models behave differently from accreting ones in a volume
      whose size is roughly comparable with the Hill sphere. 
      Within this region matter is pressure supported and thus a spherical 
      envelope builds up.
      Except for the case $\Mp=20\;\MEarth$, the total torque 
      $\mathcal{T}_\mathrm{D}$ exerted by the disk is on the same order of
      magnitude as that measured in accreting models.
\item According to Ward's theory \citep{ward1997}, the migration speed settles 
      to a constant value when the planet-to-star mass ratio 
      $q \gtrsim 4\times10^{-4}$.
      Our numerical results give a similar trend at a slightly different 
      magnitude though.
      Most of these simulations predict an inward migration except the one
      where a $20\;\MEarth$, non-accreting, protoplanet is involved.
      In the mass range $[7\;\MEarth, 15\;\MEarth]$ migration speeds can be
      30 times as slow as those predicted by \citet{tanaka2002} although,
      outside of this range,
      the agreement between our computational data and the type~I
      migration by the same authors is remarkably good.
      We suspect that
      this surprising outcome may be caused by the onset of non-linear
      effects appearing around ten Earth's masses, which conspire
      to give such long migration time scales. 
      If correct, this much slower inward motion may help to
      solve the problem of the too rapid drift of planets
      toward their host stars.
\item In agreement with studies on planet formation
      \citep{bodenheimer1986,tajima1997}, the growth time scale
      shortens as the protoplanet's mass increases. The minimum is found
      at $\Mp=20\;\MEarth$. 
      Albeit the feeding process slows down as soon as angular momentum 
      transferred by the planet to the surrounding material is large enough 
      to dig a density gap.
      Then, at $\Mp\approx 1\;\MJup$, the accretion rate greatly reduces
      and the growth time scale becomes consistently very long.
      We present an analytical formula for the growth rate which 
      may be useful for global studies in planet formation.
\item As long as migration and mass accretion are considered, two-dimensional
      computations still yields reliable results when the mass ratio 
      $q\gtrsim10^{-4}$ ($\Mp\gtrsim 30\;\MEarth$ if $M_{\bigstar}=1\;\MSun$).
      In practice, 2D geometry is applicable whenever the Hill radius $\RH$
      exceeds the 60\% of the local pressure scale-height of the disk $H$.
      But for smaller masses three-dimensional calculations have to be
      considered.
\end{enumerate}

The three-dimensional calculations presented here achieve a
new level of accuracy by using a sophisticated nested-grid technique.
This numerical feature allows a global and local resolution
not obtained hitherto.
However, similar to
all of the previous calculations, the models presented here have one
principal limitation: the lack of an \textit{appropriate} energy equation.
Because of this, we could not couple the thermal and the hydrodynamical
evolution of the system. 
If one wishes to do that in three dimensions, the energy equation has 
to include radiation and convective transfer. 
Yet, only with massive parallel computations one can hope to pursue this goal. 
Thereupon, the direction of future developments and improvements is already 
marked.

\acknowledgments

We are grateful to Udo Ziegler for having made available to us the FORTRAN
Version of his code 
\textsc{Nirvana}\footnote{\url{http://www.aip.de/$\tilde{~}$ziegler/}.}.
G.~D.\ wishes to thank sincerely Dr.~G. Wuchterl for the time spent on 
intriguing discussions and for his precious suggestions about the protoplanet 
envelope solution by D.~J.~Stevenson. 
We are indebted to P.~Bodenheimer who provided us with the estimates of the
protoplanets' radii.
Remarks by an anonymous referee helped to clarify and improve many
parts of this article. We much appreciated his job.
This work was supported by the German Science Foundation (DFG) under grant
KL 650/1-1.
The numerical computations were carried out at the Computer Center of the 
University of Jena and at the Institute of Astronomy and Astrophysics of 
the University of T\"ubingen.



\appendix
\normalsize
\section{Interpolation Formulas for the Fine-coarse Grid Interaction}
\label{appendixa}
The aim of this appendix is to furnish some algorithms useful for
updating the values of scalars and momenta on a coarse grid with 
those computed on the hosted (finer) grid, in spherical polar coordinates.

For the purpose,
we indicate with $\rho^{\mathrm{C}}$ the mass density to be
interpolated on the coarse level.
The interpolating values, on the finer subgrid level, are
indicated simply as $\rho(i,j,k)$.
For the density average, the $i$-index varies between $i$ and $i+1$ 
and so do the other two indexes.

Accordingly, $U^{\mathrm{C}}_R$, $U^{\mathrm{C}}_\theta$ and 
$U^{\mathrm{C}}_\varphi$ are the coarse linear, meridional and
azimuthal angular momentum.
The finer values from which they are reset will be denoted as
$U_{R}(i,j,k)$, $U_{\theta}(i,j,k)$, and $U_{\varphi}(i,j,k)$,
respectively. 
For the average of the radial vector component, the $i$-index varies
between $i-1$ and $i+1$, while the other two indexes vary between $j$ ($k$)
and $j+1$ ($k+1$).
This is related to the locations where velocity
components are defined on the mesh.
In case of the
meridional angular momentum, the $j$-index ranges from $j-1$ to $j+1$,
while the others are $i$ ($k$) and $i+1$ ($k+1$). 
For the azimuthal angular momentum, it is the $k$-index to extend over
the wider range.

\begin{figure}[!t]
\epsscale{0.5}
\plotone{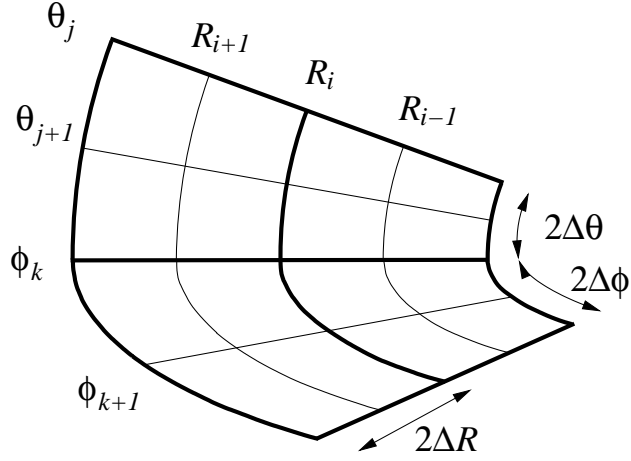}
\caption{%
       Sketch of a spherical sector composed of two coarse grid cells
       (thick lines) and the set of nested cells (thin lines). 
       The discretized spherical 
       coordinates $(R_i, \theta_j, \varphi_k)$ are 
       relative to the fine grid, and so is the resolution
       $(\Delta R, \Delta \theta, \Delta \varphi)$. Scalars are
       cell-centered on each grid level, whereas vector components are
       face-centered, each on a different face of the volume element.
       \label{fig:grid}}
\end{figure}

The finer grid coordinates are $(R_i, \theta_j, \varphi_k)$,
and the grid spacing is $(\Delta R, \Delta \theta, \Delta \varphi)$
(see \refFgp{fig:grid}), which is constant. 
Since the grid has a staggered structure, scalars are volume-centered,
i.e., $\rho(i,j,k)$ lies at 
$(R_i + \Delta R/2, \theta_j + \Delta\theta/2, \varphi_k +
\Delta\varphi/2)$, while $\rho^{\mathrm{C}}$ resides at
$(R_i + \Delta R, \theta_j + \Delta\theta, \varphi_k +
\Delta\varphi)$, because the linear resolution doubles from a grid to
the hosted one.
Instead, vector components are
centered each on a different face of the volume element. For example,
the radial component $U_{R}(i,j,k)$ is located at 
$(R_i, \theta_j + \Delta \theta/2, \varphi_k + \Delta \varphi/2)$, whereas
the coarse radial momentum $U^{\mathrm{C}}_R$ is defined at
$(R_i, \theta_j + \Delta \theta, \varphi_k + \Delta \varphi)$.
The locations of the other components follow by similarity.

The interpolation is basically a volume-weighted average.
Eight volumes are necessary to carry out a scalar
interpolation. Momentum interpolations require that twelve spherical
sectors must be employed.
Yet, since the metric in a spherical polar topology is independent of
the azimuthal angle $\varphi$, some of them actually coincide.
In order to distinguish
among the four volume sets, we introduce the notations $V^{(\rho)}$, 
$V^{(R)}$, $V^{(\theta)}$ and $V^{(\varphi)}$,
according to the quantity to average ($\rho^{\mathrm{C}}$, 
$U^{\mathrm{C}}_{R}$,
$U^{\mathrm{C}}_{\theta}$, and $U^{\mathrm{C}}_{\varphi}$). 
These four sets differ because of the space metric and the staggered mesh.

Once the correct elements have been identified,
the coarse mass density can be replaced by
\begin{equation}
\rho^{\mathrm{C}}=\frac{\sum_{ijk}\rho(i,j,k)\,V^{(\rho)}(i,j)}%
                       {2\,\sum_{ij}V^{(\rho)}(i,j)}.
\label{rhoc}
\end{equation}
In fact, sectors $V^{(\rho)}$ are $\varphi$-independent, thus only four
volumes enter this average. In a similar fashion, corrected momenta can be
written, in a concise form, as
\begin{equation}
U_{\Xi}^{\mathrm{C}}=\frac{\sum_{ijk}U_{\Xi}(i,j,k)\,V^{(\Xi)}(i,j,k)}%
                       {\sum_{ijk}V^{(\Xi)}(i,j,k)},
\label{uXi}
\end{equation}
where $\Xi=R$, $\theta$, and $\varphi$.
For computational purposes, a volume element is preferentially cast
into the form
\begin{equation}
V=\left(\Delta R^3/3\right)\,%
  \left(-\Delta \cos{\theta}\right)\,\left(\Delta \varphi\right).
\end{equation}
The four sectors $V^{(\rho)}$ required in \refeqt{rhoc} are the following
\begin{eqnarray}
V^{(\rho)}(i,j)&=&\frac{1}{3}\left(R_{i+1}^3 - R_{i}^3\right)\,
        \left[\cos{(\theta_j)}-\cos{(\theta_{j+1})}\right]\,%
         \Delta\varphi \nonumber \\
V^{(\rho)}(i+1,j)&=&\frac{1}{3}\left[(R_{i+1}+\Delta R)^3-R_{i+1}^3\right]\,
        \left[\cos{(\theta_j)}-\cos{(\theta_{j+1})}\right]\,%
         \Delta\varphi \nonumber \\
V^{(\rho)}(i,j+1)&=&\frac{1}{3}\left(R_{i+1}^3 - R_{i}^3\right)\,
        \left[\cos{(\theta_{j+1})}-\cos{(\theta_{j+1}+\Delta \theta)}\right]\,%
         \Delta\varphi \\
V^{(\rho)}(i+1,j+1)&=&\frac{1}{3}\left[(R_{i+1}+\Delta R)^3-R_{i+1}^3\right]
         \nonumber \\
      & &\left[\cos{(\theta_{j+1})}-\cos{(\theta_{j+1}+\Delta \theta)}\right]\,%
         \Delta\varphi.
         \nonumber
\end{eqnarray}

Also for the radial and meridional directions, the denominator of 
\refeqt{uXi} reduces to $2\,\sum_{ij}V^{(\Xi)}(i,j)$, though the
summation includes six terms, this time.
Therefore, the set of volume elements necessary for the interpolation 
of the radial momentum $U^{\mathrm{C}}_R$ is:
\begin{eqnarray}
V^{(R)}(i-1,j)&=&\frac{1}{3}\left[\frac{1}{8}\,(R_{i-1}+R_i)^3 - R_{i-1}^3\right]
           \nonumber \\
        & &\left[\cos{(\theta_j)}-\cos{(\theta_{j+1})}\right]\,%
         \Delta\varphi \nonumber \\
V^{(R)}(i,j)  &=&\frac{1}{24}\left[(R_{i}+R_{i+1})^3-(R_{i-1}+R_i)^3\right]
           \nonumber \\
        & &\left[\cos{(\theta_j)}-\cos{(\theta_{j+1})}\right]\,%
         \Delta\varphi \nonumber \\
V^{(R)}(i+1,j)&=&\frac{1}{3}\left[R_{i+1}^3 - \frac{1}{8}\,(R_i+R_{i+1})^3\right]
           \nonumber \\
        & &\left[\cos{(\theta_j)}-\cos{(\theta_{j+1})}\right]\,%
         \Delta\varphi \\
V^{(R)}(i-1,j+1)&=&\frac{1}{3}\left[\frac{1}{8}\,(R_{i-1}+R_i)^3 - R_{i-1}^3\right]
             \nonumber \\
          & &\left[\cos{(\theta_{j+1})}-\cos{(\theta_{j+1}+\Delta\theta)}\right]\,%
         \Delta\varphi \nonumber \\
V^{(R)}(i,j+1)&=&\frac{1}{24}\left[(R_{i}+R_{i+1})^3-(R_{i-1}+R_i)^3\right]
           \nonumber \\
        & &\left[\cos{(\theta_{j+1})}-\cos{(\theta_{j+1}+\Delta\theta)}\right]\,%
         \Delta\varphi \nonumber \\
V^{(R)}(i+1,j+1)&=&\frac{1}{3}\left[R_{i+1}^3 - \frac{1}{8}\,(R_i+R_{i+1})^3\right]
           \nonumber \\
        & &\left[\cos{(\theta_{j+1})}-\cos{(\theta_{j+1}+\Delta\theta)}\right]\,%
         \Delta\varphi.
         \nonumber
\end{eqnarray}
The group of elements involved in the updating process of the
meridional angular momentum $U^{\mathrm{C}}_\theta$ is:
\begin{eqnarray}
V^{(\theta)}(i,j-1)&=&\frac{1}{3}\left(R_{i+1}^3- R_i^3\right)\,
        \left\{\cos{(\theta_{j-1})}-\cos{[(\theta_{j-1}+\theta_j)/2]}\right\}\,%
         \Delta\varphi \nonumber \\
V^{(\theta)}(i,j)  &=&\frac{1}{3}\left(R_{i+1}^3- R_i^3\right)
           \nonumber \\
 & &\left\{\cos{[(\theta_{j-1}+\theta_j)/2]}-\cos{[(\theta_j+\theta_{j+1})/2]}\right\}\,%
         \Delta\varphi \nonumber \\
V^{(\theta)}(i,j+1)&=&\frac{1}{3}\left(R_{i+1}^3- R_i^3\right)\,
         \left\{\cos{[(\theta_j+\theta_{j+1})/2]}-\cos{(\theta_{j+1})}\right\}\,%
         \Delta\varphi \nonumber \\
V^{(\theta)}(i+1,j-1)&=&\frac{1}{3}\left[(R_{i+1}+\Delta R)^3 - R_{i+1}^3\right]
                     \\
          & &\left\{\cos{(\theta_{j-1})}-\cos{[(\theta_{j-1}+\theta_j)/2]}\right\}\,%
         \Delta\varphi \nonumber \\
V^{(\theta)}(i+1,j)&=&\frac{1}{3}\left[(R_{i+1}+\Delta R)^3 - R_{i+1}^3\right]
           \nonumber \\
 & &\left\{\cos{[(\theta_{j-1}+\theta_j)/2]}-\cos{[(\theta_j+\theta_{j+1})/2]}\right\}\,%
         \Delta\varphi \nonumber \\
V^{(\theta)}(i+1,j+1)&=&\frac{1}{3}\left[(R_{i+1}+\Delta R)^3 - R_{i+1}^3\right]
           \nonumber \\
 & &\left\{\cos{[(\theta_j+\theta_{j+1})/2]}-\cos{(\theta_{j+1})}\right\}\,%
         \Delta\varphi.
         \nonumber
\end{eqnarray}

In order to perform the interpolation of the azimuthal angular
momentum $U^{\mathrm{C}}_\varphi$, the following set of volume
elements is required:
\begin{eqnarray}
V^{(\varphi)}(i,j,k-1)&=&\frac{1}{3}\left(R_{i+1}^3 - R_{i}^3\right)\,
        \left[\cos{(\theta_j)}-\cos{(\theta_{j+1})}\right]\,%
         \Delta\varphi/2 \nonumber \\ 
V^{(\varphi)}(i+1,j,k-1)&=&\frac{1}{3}\left[(R_{i+1}+\Delta R)^3-R_{i+1}^3\right]
                 \nonumber \\
      & &\left[\cos{(\theta_j)}-\cos{(\theta_{j+1})}\right]\,%
         \Delta\varphi/2 \nonumber \\
V^{(\varphi)}(i,j+1,k-1)&=&\frac{1}{3}\left(R_{i+1}^3 - R_{i}^3\right)
                 \nonumber \\
      & &\left[\cos{(\theta_{j+1})}-\cos{(\theta_{j+1}+\Delta \theta)}\right]\,%
         \Delta\varphi/2 \nonumber \\
V^{(\varphi)}(i+1,j+1,k-1)&=&\frac{1}{3}\left[(R_{i+1}+\Delta R)^3-R_{i+1}^3\right]
         \nonumber \\
      & &\left[\cos{(\theta_{j+1})}-\cos{(\theta_{j+1}+\Delta \theta)}\right]\,%
         \Delta\varphi/2 \nonumber \\
V^{(\varphi)}(i,j,k)    &=&2\,V^{(\varphi)}(i,j,k-1) \label{Vphi} \\
V^{(\varphi)}(i,j+1,k)  &=&2\,V^{(\varphi)}(i,j+1,k-1) \nonumber \\
V^{(\varphi)}(i+1,j,k)  &=&2\,V^{(\varphi)}(i+1,j,k-1) \nonumber \\
V^{(\varphi)}(i+1,j+1,k)&=&2\,V^{(\varphi)}(i+1,j+1,k-1) \nonumber \\
V^{(\varphi)}(i,j,k+1)  &=&V^{(\varphi)}(i,j,k-1) \nonumber \\
V^{(\varphi)}(i,j+1,k+1)&=&V^{(\varphi)}(i,j+1,k-1) \nonumber \\
V^{(\varphi)}(i+1,j,k+1)&=&V^{(\varphi)}(i+1,j,k-1) \nonumber \\
V^{(\varphi)}(i+1,j+1,k+1)&=&V^{(\varphi)}(i+1,j+1,k-1). \nonumber
\end{eqnarray}
Equations~\ref{Vphi} imply that
the denominator of \refeqt{uXi} is also equivalent to
$2\,\sum_{ij}V^{(\varphi)}(i,j,k)$.

Velocities are retrieved from momenta and density. Fine-coarse
interaction also involves the correction of momentum flux components
across the boundaries between two neighboring grids.
This is accomplished via time and surface-weighted means (for details,
see \DHK).

\end{document}